\def\ion#1#2{{\rm #1}\,{\sc #2}}
\documentclass[useAMS]{mn2e}

\usepackage{supertabular}
\usepackage{subfigure}
\usepackage{lscape}
\usepackage{longtable}
\usepackage{psfig}
\usepackage{rotating}

\usepackage{graphicx}

\title[A Homogeneous Sample of Sub-DLAs I: Sample and Chemical Abundance Measurements]
{A Homogeneous Sample of Sub-DLAs I: Construction of the Sample and Chemical Abundance Measurements
\thanks{Based on ESO public data released from July 2001 obtained with UVES at the VLT Kueyen 
telescope, Paranal, Chile}
}

\author[M. Dessauges-Zavadsky et al.]
{M.~Dessauges-Zavadsky$^{1,3}$\thanks{email: miroslava.dessauges@obs.unige.ch},
C.~P\'eroux$^{2,4}$,
T.-S.~Kim$^{3,4}$,
S.~D'Odorico$^3$ and 
\newauthor
R.~G.~McMahon$^4$\\
$^1$Observatoire de Gen\`eve, 1290 Sauverny, Switzerland\\
$^2$Osservatorio Astronomico di Trieste, Via G. B. Tiepolo 11, 34131 Trieste, Italy\\
$^3$European Southern Observatory, Karl-Schwarzchild-Str. 2, 85748 Garching bei M\"unchen, Germany\\
$^4$Institute of Astronomy, Madingley Road, Cambridge CB3 0HA, UK\\
}

\begin{document}

\date{Accepted. Received}

\pagerange{\pageref{firstpage}--\pageref{lastpage}} \pubyear{2002}

\maketitle

\label{firstpage}
%

\begin{abstract}
In this first paper of a series, we report on the use of quasar spectra obtained with the UVES high 
resolution spectrograph and available through the ESO VLT archive to build the first homogeneous and
unbiased sample of sub-DLA systems, absorbers with \ion{H}{i} column densities $> 10^{19}$ cm $^{-2}$ 
but lower than the classical definition of damped Ly$\alpha$ systems (DLAs) $2\times 10^{20}$ 
cm$^{-2}$. A systematic investigation of the properties of these systems and a comparison with those 
of the DLAs (Paper~II of this series; P\'eroux et~al. 2003b) is expected to provide new clues on the 
association of high column density absorbers with galaxies and on the overall evolution of the 
neutral hydrogen gas mass and metal content in the Universe. In the spectra of 22 quasars which were 
found suitable for a sub-DLA search, we identified 12 sub-DLAs and 1 borderline case between the DLA 
and sub-DLA systems in the redshift interval $z = 1.8-4.3$. We measured the column densities of 
\ion{H}{i} and of up to 16 additional ions of low-, intermediate- and high-ionization, \ion{O}{i}, 
\ion{C}{ii}, \ion{C}{iv}, \ion{Si}{ii}, \ion{Si}{iv}, \ion{N}{i}, \ion{S}{ii}, \ion{Mg}{i}, 
\ion{Mg}{ii}, \ion{Al}{ii}, \ion{Al}{iii}, \ion{Fe}{ii}, \ion{Fe}{iii}, \ion{Ni}{ii}, \ion{Zn}{ii}, 
and \ion{Cr}{ii}. We further investigated the significance of the ionization corrections in the 
determination of the chemical abundances from the low-ionization ions in the sub-DLA \ion{H}{i} 
column density range. Using the predictions of different ion ratios as a function of the ionization 
parameter computed with the CLOUDY software package, we have estimated that with the exception of 
one case, the ionization corrections to the abundances of 9 systems for which we were able to 
constrain the ionization parameter, are lower than 0.2~dex for all of the elements except 
\ion{Al}{ii} and \ion{Zn}{ii} down to \ion{H}{i} column densities of $\log N$(\ion{H}{i}) $= 19.3$ 
cm$^{-2}$. We finally present the first sub-DLA chemical abundance database which contains the 
abundance measurements of 11 different elements (O, C, Si, N, S, Mg, Al, Fe, Ni, Zn, and Cr). We 
took advantage of the lower \ion{H}{i} column densities in sub-DLAs to measure, in particular, the O 
and C abundances using lines which are normally saturated in DLAs.
\end{abstract}
%

\begin{keywords}
galaxies: abundances -- galaxies: high-redshift -- quasars: absorption lines -- quasars. 
\end{keywords}
%

\section{Introduction}

Galaxy formation and evolution remain important issues in our understanding of the early epoch of 
the Universe. Fundamental steps towards these issues would be the reconstruction of the chemical 
histories of galaxies and the metal census of the Universe. They are investigated both via the 
traditional approach of observing the galaxy starlight, and via the absorption line systems detected 
along the quasar lines of sight independently of their distance, luminosity and morphology. The 
absorption line systems, in particular, provide a very accurate observational method to measure the 
neutral gas and metal content of the Universe up to very high redshifts, $z>4$ (e.g. Lu et~al. 1996a; 
Prochaska, Gawiser \& Wolfe 2001; Dessauges-Zavadsky et~al. 2001a; Songaila \& Cowie 2002).

The quasar absorption line systems are divided into three classes according to their neutral hydrogen
column density: the Ly$\alpha$ forest with $N$(\ion{H}{i}) ranging from $\sim 10^{12}$ to $1.6\times
10^{17}$ cm$^{-2}$, the Lyman limit systems (LLs) with $N$(\ion{H}{i}) $> 1.6\times 10^{17}$
cm$^{-2}$, and the damped Ly$\alpha$ systems (DLAs) with $N$(\ion{H}{i}) $> 2.0\times 10^{20}$ 
cm$^{-2}$ (Wolfe et~al. 1986). They thus probe different media from voids to halos and disks of both 
dwarf and normal (proto-)galaxies (Prochaska \& Wolfe 1998; Ledoux et~al. 1998; Haehnelt, Steinmetz 
\& Rauch 1998; Matteucci, Molaro \& Vladilo 1997; Jimenez, Bowen \& Matteucci 1999; Calura, Matteucci
\& Vladilo 2003). The DLAs, in particular, are a cosmologically important population, since they 
dominate the neutral hydrogen content of the Universe (e.g. Wolfe et~al. 1995; Storrie-Lombardi \& 
Wolfe 2000) and are believed to be the progenitors of the present-day galaxies.
%

\begin{table*}
\caption{Observational properties of the selected quasar spectra from the ESO UVES archives and 
their intervening DLAs and sub-DLAs}
\label{sub-DLAs}
\begin{tabular}{lclllccl}
\hline
Quasar name & $z_{\rm em}$ & ESO prog. No.       & Settings             & Wavelength coverage$^*$   & Exposure & $z_{\rm DLA}$ & $z_{\rm sub-DLA}$ \\
            &              &                     &                      & \phantom{Wavelen}\AA      & s        &               & +(Ref.)
\\
\hline
Q0000$-$2620       & 4.11 & Commissioning	 & B437+R860            & 3770$-$5000, 6700$-$10500 & 19700           & 3.390 & ...   \\
APM\,BR\,J0307$-$4945& 4.75 & Commissioning	 & R600, R760, R800     & 5000$-$9900               & 26300	      & 4.466 & ?$^a$   \\
Q0347$-$3819       & 3.23 & Commissioning	 & B427+R860            & 3655$-$4888, 6722$-$10000 & \phantom{0}9500 & 3.025 & ...   \\
Q0841$+$129        & 2.50 & 65.O-0063		 & B412+R860            & 3550$-$4740, 6710$-$10000 & \phantom{0}5400 & 2.375 & ...   \\
...                & ...  & ... 		 &  ...			& ...                       & ...	      & 2.476 & ...   \\
HE 0940$-$1050     & 3.05 & 65.O-0474		 & B346+R580            & 3600$-$3870, 4780$-$6809  & \phantom{0}3600 & ...   & ...   \\
Q1038$-$272        & 2.32 & 65.O-0063		 & B437+R860            & 3765$-$4998, 6707$-$10000 & \phantom{0}5400 & ...   & ...   \\
Q1101$-$264$^b$    & 2.14 & Science Verification & B346+R580, B437+R860 & 3050$-$8530               & 45000           & ...   & 1.838$^b$ (1)\\
HE 1104$-$1805$^c$ & 2.31 & Commissioning	 & B360, R670           & 3184$-$4024, 5698$-$7714  & 10880	      & 1.662 & ...   \\
Q1151$+$068        & 2.76 & 65.O-0158		 & B346+R580, B437+R860 & 3100$-$10000              & 21600	      & 1.774 & ...   \\  
PKS 1157$+$014     & 1.99 & 65.O-0063		 & B346+R550            & 3050$-$3870, 4500$-$6500  & \phantom{0}7200 & 1.944 & ...   \\
Q1223$+$1753       & 2.94 & 65.P-0038, 65.O-0158 & B390+R564, B437+R860 & 3300$-$10000              & 18000	      & 2.466 & 2.557 \\
PKS 1232$+$0815    & 2.57 & 65.P-0038		 & B390+R564            & 3300$-$6650               & 10800 	      & 2.338 & ...   \\ 
Q1409$+$095        & 2.86 & 65.O-0158		 & B437+R860            & 3769$-$4998, 6707$-$10000 & 18000           & 2.456 & 2.668 (2,3,4)\\
Q1444$+$014        & 2.21 & 65.O-0158		 & B390+R564            & 3294$-$6650               & 18000           & ...   & 2.087 (3,5)\\
Q1451$+$123        & 3.25 & 65.O-0063		 & B424+R750            & 3640$-$4855, 5900$-$9400  & \phantom{0}7200 & 2.469 & 2.255$^d$ (2)\\
...                & ...  & ... 		 &  ...			& ...                       & ...	      & ...   & 3.171$^e$ (1,2)\\
Q1511$+$090        & 2.88 & 65.P-0038		 & B390+R564, B437+R860 & 3300$-$10000              & 32400           & ...   & 2.088 \\
Q2059$-$360        & 3.09 & 65.O-0063		 & B480+R740            & 4200$-$5340, 5900$-$9300  & 10800           & 3.083 & 2.507 \\
Q2116$-$358        & 2.34 & 65.O-0158		 & B390+R564            & 3294$-$6650               & \phantom{0}7200 & ...   & 1.996 (6,7,8)\\
Q2132$-$433        & 2.42 & 65.O-0158		 & B390+R564            & 3294$-$6650               & \phantom{0}3600 & 1.914 & ...   \\
PSS J2155$+$1358   & 4.26 & 65.O-0296		 & R565, R800           & 4633$-$9800               & 11600	      & 3.316 & 3.142 \\
...                & ...  & ... 		 &  ...			& ...	                    &...	      & ...   & 3.565 \\
...                & ...  & ... 		 &  ...			& ...	                    &...	      & ...   & 4.212 \\
Q2206$-$199        & 2.56 & 65.O-0158		 & B390+R564, B437+R860 & 3294$-$10000              & 25200           & 1.920 & ...   \\
...                & ...  & ... 		 &  ...			& ...                       & ...	      & 2.076 & ...   \\
PSS J2344$+$0342   & 4.24 & 65.O-0296		 & R565, R800           & 4633$-$9800               & 13500	      & 3.220 & 3.882 \\
\hline		       
\end{tabular}
\begin{minipage}{140mm}
$^*$ With some gaps in the wavelength coverage. \\
$^a$ The detection of sub-DLAs is uncertain (see Section~\ref{Q0307}).\\
$^b$ This quasar has been observed for the study of the Ly$\alpha$ forest. Nevertheless, an analysis 
of the spectra revealed the presence of a sub-DLA. For consistency reasons, this system will {\it 
not} be included in the study of the sub-DLA statistical properties presented in Paper~II. \\
$^c$ This quasar is gravitationally lensed. Only the brightest line of sight was included in our 
study. \\
$^d$ This absorption line system is a borderline case between the DLA and sub-DLA systems with a
\ion{H}{i} column density of $\log N$(\ion{H}{i}) $= 20.30\pm 0.15$. It is not included in the sub-DLA
sample. \\
$^e$ The spectra do {\it not} cover the Ly$\alpha$ absorption line at $\lambda_{\rm obs} \sim 5070$ 
\AA\ of this sub-DLA. It has been identified thanks to the \ion{H}{i} column density measurement
reported by Petitjean, Srianand \& Ledoux (2000). \\
{\bf References to the sub-DLAs already reported in the literature:} \\
(1)~Petitjean, Srianand \& Ledoux (2000); (2)~Lanzetta et~al. (1991); (3)~Pettini et~al. (2002); 
(4)~Lu et~al. (1993); (5)~Wolfe et~al. (1995); (6)~Lanzetta, Wolfe \& Turnshek (1987); 
(7)~Wampler, Bergeron \& Petitjean (1993); (8)~M\o ller, Jakobsen \& Perryman (1994) \\
{\bf Observing programmes:} \\ 
65.P-0038: ``Probing the nuclear regions of QSOs with $z_{\rm abs} \simeq z_{\rm em}$ associated 
systems'' \\
\phantom{65} (Srianand, Petitjean \& Aracil) \\
65.O-0063: ``Molecular hydrogen at high redshift'' (Ledoux, Petitjean, Srianand \& Rauch) \\
65.O-0158: ``Abundance ratios in Damped Ly$\alpha$ systems: clues to the nucleosynthesis of N and 
Mn'' \\
\phantom{65} (Pettini, Bergeron \& Petitjean) \\
65.O-0296: ``A study of the IGM-galaxy connection at $z=3$ using the UVES high quality absorption 
spectra'' \\
\phantom{65} (D'Odorico, Cristiani, Fontana, Giallongo \& Savaglio) \\
65.O-0474: ``Deuterium abundance in the high redshift QSO absorption system towards HE 0940$-$1050'' \\
\phantom{65} (Molaro, Centuri\'on \& Bonifacio) 
\end{minipage}
\end{table*}
%

A recent study by P\'eroux et~al. (2003a) has, however, suggested that the relative importance of
lower \ion{H}{i} column density systems, the so-called `sub-DLA systems' with \ion{H}{i} column
densities between $10^{19}$ and $2\times 10^{20}$ cm$^{-2}$, increases with redshift. Indeed, this 
until now poorly studied class of absorbers should contain about 45\% of the neutral hydrogen mass 
at $z>3.5$, and may thus play an important role at high redshift in the study of the redshift 
evolutions of both the total amount of neutral gas and the metal content in the Universe. The 
sub-DLAs, similarly to the DLAs, present the advantage that the \ion{H}{i} column density of these 
systems can accurately be estimated thanks to the presence of damping wings in their Ly$\alpha$ 
lines, observed down to \ion{H}{i} column densities of $10^{19}$ cm$^{-2}$.

At the present time only the chemical abundances of DLAs have been used to trace the metal content 
in the Universe, but this has shown unexpected results. Contrary to the cosmic chemical evolution 
model predictions (e.g Pei \& Fall 1995; Pei, Fall \& Hauser 1999; Cen \& Ostriker 1999; Cen et~al. 
2003), the most recent observations indicate only a mild evolution of the metal content with redshift 
(Pettini et~al. 1999; Prochaska \& Wolfe 2000; Vladilo et~al. 2000; Dessauges-Zavadsky et~al. 2001b; 
Prochaska \& Wolfe 2002). Only Savaglio (2000), who focused on lower \ion{H}{i} column density DLA 
systems, claims evidence for redshift evolution. If the P\'eroux et~al. (2003a) predictions are
confirmed, the sub-DLA metallicities should be taken into account in order to obtain a complete 
picture of the redshift evolution of the Universe's metal content. Indeed, the current metallicity 
studies which focused only on the higher column density absorption line systems, may have provided a 
biased and/or incomplete view of the global metal evolution in the Universe at $z>3.5$. The 
observational confirmation of the impact of the sub-DLA systems at high redshift thus is crucial. In 
addition, the investigations of the chemical, kinematic and clustering properties of the sub-DLAs 
to see whether we are dealing with a different class of objects with a lower \ion{H}{i} content/mass 
relative to the DLAs are our further motivations for studying these absorbers. If sub-DLAs are the 
basic building blocks of the hierarchical growth of galactic structures as suggested by P\'eroux 
et~al. (2003a), their abundance analysis may provide an insight into the very early stages of the 
chemical enrichment of galaxies.

The work presented in this paper therefore aims at constructing the first homogeneous sample of 
sub-DLA systems. For this purpose, we have used all of the high-resolution quasar spectra available 
to us in July 2001 in the ESO UVES-VLT archives. Section~2 explains how the quasar spectra were 
selected and how the sub-DLA systems were identified. We provide a detailed chemical analysis for 
each sub-DLA system in Section~3, and we address the issues of the photoionization effects in 
Section~4. Finally, in Section~5, we summarize all the ionic column density and the absolute 
abundance measurements. In Paper~II of this series (P\'eroux et~al. 2003b), we use the database
presented here in combination with the data on DLAs and other sub-DLAs from the literature to 
discuss the overall properties of these objects.
%

\section{Construction of a homogeneous sub-DLA sample}

All the observations presented in this paper were acquired with the Ultraviolet-Visual Echelle 
Spectrograph (UVES; see D'Odorico et~al. 2000) on the VLT 8.2\,m ESO telescope at Paranal, Chile, 
between the Commissioning and the Science Verification time of the instrument at the end of 1999 and 
beginning of 2000, respectively, and July 2000. All these data reached in July 2001 the end of their 
one-year long propriety period and could be retrieved from the ESO archives. A total of 35 high 
resolution, high signal-to-noise ratio (S/N) quasar spectra was available, illustrating the success 
of both the instrument and the data distribution policy of the ESO archives.
%

\subsection{Quasar Selection}\label{QSO-selection}

The goal of our study being well defined, we have selected from the 35 quasars available in the 
ESO UVES archives only the quasars suitable for the sub-DLA search. First, we took the conservative 
approach of excluding all quasars exhibiting Broad Absorption Lines (BALs), since the nature of these 
objects and their characteristic features are not fully understood yet, rendering the distinction 
between true intervening absorption systems and other quasar spectral features difficult. However, we 
have not excluded the quasars selected for the study of associated systems, $z_{\rm abs}\simeq z_{\rm 
em}$, since the quasars considered in the ESO programme 65.P-0038 (see Table~\ref{sub-DLAs}) do not 
exhibit BALs which are often confused with associated absorbers. Secondly, we have excluded the 
quasars which were observed for the sole purpose of studying the Ly$\alpha$ forest (e.g. for the 
analysis of voids, low column density distribution). Indeed, these quasars have been pre-selected for 
having no high column density systems in their spectrum, precisely the type of systems we are looking 
for in the present study. In one case, however, a sub-DLA has been found in the later sample of 
quasars, the quasar Q1101$-$264 at $z_{\rm em} = 2.12$ observed during the UVES Science Verification. 
This absorber is included in the present abundance analysis, but for consistency reasons it will be 
excluded from the study of the sub-DLA statistical properties (see Paper~II).

The above selection process resulted in a sample of 22 quasars. A large majority of these quasars
were observed with UVES because of their intervening DLA systems. There is {\it a priori} no reason
why this should create a bias in our sample. We assume that the sub-DLA systems are not 
preferentially found close to the DLA systems. An analysis of the clustering properties of the 
sub-DLA systems identified in our sample is presented in Paper~II. 

Table~\ref{sub-DLAs} summarizes the observing programme details (i.e. the ESO programme number, the 
exposure time, the settings), the wavelength coverages, the redshifts of the 22 selected quasars 
and the redshifts of their intervening DLA systems. The sample of the selected quasars should result
in the construction of an unbiased and homogeneous sample of sub-DLA systems.
%

\subsection{Data Reduction}
%

\begin{figure*}
   \includegraphics[width=165mm]{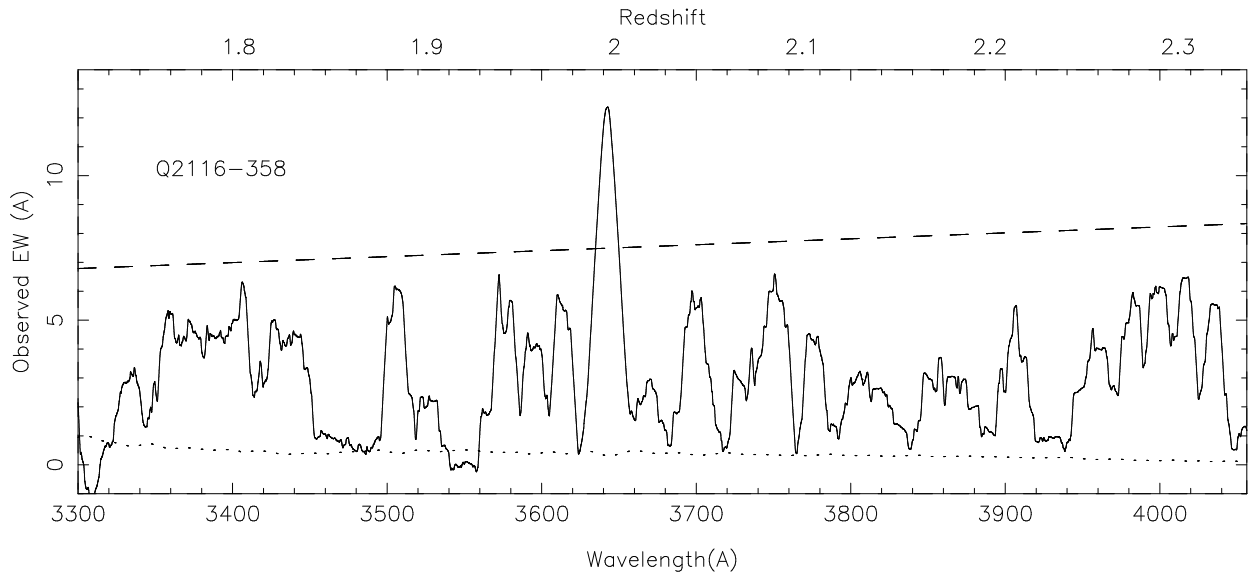}
\caption{Output of the automated sub-DLA search algorithm for the quasar Q2116$-$358. The solid line 
corresponds to the equivalent width spectrum and the dotted line to the error array. The straight 
dashed line represents the threshold above which the sub-DLA systems are detected. In this case a 
sub-DLA at $\sim 3640$ \AA\ ($z_{\rm abs} = 1.996$) is clearly identified.}
\label{sub-DLA-search}
\end{figure*}
%

The constructed sample of 22 quasars has benefited from a homogeneous data reduction. All of the 
quasar spectra were reduced using the ESO data reduction package {\tt MIDAS} and the UVES pipeline 
in an interactive mode available as a {\tt MIDAS} context. A detailed description of the pipeline 
can be found in Ballester et~al. (2000). To optimize the results we made a systematic check of each 
step of the pipeline reduction. Once reduced, the wavelengths of the one-dimensional spectra were 
converted to a vacuum-heliocentric scale, the individual spectra were co-added using their S/N ratio 
as weights to get the maximum signal-to-noise ratio, and the various settings were combined to 
obtain the maximum wavelength spectral coverage. An average resolution of $5-7$ km~s$^{-1}$ FWHM was 
achieved in the spectra, and an average S/N per pixel between 10 and 150 is observed from spectrum 
to spectrum according to the exposure time and the magnitude of the quasar.

The final step was the normalization of the resulting spectra obtained by dividing them by a spline 
function fitted to smoothly connect the regions free from absorption features. The continuum in the 
Ly$\alpha$ forest was fitted by using small regions deemed to be free of absorptions and by 
interpolating between these regions with a spline. The higher the redshift of the quasar, the more 
difficult is the normalization in the Ly$\alpha$ forest due to the higher density of absorption 
lines (e.g. Kim, Cristiani \& D'Odorico 2001). The procedure we applied for the normalization is the 
same as the one we used for two high-redshift and four intermediate-redshift quasars we have 
previously analysed for a detailed chemical abundance study of their intervening DLA systems 
(Dessauges-Zavadsky et~al. 2001a; Levshakov et~al. 2002; Dessauges-Zavadsky et~al. 2002). This 
procedure has proven to lead to an accuracy of the normalization level of about $5-10$\% in the 
Ly$\alpha$ forest and better than 5\% redwards of the Ly$\alpha$ emission of the quasar (according to 
the signal-to-noise ratio of the spectra). For each quasar we tested the derived normalization first 
by selecting different regions in the continuum and by varying the number of these regions used for 
the fit of the continuum with a spline. In the Ly$\alpha$ forest, in particular, some regions are 
key regions in the continuum definition, when omitted, the shape and level of the continuum are 
drastically modified. We tried to qualitatively and quantitatively estimate these modifications. 
Secondly, we checked whether the adopted continuum fit is within the 1~$\sigma$ error of the spectra 
along the whole wavelength range sampled, and we estimated the normalization accuracy by determining 
the mean deviation of the adopted continuum fit from the observed continuum within its 1~$\sigma$ 
error and the maximum variation of this fit tolerated within the 1~$\sigma$ error of the spectra. 
%

\subsection{Sub-DLA Identification}\label{sub-DLA-ident}

The sample of quasars to be analysed being defined and their high resolution spectra being reduced,
we used these spectra to identify their intervening sub-DLA systems. In order to select all the
sub-DLAs, we have used a detection algorithm equivalent to the one applied by P\'eroux et~al. (2001) 
to detect the DLAs in medium resolution spectra ($\sim 2$ \AA\ pixel$^{-1}$), and which was 
previously used by Lanzetta et~al. (1991), Wolfe et~al. (1995), and Storrie-Lombardi \& Wolfe (2000). 
This technique consists in building an equivalent width spectrum over 500 pixel wide boxes for each 
quasar. The analysis was done 3000 km~s$^{-1}$ bluewards from the Ly$\alpha$ emission of the quasar 
in order to avoid possible contaminations from systems associated with the quasar itself, and it 
ended where the S/N was too low to detect absorption features at the sub-DLA threshold, i.e. ${\rm 
EW}_{\rm rest} = 2.5$ \AA. 

Fig.~\ref{sub-DLA-search} displays an example of the output of the sub-DLA search algorithm in the 
case of the quasar Q2116$-$358. The solid line is the equivalent width spectrum and the dotted line 
is the error array. The straight dashed line corresponds to the threshold above which the sub-DLA
systems are detected. In this example, the sub-DLA system at $\sim 3640$ \AA\ ($z_{\rm abs} = 1.996$) 
is clearly identified.
%

\begin{figure}
   \includegraphics[width=85mm]{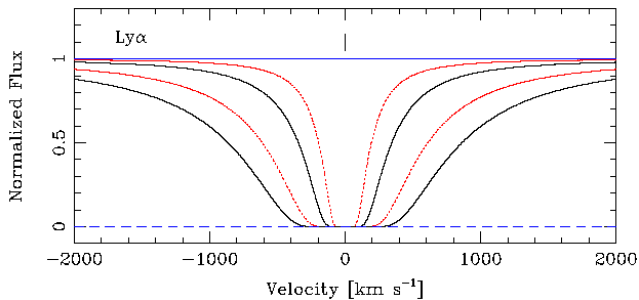}
\caption{Ly$\alpha$ absorption line profiles for four different \ion{H}{i} column densities in the 
sub-DLA range: from left to right, $\log N$(\ion{H}{i}) $= 20.3$ (solid line), 20.0 (dotted line), 
19.5 (solid line) and 19.0 (dotted line) with the Doppler parameter $b$ fixed at 20 km~s$^{-1}$. This 
illustrates that in the sub-DLA range, $10^{19} < N$(\ion{H}{i}) $< 2\times 10^{20}$ cm$^{-2}$, the 
damping wings are clearly observed.}
\label{sub-DLA-HI}
\end{figure}
%

We then supplemented the sub-DLA candidates found by the algorithm with a visual identification. 
Indeed, since the algorithm selects any region with the required equivalent width regardless of the 
shape of the absorption feature and whether or not the absorption goes to zero, an additional check 
of the candidates is necessary to distinguish the true sub-DLAs from blends of several Ly$\alpha$ 
clouds. Three main visual criteria were used to identify a sub-DLA system: {\it i)}~the saturation 
of the absorption feature identified as the Ly$\alpha$ absorption line of the sub-DLA candidate, 
{\it ii)}~the presence of damping wings in the Ly$\alpha$ absorption lines which are clearly 
observed in the sub-DLA \ion{H}{i} column density regime between $10^{19}$ and $2\times 10^{20}$ 
cm$^{-2}$ as illustrated in Fig.~\ref{sub-DLA-HI}, and {\it iii)}~the detection of lines of higher 
members of the Lyman series with the same \ion{H}{i} column density as the one measured from the 
Ly$\alpha$ line when the wavelength coverage is available. An additional visual criterion, though 
not necessary, was the identification of associated metal lines which are very likely to occur in 
absorbers with \ion{H}{i} column densities greater than $10^{19}$ cm$^{-2}$. On the basis of these 
visual criteria and to impartially select the sub-DLAs, three of us (M.D.-Z., C.P., T.-S.K.) looked 
{\it independently} at the sub-DLA candidates found by the algorithm and closely inspected the 
original spectra to determine whether or not these absorption features correspond to sub-DLAs. 

We reached a good agreement in all but one case, the quasar APM BR J0307$-$4945 at $z_{\rm em} = 
4.78$. Given the high redshift of this quasar the Ly$\alpha$ forest is very dense, and thus at three 
different redshifts the presence of sub-DLA systems is ambiguous (see Section~\ref{Q0307}). The 
absence of lines of higher members of the Lyman series for two of them does not allow us to draw a
conclusion on the true \ion{H}{i} column density of these systems, leading us to not include them in 
the sub-DLA sample. In the third candidate, the Lyman lines are heavily blended and prevent us from 
obtaining an accurate \ion{H}{i} column density measurement, thus this candidate is also not 
analysed and not included in our sub-DLA sample. Moreover, the sub-DLA system at $z_{\rm abs} = 
3.171$ towards Q1451+123 has been identified thanks to the \ion{H}{i} column density measurement 
reported by Petitjean, Srianand \& Ledoux (2000), the spectra we had at our disposal do not cover 
the Ly$\alpha$ line of this system. With the lines of higher members of the Lyman series $-$ 
Ly$\beta$, Ly$\gamma$ and Ly8~$-$ we confirm the reported \ion{H}{i} column density measurement of 
this absorption system (see Section~\ref{Q1451-3p171}). This system is included in the abundance 
analysis, but it is excluded from the study of the sub-DLA statistical properties (see Paper~II), 
similarly to the sub-DLA system towards Q1101$-$264 (see the explanation in 
Section~\ref{QSO-selection}). Finally, the identified absorption line system at $z_{\rm abs} = 
2.255$ towards Q1451+123 is a borderline case between the DLA and sub-DLA systems with a \ion{H}{i} 
column density of $\log N$(\ion{H}{i}) $= 20.3\pm 0.15$ (see Section~\ref{Q1451-2p255}). We provide 
a detailed abundance analysis of this system, but we do not consider it as a sub-DLA system, and 
thus we do not include it in the sub-DLA sample used for the different studies undertaken in 
Paper~II.
%

\begin{figure}
\centering
   \includegraphics[width=70mm]{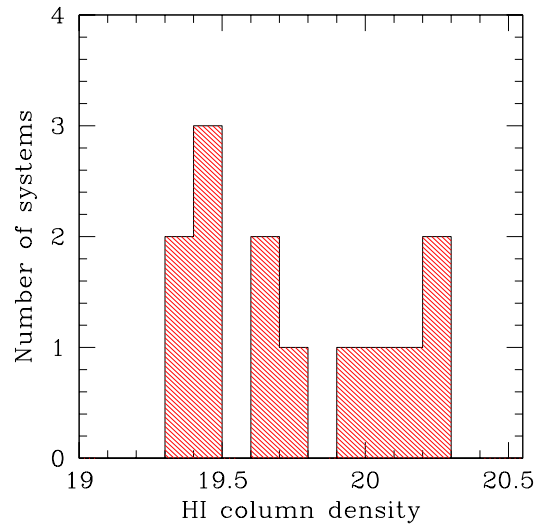}
\caption{\ion{H}{i} column density distribution of the 12 sub-DLA systems and 1 borderline case
between the DLA and sub-DLA systems detected towards the 22 selected quasars.}
\label{HI-distribution}
\end{figure}
%

In Table~\ref{sub-DLAs} we give the redshifts and the references of the 12 sub-DLAs and 1 borderline 
case between the DLA and sub-DLA systems detected towards the 22 selected quasars, and in 
Fig.~\ref{HI-distribution} we show their \ion{H}{i} column density distribution. The detected 
systems span a range of \ion{H}{i} column densities from $10^{19.3}$ to $10^{20.3}$ cm$^{-2}$. No 
particular trend can be confirmed with this small number of objects yet. The sample of the 10 
sub-DLAs, when excluding the sub-DLA towards Q1101$-$264 and towards Q1451+123, constitutes an 
unbiased and homogeneous sample of sub-DLAs, ideal for a statistical analysis (see Paper~II).
%

\begin{table*}
\caption{Component structure of the $z_{\rm abs} = 1.838$ sub-DLA system towards Q1101$-$264} 
\label{Q1101-values}
\begin{tabular}{l c c c l c | l c c c l c}
\hline
No & $z_{\rm abs}$ & $v_{\rm rel}^*$ & $b (\sigma _b)$ & Ion & $\log N (\sigma_{\log N})$ & No & $z_{\rm abs}$ & $v_{\rm rel}^*$ & $b (\sigma _b)$ & Ion & $\log N (\sigma_{\log N})$ \\
   &               & km~s$^{-1}$     & km~s$^{-1}$     &     & cm$^{-2}$                  &    &               & km~s$^{-1}$     & km~s$^{-1}$     &     & cm$^{-2}$                  
\\     
\hline
\multicolumn{4}{l}{Low-ion transitions} & & & & & & & & \\
\hline
1 & 1.837407 & $-$158           & 3.0{\scriptsize (0.9)} & \ion{C}{ii}  & 12.30{\scriptsize (0.03)} & 8  & 1.838902 & \phantom{0}0 & 5.3{\scriptsize (0.1)} & \ion{Si}{ii} & 13.61{\scriptsize (0.01)} \\
  &          &                  &                        & \ion{Mg}{ii} & 11.45{\scriptsize (0.04)} &	 &	    &		   &			    & \ion{O}{i}   & 14.10{\scriptsize (0.15)} \\
2 & 1.837507 & $-$147           & 4.3{\scriptsize (0.5)} & \ion{C}{ii}  & 12.65{\scriptsize (0.01)} &	 &	    &		   &			    & \ion{S}{ii}  & 13.23{\scriptsize (0.08)} \\
  &          &                  &                        & \ion{Mg}{ii} & 11.70{\scriptsize (0.04)} &	 &	    &		   &			    & \ion{C}{ii}  & 14.70{\scriptsize (0.08)} \\
3 & 1.837711 & $-$126           & 6.9{\scriptsize (0.1)} & \ion{Si}{ii} & 12.23{\scriptsize (0.05)} &	 &	    &		   &			    & \ion{Al}{ii} & 12.26{\scriptsize (0.05)} \\
  &          &                  &                        & \ion{C}{ii}  & 13.29{\scriptsize (0.02)} &	 &	    &		   &			    & \ion{Fe}{ii} & 13.14{\scriptsize (0.01)} \\
  &          &                  &                        & \ion{Al}{ii} & 11.50{\scriptsize (0.05)} &	 &	    &		   &			    & \ion{Mg}{ii} & 13.69{\scriptsize (0.02)} \\
  &          &                  &                        & \ion{Fe}{ii} & 11.48{\scriptsize (0.04)} &	 &	    &		   &			    & \ion{Mg}{i}  & 11.49{\scriptsize (0.01)} \\
  &          &                  &                        & \ion{Mg}{ii} & 12.40{\scriptsize (0.02)} & 9  & 1.839144 & $+$26	   & 8.4{\scriptsize (0.3)} & \ion{Si}{ii} & 12.86{\scriptsize (0.02)} \\
4 & 1.838201 & $-$74            & 7.8{\scriptsize (0.2)} & \ion{Si}{ii} & 12.55{\scriptsize (0.03)} &	 &	    &		   &			    & \ion{O}{i}   & 13.80{\scriptsize (0.11)} \\
  &	     &  	        &                        & \ion{C}{ii}  & 13.44{\scriptsize (0.08)} &	 &	    &		   &			    & \ion{C}{ii}  & 13.55{\scriptsize (0.05)} \\
  &	     &  	        &                        & \ion{Al}{ii} & 11.70{\scriptsize (0.02)} &	 &	    &		   &			    & \ion{Al}{ii} & 11.61{\scriptsize (0.04)} \\
  &	     &  	        &                        & \ion{Fe}{ii} & 11.82{\scriptsize (0.02)} &	 &	    &		   &			    & \ion{Fe}{ii} & 12.58{\scriptsize (0.04)} \\
  &	     &  	        &                        & \ion{Mg}{ii} & 12.59{\scriptsize (0.01)} &	 &	    &		   &			    & \ion{Mg}{ii} & 12.64{\scriptsize (0.04)} \\
5 & 1.838326 & \phantom{0}$-$61 & 5.9{\scriptsize (0.2)} & \ion{Si}{ii} & 12.49{\scriptsize (0.03)} &	 &	    &		   &			    & \ion{Mg}{i}  & 11.01{\scriptsize (0.02)} \\
  &	     &  	        &                        & \ion{C}{ii}  & 13.24{\scriptsize (0.05)} & 10 & 1.839249 & $+$37	   & 2.0{\scriptsize (0.3)} & \ion{Si}{ii} & 12.46{\scriptsize (0.03)} \\
  &	     &  	        &                        & \ion{Al}{ii} & 11.46{\scriptsize (0.03)} &	 &	    &		   &			    & \ion{O}{i}   & 13.34{\scriptsize (0.11)} \\
  &	     &  	        &                        & \ion{Fe}{ii} & 11.67{\scriptsize (0.02)} &	 &	    &		   &			    & \ion{C}{ii}  & 13.12{\scriptsize (0.04)} \\
  &	     &  	        &                        & \ion{Mg}{ii} & 12.30{\scriptsize (0.02)} &	 &	    &		   &			    & \ion{Al}{ii} & 11.20{\scriptsize (0.03)} \\
6 & 1.838542 & \phantom{0}$-$38 & 6.0{\scriptsize (0.1)} & \ion{Si}{ii} & 13.37{\scriptsize (0.01)} &	 &	    &		   &			    & \ion{Fe}{ii} & 12.20{\scriptsize (0.04)} \\
  &	     &  	        &                        & \ion{O}{i}   & 13.77{\scriptsize (0.02)} &	 &	    &		   &			    & \ion{Mg}{ii} & 12.36{\scriptsize (0.04)} \\
  &	     &  	        &                        & \ion{S}{ii}  & 13.13{\scriptsize (0.10)} &	 &	    &		   &			    & \ion{Mg}{i}  & 10.88{\scriptsize (0.03)} \\
  &	     &  	        &                        & \ion{C}{ii}  & 14.60{\scriptsize (0.08)} & 11 & 1.839334 & $+$46	   & 5.2{\scriptsize (0.1)} & \ion{Si}{ii} & 12.48{\scriptsize (0.03)} \\
  &	     &  	        &                        & \ion{Al}{ii} & 12.29{\scriptsize (0.08)} &	 &	    &		   &			    & \ion{O}{i}   & 13.30{\scriptsize (0.10)} \\
  &	     &  	        &                        & \ion{Fe}{ii} & 12.83{\scriptsize (0.01)} &	 &	    &		   &			    & \ion{C}{ii}  & 13.29{\scriptsize (0.03)} \\
  &	     &  	        &                        & \ion{Mg}{ii} & 13.58{\scriptsize (0.06)} &	 &	    &		   &			    & \ion{Al}{ii} & 11.50{\scriptsize (0.02)} \\
  &	     &  	        &                        & \ion{Mg}{i}  & 11.29{\scriptsize (0.02)} &	 &	    &		   &			    & \ion{Fe}{ii} & 11.97{\scriptsize (0.02)} \\
7 & 1.838712 & \phantom{0}$-$20 & 8.0{\scriptsize (0.3)} & \ion{Si}{ii} & 13.19{\scriptsize (0.01)} &	 &	    &		   &			    & \ion{Mg}{ii} & 12.40{\scriptsize (0.01)} \\
  &	     &  	        &                        & \ion{O}{i}   & 13.83{\scriptsize (0.02)} & & & & & & \\
  &	     &  	        &                        & \ion{S}{ii}  & 13.19{\scriptsize (0.15)} & & & & & & \\
  &	     &  	        &                        & \ion{C}{ii}  & 13.99{\scriptsize (0.06)} & & & & & & \\
  &	     &  	        &                        & \ion{Al}{ii} & 12.10{\scriptsize (0.08)} & & & & & & \\
  &	     &  	        &                        & \ion{Fe}{ii} & 12.64{\scriptsize (0.01)} & & & & & & \\
  &	     &  	        &                        & \ion{Mg}{ii} & 13.08{\scriptsize (0.02)} & & & & & & \\
  &	     &  	        &                        & \ion{Mg}{i}  & 10.97{\scriptsize (0.02)} & & & & & & \\
\hline
\multicolumn{4}{l}{Intermediate-ion transitions} & & & & & & & & \\
\hline 
1 & 1.838240 & $-$70 & 13.0{\scriptsize (1.4)} & \ion{Al}{iii} & 11.75{\scriptsize (0.04)} & 3 & 1.838865 & $-$4 & 14.5{\scriptsize (1.2)} & \ion{Al}{iii} & 11.82{\scriptsize (0.04)} \\
2 & 1.838574 & $-$35 & 14.9{\scriptsize (1.3)} & \ion{Al}{iii} & 11.99{\scriptsize (0.03)} & & & & & & \\
\hline
\multicolumn{4}{l}{High-ion transitions} & & & & & & & & \\
\hline 
1 & 1.837500 & $-$148           &           10.5{\scriptsize (0.4)} & \ion{Si}{iv} & 12.59{\scriptsize (0.02)} & 4 & 1.838591 & $-$33		&	    14.5{\scriptsize (0.2)} & \ion{Si}{iv} & 13.34{\scriptsize (0.03)} \\
  & 1.837485 & $-$150           & \phantom{0}9.2{\scriptsize (0.2)} & \ion{C}{iv}  & 12.86{\scriptsize (0.03)} &   & 1.838570 & $-$35		&	    16.0{\scriptsize (0.2)} & \ion{C}{iv}  & 13.45{\scriptsize (0.02)} \\
2 & 1.837715 & $-$125           & \phantom{0}8.5{\scriptsize (0.1)} & \ion{Si}{iv} & 13.02{\scriptsize (0.02)} & 5 & 1.838942 & \phantom{0}$+$4 &	    15.6{\scriptsize (0.2)} & \ion{Si}{iv} & 13.13{\scriptsize (0.04)} \\
  & 1.837718 & $-$125           &           12.1{\scriptsize (0.1)} & \ion{C}{iv}  & 13.41{\scriptsize (0.02)} &   & 1.838970 & \phantom{0}$+$7 &	    14.7{\scriptsize (0.1)} & \ion{C}{iv}  & 13.76{\scriptsize (0.03)} \\
3 & 1.838210 & \phantom{0}$-$73 &           15.5{\scriptsize (0.2)} & \ion{Si}{iv} & 13.22{\scriptsize (0.12)} & 6 & 1.839337 & $+$46		& \phantom{0}9.2{\scriptsize (0.4)} & \ion{Si}{iv} & 12.40{\scriptsize (0.02)} \\
  & 1.838176 & \phantom{0}$-$77 &           20.7{\scriptsize (0.2)} & \ion{C}{iv}  & 13.62{\scriptsize (0.08)} &   & 1.839365 & $+$49		&	    17.6{\scriptsize (0.8)} & \ion{C}{iv}  & 12.67{\scriptsize (0.03)} \\  
\hline
\end{tabular}
\begin{minipage}{140mm}
$^*$ Velocity relative to $z = 1.838902$
\end{minipage}
\end{table*}
%

\section{Ionic Column Densities}

In this Section we present the ionic column density measurements of the 12 detected sub-DLA 
systems and 1 borderline case between the DLA and sub-DLA systems. All of the column densities of 
the metal species were derived with the Voigt profile fitting technique. This technique consists in 
fitting theoretical Voigt profiles to the observed absorption metal lines well described as a 
complex of components, each defined by a redshift $z$, a Doppler parameter $b$, a column density $N$ 
and the corresponding errors. The fits were performed using a $\chi^2$ minimization routine {\tt 
fitlyman} in {\tt MIDAS} (Fontana \& Ballester 1995).

We assumed that all of the metal species with {\it similar ionization potentials} can be fitted using 
identical component fitting parameters, i.e. the same $b$ (which means that macroturbulent motions 
dominate over thermal broadening) and the same $z$ in the same component, and allowing for variations 
from metal species to metal species in $N$ only. We distinguish three categories of metal species 
with similar ionization potentials: the low-ion transitions (i.e. the neutral and singly ionized 
species), the intermediate-ion transitions (e.g. \ion{Fe}{iii}, \ion{Al}{iii}) and the high-ion 
transitions (e.g. \ion{C}{iv}, \ion{Si}{iv}). By using relatively strong lines (but not saturated) 
to fix the component fitting parameters (the $b$ and $z$ values for each component), we then obtain 
excellent fit results even for weak metal lines and for metal lines in the Ly$\alpha$ forest where 
the probability of blending is high, by allowing only the column density to vary. For all of the 
analysed systems exhibiting multicomponent velocity structures, we had a sufficient number of 
relatively strong metal lines to constrain well the fitting parameters.

\begin{figure*}
   \includegraphics[width=170mm]{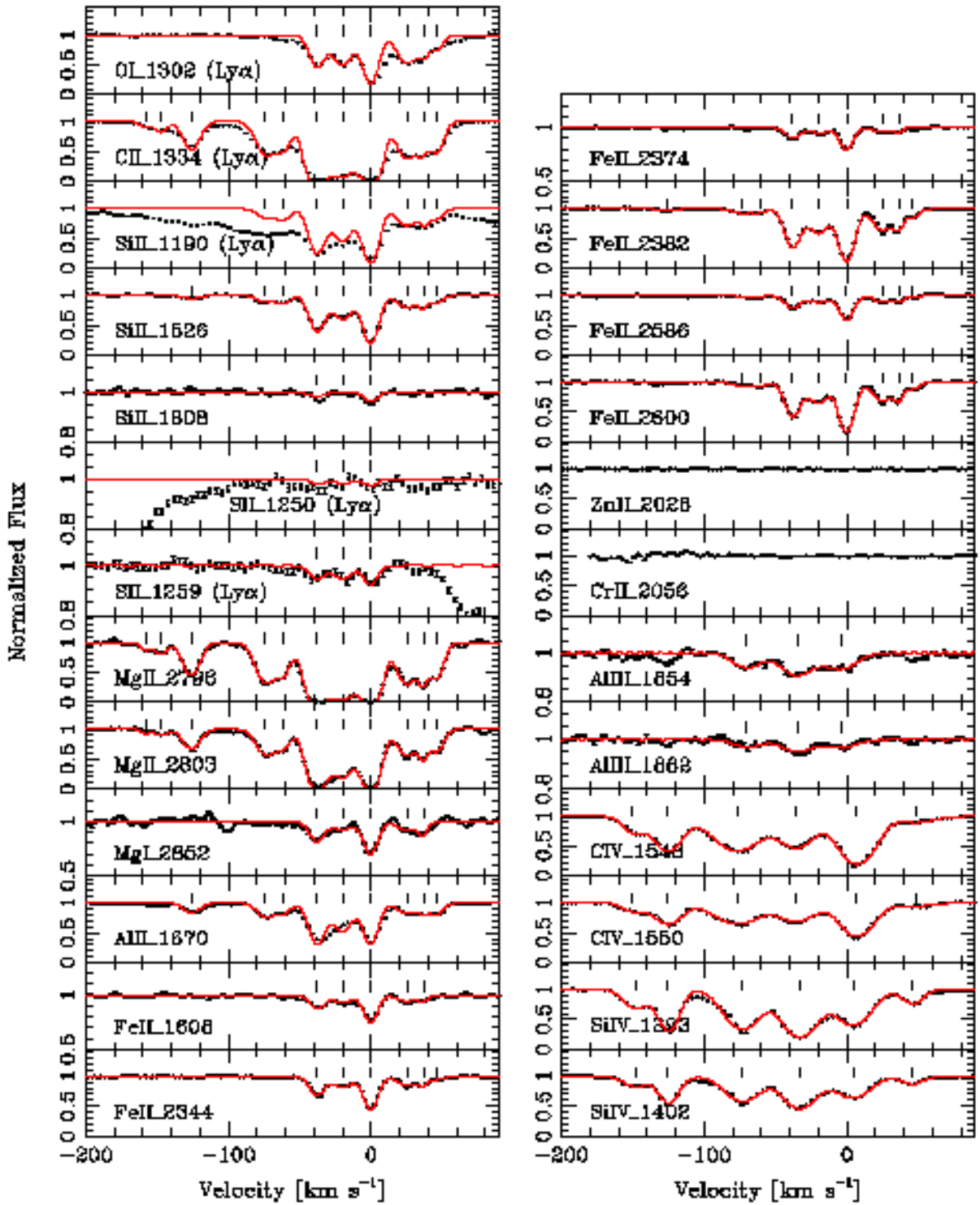}
\caption{Velocity plots of the low-, intermediate- and high-ion transitions (normalized intensities
shown by dots with 1 $\sigma$ error bars) for the sub-DLA towards Q1101$-$264. The zero velocity is
fixed at $z = 1.838902$. The vertical bars mark the positions of the fitted velocity components (see
Table~\ref{Q1101-values}). For this and all the following figures, the notation ``(Ly$\alpha$)'' near 
the name of a transition means that the line is located in the Ly$\alpha$ forest, and the symbols 
$\oplus$ correspond to telluric lines.}
\label{Q1101-fits}
\end{figure*}

Tables~\ref{Q1101-values}$-$\ref{J2344-values} present the results of the component per component 
ionic column density measurements for the fitting model solutions of the low-, intermediate- and 
high-ion transitions for all of the analysed systems. The reported errors are the 1 $\sigma$ errors 
computed by {\tt fitlyman}. These errors do not take into account the uncertainties on the continuum 
level determination. For the saturated components, the column densities are listed as lower limits. 
The values reported as upper limits correspond to cases contaminated by significant line 
blendings due to \ion{H}{i} clouds in the Ly$\alpha$ forest or telluric lines. 
Figs.~\ref{Q1101-fits}$-$\ref{J2344-fits} (odd numbers) show the best fitting solutions of the low-, 
intermediate- and high-ion transitions for all of the analysed systems. In these velocity plots, 
$v=0$ corresponds to an arbitrary component, and all the identified components are marked by small 
vertical bars. The thin solid line represents the best fit solution. We point out that the high-ion 
transitions in the bulk of sub-DLA systems present very different profiles from the low-ion 
transitions, whereas the intermediate-ion transitions show relatively similar profiles with small 
differences in the model fitting parameters. This will be discussed individually for each system in 
the following sub-sections.

\begin{figure}
   \includegraphics[width=85mm]{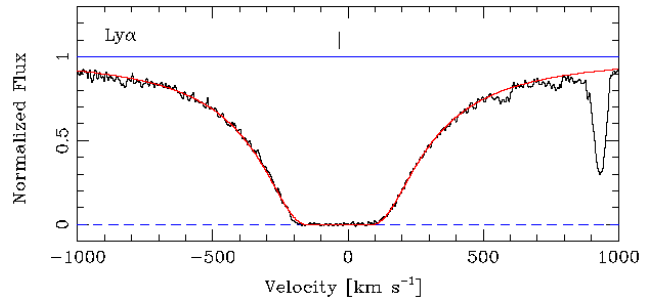}
\caption{Normalized UVES spectrum of Q1101$-$264 showing the sub-DLA Ly$\alpha$ profile with the Voigt
profile fit. The zero velocity is fixed at $z = 1.838902$. The vertical bar corresponds to the 
velocity centroid used for the best fit, $z = 1.838608$. The measured \ion{H}{i} column density is 
$\log N$(\ion{H}{i}) $= 19.50\pm 0.05$.}
\label{Q1101-Ly}
\end{figure}

The neutral hydrogen column densities were measured from the fits of the Ly$\alpha$ damping line
profiles. The $b$-value was usually fixed at 20 km~s$^{-1}$ or left as a free parameter $-$
similarly to the DLA systems the fit of the damped Ly$\alpha$ line of sub-DLA systems is independent
of the $b$-value $-$, and the redshift $z$ was fixed at the redshift of one of the strongest 
components of the low-ion metal line profiles or also left as a free parameter depending on the 
sub-DLA system (see the comments in the following sub-sections). When other lines of higher members 
of the Lyman series were covered in the spectra, we used them to check the \ion{H}{i} column density 
obtained from the fit of the Ly$\alpha$ line. Figs.~\ref{Q1101-Ly}$-$\ref{J2344-Ly} (even numbers) 
show the results of the \ion{H}{i} fitting solutions for all of the analysed systems. The small 
vertical bar corresponds to the redshift used in the best fit solution and the thin solid line 
represents the best fit solution. In one case, the sub-DLA at $z_{\rm abs} = 3.171$ towards 
Q1451+123, the Ly$\alpha$ line is not covered in the available spectra, thus the \ion{H}{i} column 
density measurement was performed using the Ly$\beta$, Ly$\gamma$ and Ly8 lines, the other members 
of the Lyman series being heavily blended.

Throughout the analysis we have adopted the list of the atomic data $-$ laboratory wavelengths and 
oscillator strengths $-$ compiled by J.~X. Prochaska and collaborators, and presented on the web site 
``The HIRES Damped Ly$\alpha$ Abundance Database''\footnote{http://kingpin.ucsd.edu/$\sim$hiresdla/}. 
The most recent measurements of the $\lambda$ and $f$-values of the metal-ions that impact the 
abundances of high-redshift absorption systems and their references are reported there.

We now comment on the individual sub-DLA systems. 
%

\subsection{APM BR J0307$-$4945}\label{Q0307}

The spectra of this quasar (Storrie-Lombardi et~al. 2001) as well as the chemical abundances of one 
of the highest redshifted damped Ly$\alpha$ system discovered by P\'eroux et~al. (2001) have been 
analyzed in detail by Dessauges-Zavadsky et~al. (2001a). Given the high redshift of this quasar 
($z_{\rm em} = 4.75$) several intervening sub-DLA systems are expected. Three candidates at $z_{\rm 
abs} =$ 3.356, 3.589 and 4.214 have been detected. However, the high density of absorption lines in 
the Ly$\alpha$ forest of such distant objects makes the blends of several absorption clouds very 
likely to occur. The lines of higher members of the Lyman series are necessary to distinguish blends 
of clouds from high \ion{H}{i} column density absorbers. As discussed below, none of the three 
sub-DLA candidates can be confirmed in the considered spectra and they probably never will be. They
thus are not included in any of our further analysis (this paper and Paper~II).

For the two candidates at lower redshifts, $z_{\rm abs} =$ 3.356 and 3.589, we do not have any 
information on the higher members of the Lyman series, these lines being beyond the quasar flux 
cut-off. Moreover, for the system at $z_{\rm abs} = 3.356$ no associated metal lines are observed. 
Thus, since the Ly$\alpha$ absorption lines of these two candidates do not show clearly any damping 
wing and since they can be fitted with one single system with $N$(\ion{H}{i}) $> 10^{19}$ cm$^{-2}$ 
as well as with several clouds of lower \ion{H}{i} column densities, the identification of these 
sub-DLA candidates is doubtful, and we choose to consider these two detected absorption line systems 
as unlikely sub-DLAs.

Several metal lines (\ion{C}{ii}, \ion{C}{iii}, \ion{C}{iv}, \ion{Si}{ii}, \ion{Si}{iii}, 
\ion{Si}{iv}, \ion{N}{iii}, \ion{Al}{iii}) and the hydrogen lines Ly$\alpha$, Ly$\beta$ and 
Ly$\gamma$ have been identified in the candidate at $z_{\rm abs} = 4.214$. From the metal line 
profiles two main sub-systems can be clearly distinguished, separated by $\simeq 400$ km~s$^{-1}$. 
These two sub-systems have already been carefully analysed by Levshakov et~al. (2003). From their 
Monte Carlo Inversion analysis they determined the \ion{H}{i} column densities of $1.4\times 10^{17}$ 
and $2.3\times 10^{17}$ cm$^{-2}$ in the bluest and reddest sub-systems, respectively, corresponding 
to values characterizing the class of the Lyman limit absorbers. Although these recovered values are 
self-consistent, they are not necessarily a unique solution. Indeed, the $N$(\ion{H}{i}) measurement 
in the reddest system is well constrained by the three observed hydrogen lines. However, all of the 
three hydrogen lines in the bluest system are highly blended, hence the real $N$(\ion{H}{i}) value 
may be higher and can possibly be above the sub-DLA $N$(\ion{H}{i}) definition threshold. Given this 
uncertainty, we prefer again to consider this absorption line system only as a sub-DLA candidate.
%

%

%

\begin{table*}
\caption{Component structure of the $z_{\rm abs} = 2.557$ sub-DLA system towards Q1223+1753} 
\label{Q1223-values}
\begin{tabular}{l c c c l c | l c c c l c}
\hline
No & $z_{\rm abs}$ & $v_{\rm rel}^*$ & $b (\sigma _b)$ & Ion & $\log N (\sigma_{\log N})$ & No & $z_{\rm abs}$ & $v_{\rm rel}^*$ & $b (\sigma _b)$ & Ion & $\log N (\sigma_{\log N})$ \\
   &               & km~s$^{-1}$     & km~s$^{-1}$     &     & cm$^{-2}$                  &    &               & km~s$^{-1}$     & km~s$^{-1}$     &     & cm$^{-2}$                  
\\     
\hline
\multicolumn{5}{l}{Low- and intermediate-ion transitions} & & & & & & & \\
\hline
1 & 2.556532 &        $-$93 & \phantom{0}4.8{\scriptsize (0.5)} & \ion{Si}{ii}  & 13.03{\scriptsize (0.04)} &  7 & 2.558001 & \phantom{0}$+$30 & \phantom{0}4.6{\scriptsize (0.7)} & \ion{Si}{ii}  & 12.79{\scriptsize (0.03)} \\
  &	     &  	    &                                   & \ion{O}{i}    &  $< 13.98$	            &	 &	    &		       &				   & \ion{O}{i}    &  $< 13.93$  \\
  &	     &  	    &                                   & \ion{Al}{ii}  & 11.80{\scriptsize (0.03)} &	 &	    &		       &				   & \ion{Al}{ii}  & 11.83{\scriptsize (0.05)} \\
  &	     &  	    &                                   & \ion{Fe}{ii}  & 12.62{\scriptsize (0.02)} &	 &	    &		       &				   & \ion{Fe}{ii}  & 12.09{\scriptsize (0.10)} \\
2 & 2.556755 &        $-$75 & \phantom{0}8.3{\scriptsize (0.5)} & \ion{Si}{ii}  & 13.02{\scriptsize (0.04)} &	 &	    &		       &				   & \ion{Fe}{iii} & 13.16{\scriptsize (0.15)} \\
  &	     &  	    &                                   & \ion{O}{i}    &  $< 14.62$	            &  8 & 2.558186 & \phantom{0}$+$46 &	   11.1{\scriptsize (1.1)} & \ion{Si}{ii}  & 13.21{\scriptsize (0.03)} \\
  &	     &  	    &                                   & \ion{Al}{ii}  & 11.90{\scriptsize (0.03)} &	 &	    &		       &				   & \ion{O}{i}    &  $< 14.51$  \\
  &	     &  	    &                                   & \ion{Fe}{ii}  & 12.64{\scriptsize (0.02)} &	 &	    &		       &				   & \ion{Al}{ii}  & 12.04{\scriptsize (0.04)} \\
  &	     &  	    &                                   & \ion{Fe}{iii} & 12.89{\scriptsize (0.14)} &	 &	    &		       &				   & \ion{Fe}{ii}  & 12.65{\scriptsize (0.03)} \\
3 & 2.557023 &        $-$52 & \phantom{0}8.8{\scriptsize (0.8)} & \ion{Si}{ii}  & 13.45{\scriptsize (0.01)} &	 &	    &		       &				   & \ion{Fe}{iii} & 13.24{\scriptsize (0.13)} \\
  &	     &  	    &                                   & \ion{O}{i}    &  $< 14.55$	            &  9 & 2.558559 & \phantom{0}$+$77 & \phantom{0}5.6{\scriptsize (2.0)} & \ion{Si}{ii}  & 12.91{\scriptsize (0.04)} \\
  &	     &  	    &                                   & \ion{Al}{ii}  & 12.19{\scriptsize (0.02)} &	 &	    &		       &				   & \ion{O}{i}    &  $< 13.90$  \\
  &	     &  	    &                                   & \ion{Fe}{ii}  & 13.09{\scriptsize (0.03)} &	 &	    &		       &				   & \ion{Al}{ii}  & 11.44{\scriptsize (0.12)} \\
  &	     &  	    &                                   & \ion{Fe}{iii} & 13.28{\scriptsize (0.13)} &	 &	    &		       &				   & \ion{Fe}{ii}  & 12.32{\scriptsize (0.08)} \\
4 & 2.557174 &        $-$39 & \phantom{0}2.6{\scriptsize (0.3)} & \ion{Si}{ii}  & 13.27{\scriptsize (0.02)} &	 &	    &		       &				   & \ion{Fe}{iii} & 12.83{\scriptsize (0.14)} \\
  &	     &  	    &                                   & \ion{O}{i}    &  $< 14.31$	            & 10 & 2.558752 & \phantom{0}$+$94 & \phantom{0}6.1{\scriptsize (1.0)} & \ion{Si}{ii}  & 13.32{\scriptsize (0.02)} \\
  &	     &  	    &                                   & \ion{Al}{ii}  & 11.89{\scriptsize (0.04)} &	 &	    &		       &				   & \ion{O}{i}    &  $< 13.95$  \\
  &	     &  	    &                                   & \ion{Fe}{ii}  & 13.16{\scriptsize (0.05)} &	 &	    &		       &				   & \ion{Al}{ii}  & 12.15{\scriptsize (0.05)} \\
  &	     &  	    &                                   & \ion{Fe}{iii} & 12.40{\scriptsize (0.07)} &	 &	    &		       &				   & \ion{Fe}{ii}  & 12.90{\scriptsize (0.04)} \\
5 & 2.557317 &        $-$27 & \phantom{0}7.2{\scriptsize (0.9)} & \ion{Si}{ii}  & 13.17{\scriptsize (0.02)} &	 &	    &		       &				   & \ion{Fe}{iii} & 13.39{\scriptsize (0.14)} \\
  &	     &  	    &                                   & \ion{O}{i}    &  $< 14.04$	            & 11 & 2.558908 &		$+107$ & \phantom{0}4.5{\scriptsize (1.2)} & \ion{Si}{ii}  & 12.93{\scriptsize (0.05)} \\
  &	     &  	    &                                   & \ion{Al}{ii}  & 12.04{\scriptsize (0.02)} &	 &	    &		       &				   & \ion{O}{i}    &  $< 13.60$  \\
  &	     &  	    &                                   & \ion{Fe}{ii}  & 12.67{\scriptsize (0.04)} &	 &	    &		       &				   & \ion{Al}{ii}  & 11.67{\scriptsize (0.11)} \\
  &	     &  	    &                                   & \ion{Fe}{iii} & 13.03{\scriptsize (0.14)} &	 &	    &		       &				   & \ion{Fe}{ii}  & 12.46{\scriptsize (0.09)} \\
6 & 2.557640 & \phantom{0}0 &           14.8{\scriptsize (0.3)} & \ion{Si}{ii}  & 14.00{\scriptsize (0.01)} &	 &	    &		       &				   & \ion{Fe}{iii} & 13.08{\scriptsize (0.15)} \\
  &	     &  	    &                                   & \ion{O}{i}    &  $< 14.76$	            & & & & & & \\
  &	     &  	    &                                   & \ion{Al}{ii}  & 12.85{\scriptsize (0.01)} & & & & & & \\
  &	     &  	    &                                   & \ion{Fe}{ii}  & 13.56{\scriptsize (0.01)} & & & & & & \\
  &	     &  	    &                                   & \ion{Fe}{iii} & 13.94{\scriptsize (0.12)} & & & & & & \\
  &	     &  	    &                                   & \ion{Al}{iii} & 12.32{\scriptsize (0.13)} & & & & & & \\
\hline
\multicolumn{4}{l}{High-ion transitions} & & & & & & & & \\
\hline
1 & 2.556414 &           $-$103 & \phantom{0}8.8{\scriptsize (1.0)} & \ion{Si}{iv} & 12.43{\scriptsize (0.03)} & 6 & 2.557734 & \phantom{0}$+$8 & \phantom{0}8.9{\scriptsize (0.2)} & \ion{Si}{iv} & 13.53{\scriptsize (0.01)} \\
  &	     &  	        & 				    & \ion{C}{iv}  & 13.15{\scriptsize (0.12)} &   &	      & 		&				    & \ion{C}{iv}  & 13.84{\scriptsize (0.02)} \\
2 & 2.556744 & \phantom{0}$-$75 & 	    12.5{\scriptsize (1.1)} & \ion{Si}{iv} & 12.57{\scriptsize (0.03)} & 7 & 2.558007 & 	  $+$31 &	    13.7{\scriptsize (0.9)} & \ion{Si}{iv} & 13.04{\scriptsize (0.03)} \\
  &	     &  	        & 				    & \ion{C}{iv}  & 13.18{\scriptsize (0.12)} &   &	      & 		&				    & \ion{C}{iv}  & 13.52{\scriptsize (0.03)} \\
3 & 2.556981 & \phantom{0}$-$56 & \phantom{0}8.6{\scriptsize (0.6)} & \ion{Si}{iv} & 12.83{\scriptsize (0.05)} & 8 & 2.558402 & 	  $+$64 &	    13.0{\scriptsize (0.2)} & \ion{Si}{iv} & 13.37{\scriptsize (0.01)} \\
  &	     &  	        & 				    & \ion{C}{iv}  & 13.56{\scriptsize (0.04)} &   &	      & 		&				    & \ion{C}{iv}  & 13.75{\scriptsize (0.01)} \\
4 & 2.557225 & \phantom{0}$-$35 & 	    16.1{\scriptsize (2.9)} & \ion{Si}{iv} & 12.70{\scriptsize (0.08)} & 9 & 2.558732 & 	  $+$92 &	    11.2{\scriptsize (1.1)} & \ion{Si}{iv} & 12.48{\scriptsize (0.03)} \\
  &	     &  	        & 				    & \ion{C}{iv}  & 13.37{\scriptsize (0.08)} &   &	      & 		&				    & \ion{C}{iv}  & 13.18{\scriptsize (0.02)} \\
5 & 2.557476 & \phantom{0}$-$14 & 	    10.2{\scriptsize (0.5)} & \ion{Si}{iv} & 12.92{\scriptsize (0.03)} &  & & & & & \\
  &	     &  	        & 				    & \ion{C}{iv}  & 13.14{\scriptsize (0.07)} & & & & & & \\
\hline
\end{tabular}
\begin{minipage}{140mm}
$^*$ Velocity relative to $z = 2.557640$
\end{minipage}
\end{table*}
%

\subsection{Q1101$-$264, $z_{\rm sub-DLA} = 1.838$}\label{Q1101}

This quasar has been observed as part of the UVES Science Verification programme for the study of 
the Ly$\alpha$ forest (see Kim, Cristiani \& D'Odorico 2002). It has previously been observed and 
analysed at lower resolution, and the absorption line system at $z_{\rm abs} = 1.838$ we are 
interested in has already been identified (e.g. Carswell et~al. 1984). Thanks to the large amount of 
UVES observing time devoted to this quasar, the spectra show an extremely high S/N. The quality of 
the data allowed us to measure with high accuracy the column densities of 11 ions and 7 elements 
(Si, O, S, C, Al, Fe, Mg) in the sub-DLA system at $z_{\rm abs} = 1.838$ observed in this 
line-of-sight. 4~$\sigma$ upper limits on the column densities of \ion{Zn}{ii} and \ion{Cr}{ii} have 
been provided, since these lines are not detected in the spectra. Only the hydrogen line, Ly$\alpha$, 
is observed in the available spectral coverage, but the absence of contamination and the presence of 
well defined damping wings lead to a very reliable $N$(\ion{H}{i}) measurement of $19.50\pm 0.05$ 
obtained by leaving the redshift and the $b$-value as free parameters (see Fig.~\ref{Q1101-Ly}). 
This absorber has already been studied by Petitjean, Srianand \& Ledoux (2000) with regards to its 
molecular content. We confirm their \ion{H}{i} column density measurement and their 
[Fe/H]\footnote{[X/H] $\equiv \log$[$N$(X)/$N$(H)]$_{\scriptsize{\textrm{DLA}}}$ 
$- \log$[$N$(X)/$N$(H)]$_{\odot}$.} and [S/H] abundance measurements.
%

\begin{figure}
   \includegraphics[width=85mm]{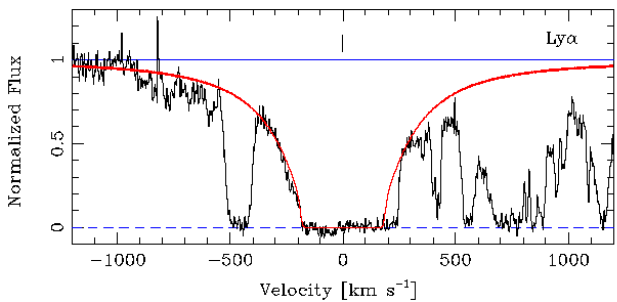}
\caption{Normalized UVES spectrum of Q1223+1753 showing the sub-DLA Ly$\alpha$ profile with the Voigt
profile fit. The zero velocity is fixed at $z = 2.557640$. The vertical bar corresponds to the 
velocity centroid of the strongest component of the low-ion transition lines used for the best fit, 
the component 6 at $z = 2.557640$. The measured \ion{H}{i} column density is $\log N$(\ion{H}{i}) 
$= 19.32\pm 0.15$.}
\label{Q1223-Ly}
\end{figure}
%

\begin{figure*}
   \includegraphics[width=170mm]{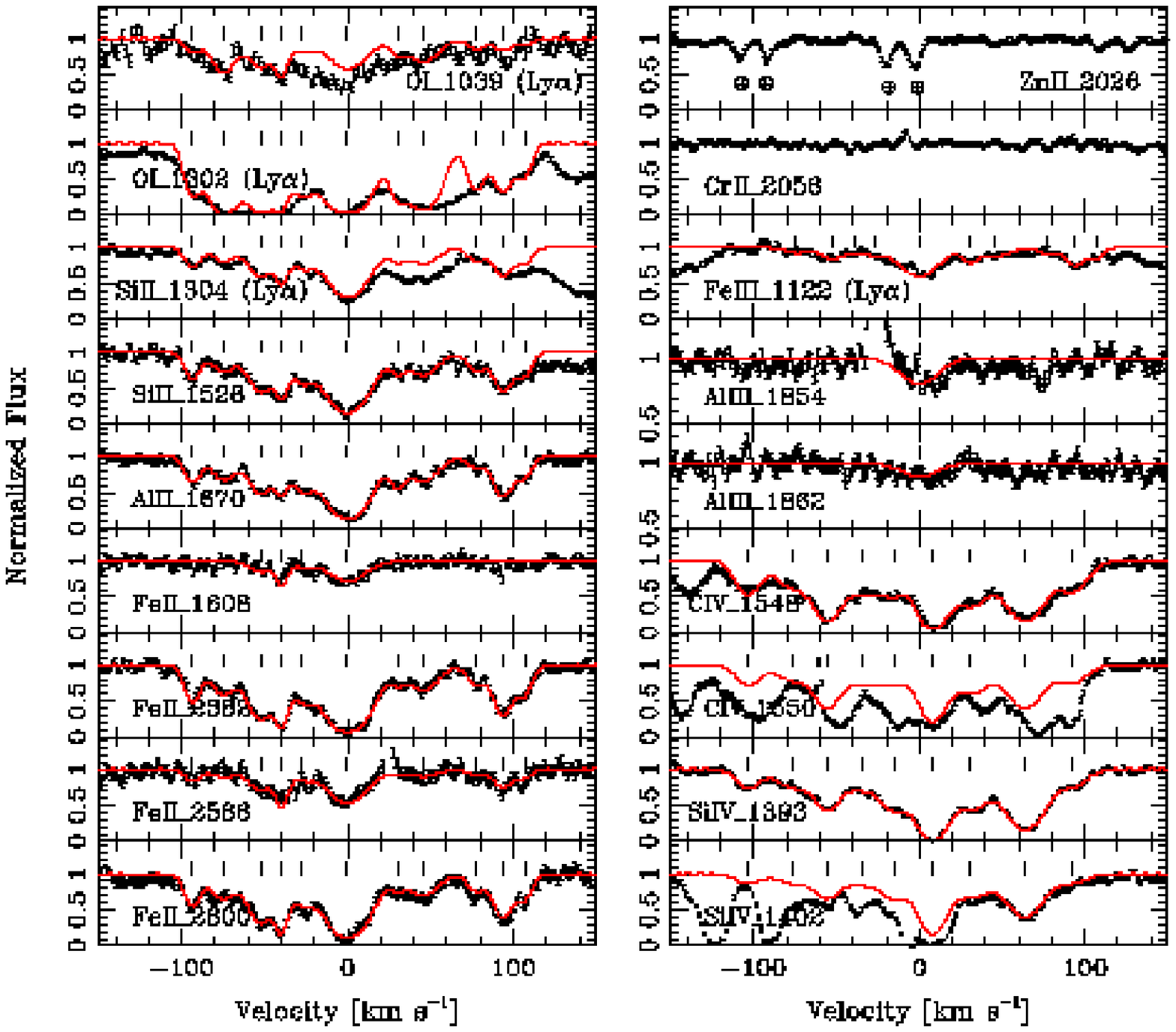}
\caption{Same as Fig.~\ref{Q1101-fits} for the sub-DLA towards Q1223+1753. The zero velocity is
fixed at $z = 2.557640$. The vertical bars mark the positions of the fitted velocity components (see
Table~\ref{Q1223-values}).}
\label{Q1223-fits}
\end{figure*}
%

The component structure of this absorber is presented in Table~\ref{Q1101-values} for the low-,
intermediate- and high-ion transitions. The low-ion profiles are described with a relatively complex 
structure of 11 components extended over $\sim 210$ km~s$^{-1}$, while the high-ion transitions are 
well fitted with only 6 components (see Fig.~\ref{Q1101-fits}). This difference in the line profiles 
suggests that the low- and high-ion lines are not emitted from the same physical region, which is a 
commonly observed situation in DLAs. Nevertheless, in this particular case, surprisingly the lines 
of C$^{+++}$ and Si$^{+++}$, two ions with the same ionization degree, do not show the same line 
profiles and require slightly different fitting parameters ($z$, $b$ and $N$). In addition, the 
\ion{Al}{iii} lines better resemble the high-ion transition profiles than the low-ion ones, contrary 
to what is usually observed in DLA studies where the intermediate-ion line profiles are very similar 
to the low-ion line profiles, indicating they are emitted from the same physical region. 
Unfortunately, no other intermediate-ion is detected in this sub-DLA system to confirm these 
observations.
%

\subsection{Q1223+1753, $z_{\rm sub-DLA} = 2.557$}

This quasar, also called LBQS 1223+1753, was discovered by Foltz et~al. (1987) and it has been 
observed on many occasions for the study of its intervening DLA system at $z_{\rm abs} = 2.466$. The 
large \ion{H}{i} column density of this absorber allows the analysis of a great number of transitions 
(over 20) as well as of the H$_2$ molecule (e.g. Pettini et~al. 1994; Centuri\'on et~al. 2000; 
Petitjean, Srianand \& Ledoux 2000; Ellison, Ryan \& Prochaska 2001; Prochaska et~al. 2001). In 
addition to this DLA, we identified a sub-DLA system at $z_{\rm abs} = 2.557$ with a 
$\log N$(\ion{H}{i}) of $19.32\pm 0.15$, and we present here for the first time its chemical 
analysis. 

This sub-DLA system was identified thanks to its Ly$\alpha$ absorption line, the only hydrogen line 
observed in the available spectra. This line shows a well defined blue damping wing, but the red
damping wing is partially blended with \ion{H}{i} clouds (see Fig.~\ref{Q1223-Ly}). The \ion{H}{i} 
column density measurement was hampered by an additional difficulty, the location of the sub-DLA 
Ly$\alpha$ line in the red wing of the DLA Ly$\alpha$ line. Therefore, we first had to locally 
renormalize the quasar spectrum with the fit of the DLA Ly$\alpha$ damping wing profile, before
measuring the $N$(\ion{H}{i}) of the sub-DLA by fixing the redshift at the position of the strongest
component of the metal line profiles (the component 6). The error on the derived hydrogen column 
density was estimated by varying the continuum level by 5\%.

As shown in Fig.~\ref{Q1223-fits} the low-ion transitions of the sub-DLA system present a complex 
velocity structure composed of 11 components spread over $\sim 210$ km~s$^{-1}$ (see 
Table~\ref{Q1223-values}). The numerous Fe$^+$ lines identified in the system (\ion{Fe}{ii}
$\lambda$1608,2382,2586,2600) allowed to obtain a very accurate $N$(\ion{Fe}{ii}) measurement. We 
also provide measurements of the \ion{Si}{ii} and \ion{Al}{ii} column densities. To be conservative, 
we prefer to consider the measured $N$(\ion{O}{i}) obtained from the \ion{O}{i} $\lambda$1039,1302 
lines located in the Ly$\alpha$ forest as an upper limit. The \ion{Zn}{ii} and \ion{Cr}{ii} lines
are not detected in the spectra, we provide 4~$\sigma$ upper limits. The \ion{C}{ii} lines are
highly saturated. 
%

\begin{table}
\caption{Component structure of the $z_{\rm abs} = 2.668$ sub-DLA system towards Q1409+095} 
\label{Q1409-values}
\begin{tabular}{l c c c l c}
\hline
No & $z_{\rm abs}$ & $v_{\rm rel}^*$ & $b (\sigma _b)$ & Ion & $\log N (\sigma_{\log N})$ \\
   &               & km~s$^{-1}$     & km~s$^{-1}$     &     & cm$^{-2}$                   
\\     
\hline
\multicolumn{5}{l}{Low- and intermediate-ion transitions} & \\
\hline
1 & 2.668092 &        $-$11 & 2.1{\scriptsize (0.9)} & \ion{Si}{ii}  & 13.32{\scriptsize (0.04)} \\
  &	     &  	    &                        & \ion{O}{i}    & 14.24{\scriptsize (0.15)} \\
  &	     &  	    &                        & \ion{C}{ii}   & $> 16.00$  \\
  &	     &  	    &                        & \ion{Fe}{ii}  & 13.63{\scriptsize (0.24)} \\
2 & 2.668227 & \phantom{0}0 & 4.8{\scriptsize (0.4)} & \ion{Si}{ii}  & 14.06{\scriptsize (0.03)} \\  
  &	     &  	    &                        & \ion{O}{i}    & 15.27{\scriptsize (0.11)} \\  
  &	     &  	    &                        & \ion{C}{ii}   & $> 15.37$  \\ 
  &	     &  	    &                        & \ion{S}{ii}   & 13.54{\scriptsize (0.06)} \\
  &	     &  	    &                        & \ion{N}{i}    & $< 13.49$  \\
  &	     &  	    &                        & \ion{Cr}{ii}  & 12.49{\scriptsize (0.15)} \\
  &	     &  	    &                        & \ion{Fe}{ii}  & 13.79{\scriptsize (0.05)} \\  
  &          &              &                        & \ion{Fe}{iii} & 13.17{\scriptsize (0.12)} \\
  &          &              &                        & \ion{Al}{iii} & 11.89{\scriptsize (0.05)} \\
\hline
\end{tabular}
\begin{minipage}{140mm}
$^*$ Velocity relative to $z = 2.668227$
\end{minipage}
\end{table}
%

In addition, we observe the \ion{C}{iv} and \ion{Si}{iv} high-ion transitions showing in this case 
very similar line profiles to the low-ion transitions. The \ion{C}{iv} $\lambda$1550 and \ion{Si}{iv} 
$\lambda$1402 lines are blended with the metal lines of other absorbers (see Fig.~\ref{Q1223-fits}). 
We also detect the \ion{Al}{iii} and \ion{Fe}{iii} intermediate-ion transitions. The intermediate-ion 
column density determination, in particular the one of Fe$^{++}$, is crucial for the estimation of 
the ionization corrections, which may not be negligible in this low \ion{H}{i} column density system 
(see Section~\ref{ionization}). Since the \ion{Al}{iii} lines are weak and the \ion{Fe}{iii} line is 
in the Ly$\alpha$ forest where blends with \ion{H}{i} clouds are possible, we cannot determine with 
high accuracy whether the intermediate-ion profiles present differences with respect to the low-ion 
profiles in this sub-DLA system. Thus, as the \ion{Al}{iii} lines show a single velocity component 
located exactly at the redshift of the strongest component of the low-ion line profiles, the 
component 6, we assume that we can adopt the same fitting parameters for the intermediate-ion 
transitions as the ones defined for the low-ion transitions. 

%

\subsection{Q1409+095, $z_{\rm sub-DLA} = 2.668$}

\begin{figure}
   \includegraphics[width=85mm]{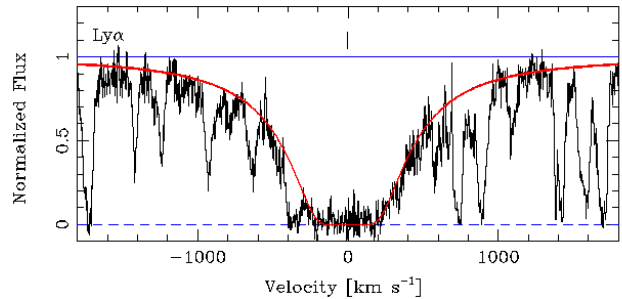}
\caption{Normalized UVES spectrum of Q1409+095 showing the sub-DLA Ly$\alpha$ profile with the Voigt
profile fit. The zero velocity is fixed at $z = 2.668227$. The vertical bar corresponds to the 
velocity centroid of the strongest component of the low-ion transition lines used for the best fit, 
the component 2 at $z = 2.668227$. The measured \ion{H}{i} column density is $\log N$(\ion{H}{i}) 
$= 19.75\pm 0.10$.}
\label{Q1409-Ly}
\end{figure}

Discovered by Hazard et~al. (1986), this quasar has first been studied at low resolution by Lanzetta 
et~al. (1991) and then by e.g. Lu et~al. (1993). They reported the identification of a DLA system at 
$z_{\rm abs} = 2.456$ and another absorption line system at $z_{\rm abs} = 2.668$ with a \ion{H}{i} 
column density below the DLA definition corresponding to a sub-DLA system. Pettini et~al. (2002) 
studied these two systems at high resolution and obtained the abundance measurements of O, Si, N and 
Fe. We identified the sub-DLA system in the UVES spectra thanks to the Ly$\alpha$ absorption line 
with clearly visible damping wings (see Fig.~\ref{Q1409-Ly}), and derived a \ion{H}{i} column density 
of $\log N$(\ion{H}{i}) $= 19.75\pm 0.10$ in agreement with the Pettini et~al. measurement.
%

%

\begin{figure*}
   \includegraphics[width=140mm]{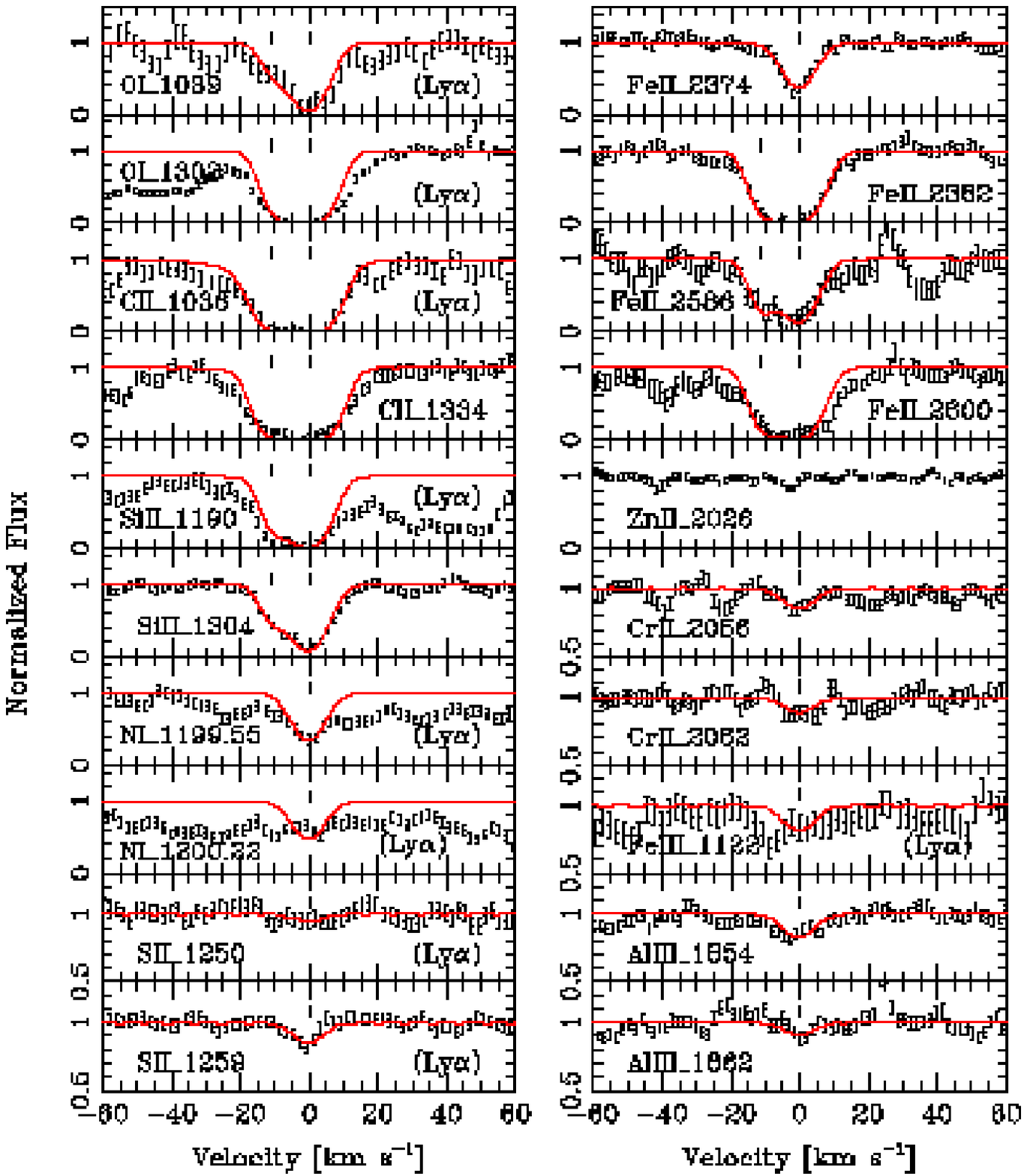}
\caption{Same as Fig.~\ref{Q1101-fits} for the sub-DLA towards Q1409+095. The zero velocity is
fixed at $z = 2.668227$. The vertical bars mark the positions of the fitted velocity components (see
Table~\ref{Q1409-values}).}
\label{Q1409-fits}
\end{figure*}
%

\begin{table*}
\caption{Component structure of the $z_{\rm abs} = 2.087$ sub-DLA system towards Q1444+014} 
\label{Q1444-values}
\begin{tabular}{l c c c l c | l c c c l c}
\hline
No & $z_{\rm abs}$ & $v_{\rm rel}^*$ & $b (\sigma _b)$ & Ion & $\log N (\sigma_{\log N})$ & No & $z_{\rm abs}$ & $v_{\rm rel}^*$ & $b (\sigma _b)$ & Ion & $\log N (\sigma_{\log N})$ \\
   &               & km~s$^{-1}$     & km~s$^{-1}$     &     & cm$^{-2}$                  &    &               & km~s$^{-1}$     & km~s$^{-1}$     &     & cm$^{-2}$                  
\\     
\hline
\multicolumn{4}{l}{Low-ion transitions} & & & & & & & & \\
\hline
1 & 2.085060 &           $-$170 &           10.9{\scriptsize (1.0)} & \ion{Si}{ii} & 12.98{\scriptsize (0.06)} &  8 & 2.086808 &    \phantom{0+}0 &	      13.3{\scriptsize (0.2)} & \ion{Si}{ii} & 14.37{\scriptsize (0.01)} \\
  &	     &  	        &                                   & \ion{O}{i}   & 13.75{\scriptsize (0.04)} &    &	       &		  &				      & \ion{O}{i}   & $> 15.81$ \\
  &	     &  	        &                                   & \ion{C}{ii}  & 13.44{\scriptsize (0.03)} &    &	       &		  &				      & \ion{C}{ii}  & $> 15.64$ \\
2 & 2.085297 &           $-$147 & \phantom{0}5.2{\scriptsize (0.6)} & \ion{Si}{ii} & 13.27{\scriptsize (0.05)} &    &	       &		  &				      & \ion{Al}{ii} & 12.66{\scriptsize (0.01)} \\
  &	     &  	        &                                   & \ion{O}{i}   & $> 14.69$                 &    &	       &		  &				      & \ion{Fe}{ii} & 13.91{\scriptsize (0.02)} \\
  &	     &  	        &                                   & \ion{C}{ii}  & $> 13.70$                 &  9 & 2.087482 & \phantom{0}$+$65 & \phantom{0}7.1{\scriptsize (0.2)} & \ion{Si}{ii} & 13.54{\scriptsize (0.02)} \\
  &	     &  	        &                                   & \ion{Al}{ii} & 11.88{\scriptsize (0.08)} &    &	       &		  &				      & \ion{O}{i}   & 14.38{\scriptsize (0.02)} \\
  &	     &  	        &                                   & \ion{Fe}{ii} & 12.81{\scriptsize (0.08)} &    &	       &		  &				      & \ion{C}{ii}  & 14.57{\scriptsize (0.03)} \\	   
3 & 2.085481 &           $-$129 &           10.4{\scriptsize (0.5)} & \ion{Si}{ii} & 14.02{\scriptsize (0.02)} &    &	       &		  &				      & \ion{Al}{ii} & 12.27{\scriptsize (0.02)} \\
  &	     &  	        &                                   & \ion{O}{i}   & $> 14.99$                 &    &	       &		  &				      & \ion{Fe}{ii} & 13.26{\scriptsize (0.04)} \\
  &	     &  	        &                                   & \ion{C}{ii}  & $> 16.07$                 & 10 & 2.087682 & \phantom{0}$+$85 & \phantom{0}6.3{\scriptsize (0.8)} & \ion{Si}{ii} & 12.73{\scriptsize (0.08)} \\
  &	     &  	        &                                   & \ion{Al}{ii} & 12.81{\scriptsize (0.02)} &    &	       &		  &				      & \ion{O}{i}   & 13.75{\scriptsize (0.03)} \\
  &	     &  	        &                                   & \ion{Fe}{ii} & 13.66{\scriptsize (0.02)} &    &	       &		  &				      & \ion{C}{ii}  & 13.65{\scriptsize (0.02)} \\
4 & 2.085711 &           $-$107 & \phantom{0}6.1{\scriptsize (0.4)} & \ion{Si}{ii} & 13.53{\scriptsize (0.02)} &    &	       &		  &				      & \ion{Al}{ii} & 11.07{\scriptsize (0.19)} \\
  &	     &  	        &                                   & \ion{O}{i}   & 14.59{\scriptsize (0.25)} & 11 & 2.087973 &	   $+$113 & \phantom{0}5.6{\scriptsize (0.4)} & \ion{Si}{ii} & 12.92{\scriptsize (0.05)} \\
  &	     &  	        &                                   & \ion{C}{ii}  & $> 14.26$                 &    &	       &		  &				      & \ion{O}{i}   & 13.86{\scriptsize (0.04)} \\
  &	     &  	        &                                   & \ion{Al}{ii} & 12.06{\scriptsize (0.05)} &    &	       &		  &				      & \ion{C}{ii}  & 13.82{\scriptsize (0.03)} \\
  &	     &  	        &                                   & \ion{Fe}{ii} & 13.28{\scriptsize (0.03)} &    &	       &		  &				      & \ion{Al}{ii} & 11.61{\scriptsize (0.07)} \\
5 & 2.085857 & \phantom{0}$-$92 & \phantom{0}8.5{\scriptsize (3.2)} & \ion{Si}{ii} & 13.16{\scriptsize (0.12)} & 12 & 2.088111 &	   $+$127 & \phantom{0}6.1{\scriptsize (0.4)} & \ion{Si}{ii} & 13.30{\scriptsize (0.03)} \\
  &	     &  	        &                                   & \ion{O}{i}   & 14.20{\scriptsize (0.12)} &    &	       &		  &				      & \ion{O}{i}   & 14.24{\scriptsize (0.02)} \\
  &	     &  	        &                                   & \ion{C}{ii}  & $> 13.95$                 &    &	       &		  &				      & \ion{C}{ii}  & 14.32{\scriptsize (0.21)} \\
  &	     &  	        &                                   & \ion{Al}{ii} & 11.75{\scriptsize (0.12)} &    &	       &		  &				      & \ion{Al}{ii} & 12.01{\scriptsize (0.04)} \\
6 & 2.086125 & \phantom{0}$-$66 &           10.2{\scriptsize (0.4)} & \ion{Si}{ii} & 14.09{\scriptsize (0.03)} &    &	       &		  &				      & \ion{Fe}{ii} & 13.13{\scriptsize (0.05)} \\
  &	     &  	        &                                   & \ion{O}{i}   & $> 15.36$                 & 13 & 2.088333 &	   $+$148 &	      13.8{\scriptsize (2.0)} & \ion{Si}{ii} & 12.97{\scriptsize (0.06)} \\
  &	     &  	        &                                   & \ion{C}{ii}  & $> 15.24$                 &    &	       &		  &				      & \ion{O}{i}   & 13.75{\scriptsize (0.05)} \\
  &	     &  	        &                                   & \ion{Al}{ii} & 12.75{\scriptsize (0.02)} &    &	       &		  &				      & \ion{C}{ii}  & 14.03{\scriptsize (0.10)} \\
  &	     &  	        &                                   & \ion{Fe}{ii} & 13.67{\scriptsize (0.02)} &    &	       &		  &				      & \ion{Al}{ii} & 11.75{\scriptsize (0.07)} \\
7 & 2.086501 & \phantom{0}$-$30 & \phantom{0}6.0{\scriptsize (1.3)} & \ion{Si}{ii} & 12.81{\scriptsize (0.07)} & 14 & 2.088617 &	   $+$176 & \phantom{0}6.0{\scriptsize (0.2)} & \ion{Si}{ii} & 13.66{\scriptsize (0.02)} \\
  &	     &  	        &                                   & \ion{O}{i}   & 13.85{\scriptsize (0.22)} &    &	       &		  &				      & \ion{O}{i}   & 14.69{\scriptsize (0.04)} \\
  &	     &  	        &                                   & \ion{C}{ii}  & 13.33{\scriptsize (0.18)} &    &	       &		  &				      & \ion{C}{ii}  & $> 14.95$ \\
  &	     &  	        &                                   & \ion{Al}{ii} & 11.36{\scriptsize (0.11)} &    &	       &		  &				      & \ion{Al}{ii} & 12.48{\scriptsize (0.02)} \\
  &          &                  &                                   &              &                           &    &	       &		  &				      & \ion{Fe}{ii} & 13.40{\scriptsize (0.02)} \\
\hline
\multicolumn{4}{l}{High-ion transitions} & & & & & & & & \\
\hline
1 & 2.088197 & $+$135 & 17.5{\scriptsize (1.8)} & \ion{Si}{iv} & 12.69{\scriptsize (0.04)} & 2 & 2.088584 & $+$173 & 10.4{\scriptsize (1.1)} & \ion{Si}{iv} & 12.58{\scriptsize (0.04)} \\
  &	     &        &	                        & \ion{C}{iv}  & 13.05{\scriptsize (0.04)} &   &	  &	   &			     & \ion{C}{iv}  & 12.72{\scriptsize (0.07)} \\
\hline
\end{tabular}
\begin{minipage}{140mm}
$^*$ Velocity relative to $z = 2.086808$
\end{minipage}
\end{table*}
%

Our determinations of the \ion{O}{i}, \ion{Si}{ii} and \ion{Fe}{ii} column densities and of the
\ion{N}{i} upper limit are also consistent with their findings, although we obtained $\sim 0.1$~dex 
higher \ion{Si}{ii} and \ion{Fe}{ii} column densities. This discrepancy may be explained by the fact 
that we used a two-component fitting model for the low-ion transitions (see 
Table~\ref{Q1409-values}), while these authors used a one-component fitting model. Indeed, weak 
metal lines, like the \ion{N}{i}, \ion{S}{ii} lines and some \ion{Fe}{ii} lines with low oscillator 
strengths, are well fitted with a single component. However, as shown in Fig.~\ref{Q1409-fits}, the 
strong \ion{Fe}{ii} $\lambda$2382,2586,2600, \ion{O}{i} $\lambda$1302, \ion{Si}{ii} $\lambda$1190 and 
\ion{C}{ii} $\lambda$1036,1334 line profiles clearly indicate the presence of a second weaker 
component at $\sim 11$ km~s$^{-1}$ bluewards of the main component. In addition to the O$^0$, Si$^+$, 
Fe$^+$ column densities and the $N$(\ion{N}{i}) upper limit~$-$ the N$^0$ triplet being blended with 
the Ly$\alpha$ forest absorptions $-$, we obtained the column density measurements of S$^+$ from the 
\ion{S}{ii} $\lambda$1250,1259 lines and Cr$^+$. The \ion{Zn}{ii} lines are not detected, thus we 
provide a 4~$\sigma$ upper limit. We also report a lower limit on $N$(\ion{C}{ii}), since the 
feature is saturated.

We detect the Al$^{++}$ and Fe$^{++}$ intermediate-ion transitions in this sub-DLA system. A single 
component is observed in the \ion{Al}{iii} $\lambda$1854 and $\lambda$1862 line profiles, and it is 
reasonably well fitted with the low-ion transition fitting parameters (see Fig.~\ref{Q1409-fits}). 
Although the fit of the \ion{Al}{iii} $\lambda$1854 component does not look optimal and could be 
shifted by a few km~s$^{-1}$ bluewards, when fitting simultaneously the two \ion{Al}{iii} lines at 
$\lambda_{\rm rest} = 1854$ and $1862$ \AA, the best fit solution is very similar to the low-ion 
transition fitting parameters. Thus, given the weak intensity of the \ion{Al}{iii} and \ion{Fe}{iii} 
absorption lines and the low S/N in the spectral regions covering these lines, we preferred to adopt 
the same fitting parameters as the ones derived for the low-ion transition lines, and we obtained 
\ion{Al}{iii} and marginal \ion{Fe}{iii} column density measurements. No high-ion transition line is 
observed, all of them being outside the available wavelength coverage.
%

\begin{figure}
   \includegraphics[width=85mm]{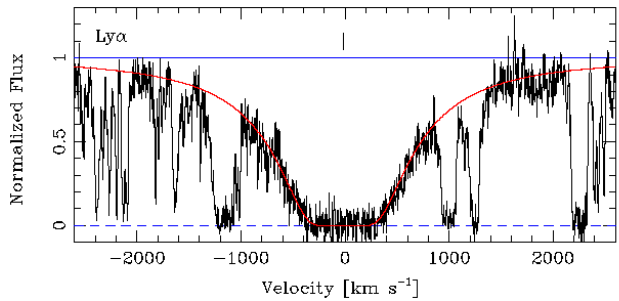}
\caption{Normalized UVES spectrum of Q1444+014 showing the sub-DLA Ly$\alpha$ profile with the Voigt
profile fit. The zero velocity is fixed at $z = 2.086808$. The vertical bar corresponds to the 
velocity centroid used for the best fit, $z = 2.086643$. The measured \ion{H}{i} column density is 
$\log N$(\ion{H}{i}) $= 20.18\pm 0.10$.}
\label{Q1444-Ly}
\end{figure}
%

\subsection{Q1444+014, $z_{\rm sub-DLA} = 2.087$}

The quasar Q1444+014 is found more commonly under the name LBQS 1444+0126 and was discovered by 
Hewett et~al. (1991). Its intervening absorption system at $z_{\rm abs} = 2.087$ was first reported 
by Wolfe et~al. (1995). Pettini et~al. (2002) analysed the \ion{H}{i}, \ion{O}{i}, \ion{Si}{ii}, 
\ion{N}{i} and \ion{Fe}{ii} lines of this absorber, loosely referring to it as a DLA system. We 
derived the \ion{H}{i} column density by fitting the weakly contaminated Ly$\alpha$ absorption line 
showing well defined damping wings (the redshift was left as a free parameter; see 
Fig.~\ref{Q1444-Ly}), and we classified the system as a sub-DLA with $\log N$(\ion{H}{i}) 
$= 20.18\pm 0.10$, in agreement with the Pettini et~al. measurement.
%

\begin{figure*}
   \includegraphics[width=170mm]{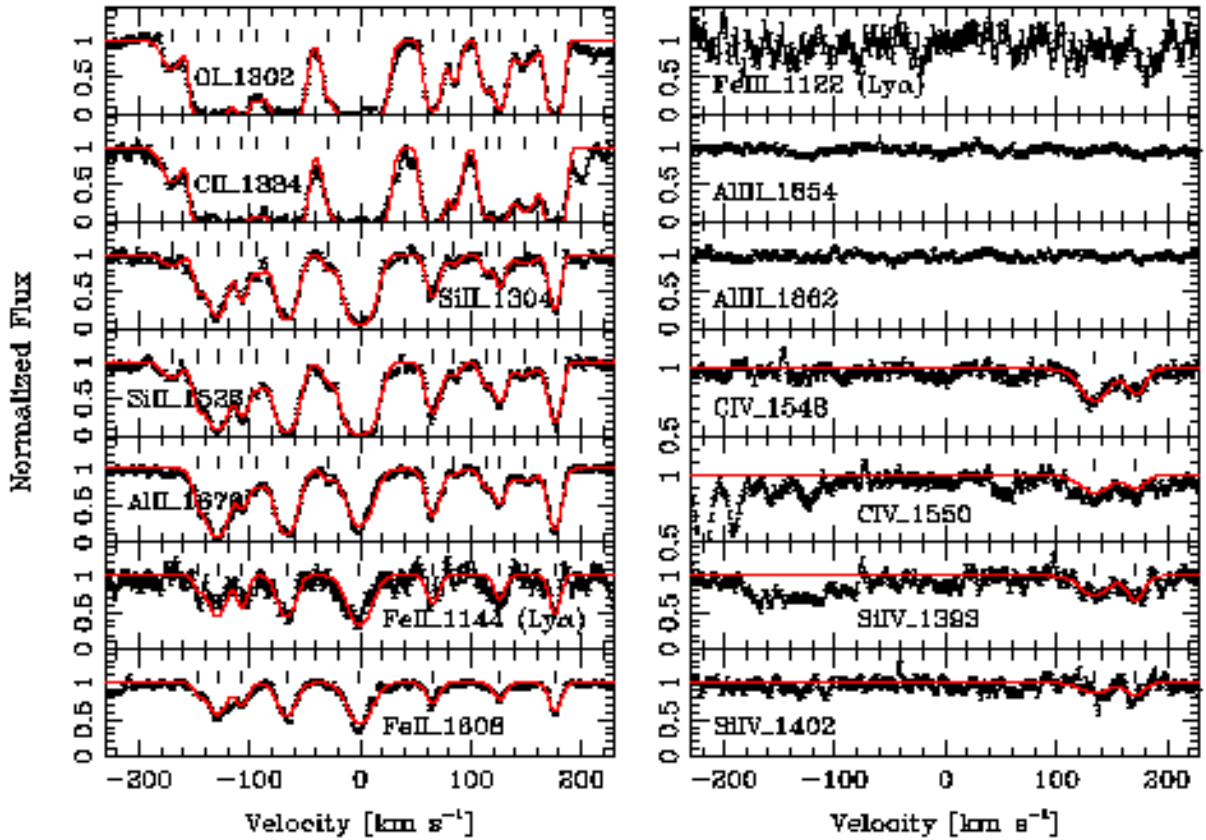}
\caption{Same as Fig.~\ref{Q1101-fits} for the sub-DLA towards Q1444+014. The zero velocity is
fixed at $z = 2.086808$. The vertical bars mark the positions of the fitted velocity components (see
Table~\ref{Q1444-values}).}
\label{Q1444-fits}
\end{figure*}
%

The low-ion transition profiles of this sub-DLA system exhibit a very complex velocity structure with 
14 components extended over $\sim 350$ km~s$^{-1}$ (see Table~\ref{Q1444-values} and
Fig.~\ref{Q1444-fits}). We obtained column density measurements of \ion{Si}{ii}, \ion{Al}{ii} and 
\ion{Fe}{ii}. Only the strongest components are observed in the \ion{Fe}{ii} $\lambda$1144,1608 line 
profiles. Unfortunately, several components between $v \simeq 0$ and $-150$ km~s$^{-1}$ of the 
\ion{O}{i} $\lambda$1302 and \ion{C}{ii} $\lambda$1334 lines are saturated, thus we could only 
provide lower limits on the column densities of these two metal transitions. No intermediate-ion 
line was detected. We therefore give only the 4~$\sigma$ upper limits on $N$(\ion{Al}{iii}) and 
$N$(\ion{Fe}{iii}) $-$ the \ion{Fe}{iii} $\lambda$1122 line is located in the Ly$\alpha$ forest, but 
the spectral region where the strongest components of the \ion{Fe}{iii} line should be detected is 
free from important \ion{H}{i} blendings, and thus we can derive a reliable 4~$\sigma$ upper limit 
on $N$(\ion{Fe}{iii}). A very surprising particularity observed in this system is the behaviour of 
the high-ion transition lines, \ion{C}{iv} and \ion{Si}{iv}. They are characterized by only two 
relatively weak components and are extended over a much narrower velocity range of $\sim 40$ 
km~s$^{-1}$ than the low-ion profiles (see Table~\ref{Q1444-values} and Fig.~\ref{Q1444-fits}). 
Most of the DLA systems show the opposite situation with the high-ion line profiles spread over a 
larger velocity range than the low-ion line profiles.
%

\subsection{Q1451+123}

The absorption line systems towards this quasar discovered by Hazard et~al. (1986) have first been 
studied by Wolfe et~al. (1986). Lanzetta et~al. (1991) identified in their low-resolution spectra a 
DLA system at $z_{\rm abs} = 2.469$ and two other absorption systems at $z_{\rm abs} = 2.255$ and 
$3.171$. Petitjean, Srianand \& Ledoux (2000) studied the intervening DLA and sub-DLA at $z_{\rm abs} 
= 3.171$ for their molecular content and report some of their elemental abundances. We present here 
for the first time the abundance analysis of the additional system at $z_{\rm abs} = 2.255$ which is a
borderline case between the DLA and sub-DLA systems.
%

\begin{figure}
   \includegraphics[width=85mm]{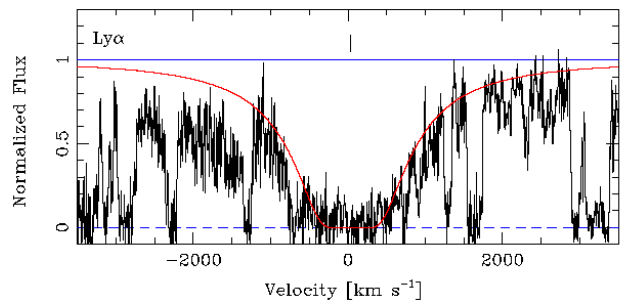}
\caption{Normalized UVES spectrum of Q1451+123 showing the Ly$\alpha$ profile with the Voigt profile 
fit of the borderline case between the DLA and sub-DLA systems. The zero velocity is fixed at $z = 
2.254710$. The vertical bar corresponds to the velocity centroid used for the best fit, $z = 
2.255151$. The measured \ion{H}{i} column density is $\log N$(\ion{H}{i}) $= 20.30\pm 0.15$.}
\label{Q1451-2p255-Ly}
\end{figure}
%

\begin{figure*}
   \includegraphics[width=140mm]{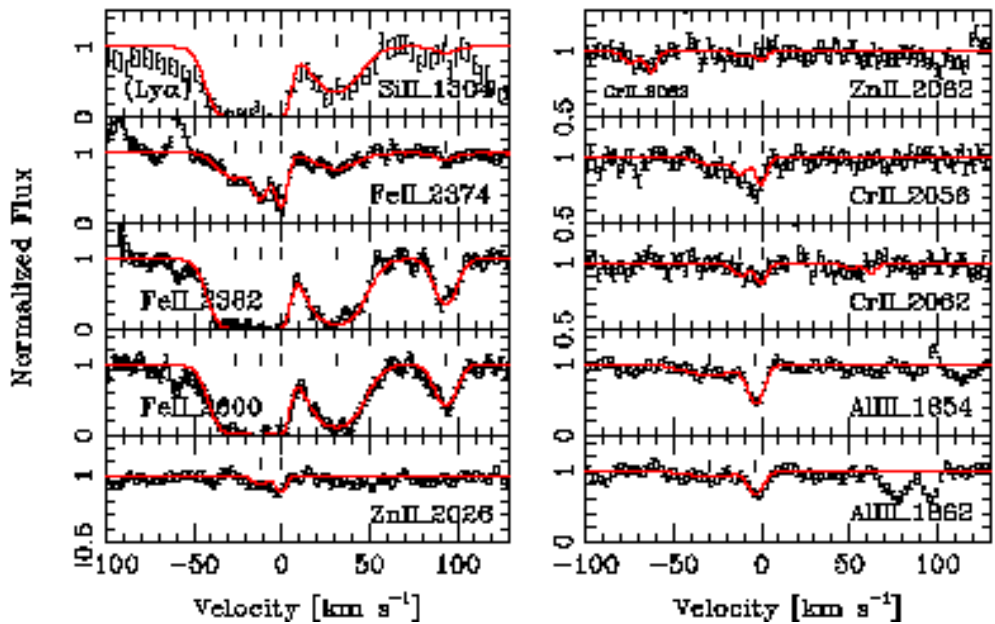}
\caption{Same as Fig.~\ref{Q1101-fits} for the borderline case between the DLA and sub-DLA systems 
towards Q1451+123. The zero velocity is fixed at $z = 2.254710$. The vertical bars mark the positions 
of the fitted velocity components (see Table~\ref{Q1451-2p255-values}).}
\label{Q1451-2p255-fits}
\end{figure*}
%

\begin{table}
\caption{Component structure of the $z_{\rm abs} = 2.255$ borderline case between the DLA and sub-DLA
systems towards Q1451+123} 
\label{Q1451-2p255-values}
\begin{tabular}{l c c c l c}
\hline
No & $z_{\rm abs}$ & $v_{\rm rel}^*$ & $b (\sigma _b)$ & Ion & $\log N (\sigma_{\log N})$ \\
   &               & km~s$^{-1}$     & km~s$^{-1}$     &     & cm$^{-2}$                   
\\     
\hline
\multicolumn{4}{l}{Low-ion transitions} & & \\
\hline
1 & 2.254420 &        $-$27 &  	        11.5{\scriptsize (0.6)} & \ion{Si}{ii} & $>14.59$		    \\ 
  &	     &              &				        & \ion{Fe}{ii} & 13.68{\scriptsize (0.02)} \\
  &          &              &                                   & \ion{Ni}{ii} & $< 13.20$                 \\ 
  &	     &              &				        & \ion{Cr}{ii} & 12.33{\scriptsize (0.15)} \\ 
2 & 2.254584 &        $-$12 & \phantom{0}5.0{\scriptsize (2.1)} & \ion{Si}{ii} & $>14.56$		    \\ 
  &	     &              &			                & \ion{Fe}{ii} & 13.73{\scriptsize (0.08)} \\ 
  &          &              &                                   & \ion{Ni}{ii} & $< 12.77$                 \\
  &	     &              &				        & \ion{Zn}{ii} & 11.49{\scriptsize (0.15)} \\ 
  &	     &              &			                & \ion{Cr}{ii} & 12.37{\scriptsize (0.13)} \\ 
3 & 2.254710 & \phantom{0}0 & \phantom{0}2.8{\scriptsize (1.4)} & \ion{Si}{ii} & $>15.31$		   \\
  &	     &  	    &					& \ion{Fe}{ii} & 13.84{\scriptsize (0.12)} \\
  &          &              &                                   & \ion{Ni}{ii} & $< 13.09$                 \\
  &	     &  	    &					& \ion{Zn}{ii} & 11.60{\scriptsize (0.08)} \\
  &	     &  	    &					& \ion{Cr}{ii} & 12.48{\scriptsize (0.15)} \\
4 & 2.255047 &        $+$31 &		15.4{\scriptsize (0.4)} & \ion{Si}{ii} & 13.95{\scriptsize (0.08)} \\
  &	     &  	    &					& \ion{Fe}{ii} & 13.56{\scriptsize (0.01)} \\
5 & 2.255725 &        $+$93 & \phantom{0}7.2{\scriptsize (0.7)} & \ion{Si}{ii} & 12.56{\scriptsize (0.21)} \\
  &	     &  	    &					& \ion{Fe}{ii} & 12.86{\scriptsize (0.03)} \\
\hline
\multicolumn{4}{l}{Intermediate-ion transitions} & & \\
\hline 
1 & 2.254395 & $-$29           & 18.2{\scriptsize (3.5)}           & \ion{Al}{iii} & 12.29{\scriptsize (0.07)} \\ 
2 & 2.254675 & \phantom{0}$-$3 & \phantom{0}5.1{\scriptsize (0.6)} & \ion{Al}{iii} & 12.48{\scriptsize (0.03)} \\
\hline
\end{tabular}
\begin{minipage}{140mm}
$^*$ Velocity relative to $z = 2.254710$
\end{minipage}
\end{table}
%

\subsubsection{$z_{\rm DLA/sub-DLA} = 2.255$}\label{Q1451-2p255}

This absorption metal-line system was identified thanks to the Ly$\alpha$ absorption line observed 
only $\sim 100$ \AA\ away from the blue edge of the quasar spectrum (the quasar flux cut-off is at 
$\sim 3835$~\AA). The normalization is very difficult in this region, but the limited number of 
Ly$\alpha$ forest absorptions leads to a clear high \ion{H}{i} column density system detection, the 
damping wings being well observed (see Fig.~\ref{Q1451-2p255-Ly}). We obtained a hydrogen column 
density measurement of $\log N$(\ion{H}{i}) $= 20.30\pm 0.15$, which shows that this absorber is a 
borderline case between the DLA and sub-DLA systems. A large error bar on $N$(\ion{H}{i}) was adopted 
to take into account the uncertainty in the continuum level. This absorption line system, although 
analysed in this paper, is not considered as a sub-DLA and is not included in the sub-DLA sample 
used for the different studies undertaken in Paper~II.
%

\begin{table}
\caption{Component structure of the $z_{\rm abs} = 3.171$ sub-DLA system towards Q1451+123} 
\label{Q1451-3p171-values}
\begin{tabular}{l c c c l c}
\hline
No & $z_{\rm abs}$ & $v_{\rm rel}^*$ & $b (\sigma _b)$ & Ion & $\log N (\sigma_{\log N})$ \\
   &               & km~s$^{-1}$     & km~s$^{-1}$     &     & cm$^{-2}$                  
\\     
\hline
\multicolumn{4}{l}{Low-ion transitions} & & \\
\hline
1 & 3.170795 &        $-$23 & 8.9{\scriptsize (1.3)} & \ion{Si}{ii} & 13.29{\scriptsize (0.04)} \\ 
  &	     &  	    &                        & \ion{O}{i}   & $< 14.46$  \\ 
  &	     &  	    &                        & \ion{C}{ii}  & 13.96{\scriptsize (0.18)} \\
  &	     &  	    &                        & \ion{Al}{ii} & 11.97{\scriptsize (0.03)} \\ 
  &	     &  	    &                        & \ion{Fe}{ii} & 12.99{\scriptsize (0.07)} \\ 
2 & 3.171109 & \phantom{0}0 & 9.2{\scriptsize (0.9)} & \ion{Si}{ii} & 13.38{\scriptsize (0.03)} \\
  &	     &  	    &                        & \ion{O}{i}   & $< 14.53$  \\
  &	     &  	    &                        & \ion{C}{ii}  & 14.04{\scriptsize (0.21)}  \\
  &	     &  	    &                        & \ion{Al}{ii} & 12.04{\scriptsize (0.03)} \\
  &	     &  	    &                        & \ion{Fe}{ii} & 13.06{\scriptsize (0.06)} \\
\hline
\multicolumn{4}{l}{High-ion transitions} & & \\
\hline
1 & 3.172149 & $+$75 & 14.3{\scriptsize (3.0)}  & \ion{C}{iv} & 13.39{\scriptsize (0.14)} \\
2 & 3.172427 & $+$95 & 12.4{\scriptsize (2.4)}  & \ion{C}{iv} & 12.78{\scriptsize (0.07)} \\
\hline
\end{tabular}
\begin{minipage}{140mm}
$^*$ Velocity relative to $z = 3.171109$
\end{minipage}
\end{table}
%

\begin{table}
\caption{Component structure of the $z_{\rm abs} = 2.087$ sub-DLA system towards Q1511+090} 
\label{Q1511-values}
\begin{tabular}{l c c c l c}
\hline
No & $z_{\rm abs}$ & $v_{\rm rel}^*$ & $b (\sigma _b)$ & Ion & $\log N (\sigma_{\log N})$ \\
   &               & km~s$^{-1}$     & km~s$^{-1}$     &     & cm$^{-2}$                  
\\     
\hline
\multicolumn{4}{l}{Low-ion transitions} & & \\
\hline
1 & 2.086977 &           $-$121 & \phantom{0}9.1{\scriptsize (0.3)} & \ion{Si}{ii} & 12.84{\scriptsize (0.01)} \\  
  &          &                  &                                   & \ion{O}{i}   & 13.14{\scriptsize (0.03)} \\
  &	     &  	        &                                   & \ion{Al}{ii} & 11.66{\scriptsize (0.18)} \\
  &	     &  	        &                                   & \ion{Fe}{ii} & 12.14{\scriptsize (0.02)} \\ 
2 & 2.087631 & \phantom{0}$-$57 &           30.0{\scriptsize (3.8)} & \ion{Si}{ii} & 12.87{\scriptsize (0.02)} \\  
  &          &                  &                                   & \ion{O}{i}   & $< 13.81$                 \\
  &	     &  	        &                                   & \ion{Al}{ii} & 12.29{\scriptsize (0.13)} \\
  &	     &  	        &                                   & \ion{Fe}{ii} & 12.12{\scriptsize (0.03)} \\ 
3 & 2.087833 & \phantom{0}$-$38 &           11.4{\scriptsize (1.0)} & \ion{Si}{ii} & 12.70{\scriptsize (0.03)} \\ 
  &          &                  &                                   & \ion{O}{i}   & $< 13.06$                 \\
  &	     &  	        &                                   & \ion{Al}{ii} & 11.81{\scriptsize (0.14)} \\ 
  &	     &  	        &                                   & \ion{Fe}{ii} & 12.05{\scriptsize (0.04)} \\
4 & 2.088223 &    \phantom{0+}0 &           13.5{\scriptsize (0.8)} & \ion{Si}{ii} & 12.98{\scriptsize (0.02)} \\  
  &          &                  &                                   & \ion{O}{i}   & $< 13.94$                 \\
  &	     &  	        &                                   & \ion{Al}{ii} & 11.78{\scriptsize (0.14)} \\ 
  &	     &  	        &                                   & \ion{Fe}{ii} & 12.19{\scriptsize (0.03)} \\ 
5 & 2.088643 & \phantom{0}$+$41 &           11.1{\scriptsize (1.0)} & \ion{Si}{ii} & 12.80{\scriptsize (0.01)} \\
  &          &                  &                                   & \ion{O}{i}   & $< 14.23$ \\
  &	     &  	        &                                   & \ion{Al}{ii} & 11.76{\scriptsize (0.16)} \\
  &	     &  	        &                                   & \ion{Fe}{ii} & 12.12{\scriptsize (0.03)} \\
6 & 2.088861 & \phantom{0}$+$62 & \phantom{0}5.6{\scriptsize (0.3)} & \ion{Si}{ii} & 13.17{\scriptsize (0.01)} \\
  &          &                  &                                   & \ion{O}{i}   & $< 14.52$ \\
  &	     &  	        &                                   & \ion{Al}{ii} & 11.99{\scriptsize (0.13)} \\
  &	     &  	        &                                   & \ion{Fe}{ii} & 12.37{\scriptsize (0.02)} \\
7 & 2.089054 & \phantom{0}$+$81 & \phantom{0}9.7{\scriptsize (0.5)} & \ion{Si}{ii} & 13.47{\scriptsize (0.18)} \\
  &          &                  &                                   & \ion{O}{i}   & $< 14.93$ \\
  &	     &  	        &                                   & \ion{Al}{ii} & 12.24{\scriptsize (0.13)} \\
  &	     &  	        &                                   & \ion{Fe}{ii} & 12.73{\scriptsize (0.03)} \\
8 & 2.089310 &           $+$106 &           17.6{\scriptsize (1.4)} & \ion{Si}{ii} & 13.50{\scriptsize (0.15)} \\
  &          &                  &                                   & \ion{O}{i}   & $<14.68$ \\ 
  &	     &  	        &                                   & \ion{Al}{ii} & 12.22{\scriptsize (0.13)} \\
  &	     &  	        &                                   & \ion{Fe}{ii} & 12.58{\scriptsize (0.04)} \\
\hline
\end{tabular}
\begin{minipage}{140mm}
$^*$ Velocity relative to $z = 2.088223$
\end{minipage}
\end{table}
%

Table~\ref{Q1451-2p255-values} and Fig.~\ref{Q1451-2p255-fits} present the velocity structure of the 
low-ion transitions in this absorber. It is characterized by 5 components spread over $\sim 130$ 
km~s$^{-1}$. We obtained the column density measurements of Zn$^+$, Cr$^+$ and Fe$^+$ from the 
\ion{Fe}{ii} $\lambda$2374,2382,2600 lines. Only the strongest components, the components 1, 2 and 3, 
are observed in the weak \ion{Zn}{ii} and \ion{Cr}{ii} lines detected thanks to the relatively high 
\ion{H}{i} column density of the system. Several interesting metal lines are unfortunately located 
in the spectral wavelength gap between 4855 and 5900 \AA. The only observed Si$^+$ line in the 
available spectra is the saturated \ion{Si}{ii} $\lambda$1304 line which is, in addition, blended 
with the damping red wing of the Ly$\alpha$ line of the DLA system at $z_{\rm abs} = 2.469$. To 
obtain a correct lower limit we first had to renormalize the portion of the quasar spectrum around 
the \ion{Si}{ii} line with the fit of the DLA Ly$\alpha$ damping wing profile. The \ion{Ni}{ii} 
$\lambda$1317,1370 lines located in the Ly$\alpha$ forest are detected at the noise level and at the 
uncertainty level of the continuum placement. We therefore considered the measured Ni$^+$ column 
density as an upper limit. The \ion{O}{i} $\lambda$1302, \ion{C}{ii} $\lambda$1334 and the 
\ion{Mg}{ii} doublet lines are also detected, but these lines are so highly saturated that no useful 
lower limit can be deduced.

From the higher ionization transitions, only the Al$^{++}$ intermediate-ion lines and the Si$^{+++}$ 
high-ion lines are covered in the available spectra. The \ion{Al}{iii} lines clearly show slightly
different profiles from the low-ion profiles, but with a very similar trend (see 
Fig.~\ref{Q1451-2p255-fits}). We provide an Al$^{++}$ column density measurement obtained with the 
fitting parameters given in Table~\ref{Q1451-2p255-values}. No column density measurement can be
obtained from the \ion{Si}{iv} doublet, since it is heavily blended with Ly$\alpha$ forest 
absorptions.
%

\begin{figure}
   \includegraphics[width=85mm]{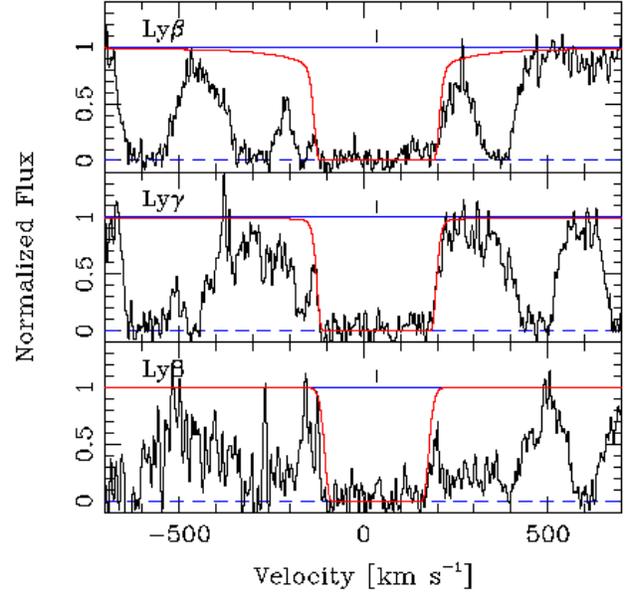}
\caption{Normalized UVES spectrum of Q1451+123 showing the sub-DLA Ly$\beta$, Ly$\gamma$ and Ly8 
profiles with the Voigt profile fits. The zero velocity is fixed at $z = 3.171109$. The vertical bar 
corresponds to the velocity centroid used for the best fit, $z = 3.171606$. The measured \ion{H}{i} 
column density is $\log N$(\ion{H}{i}) $= 19.70\pm 0.15$.}
\label{Q1451-3p171-Ly}
\end{figure}
%

\begin{figure*}
   \includegraphics[width=140mm]{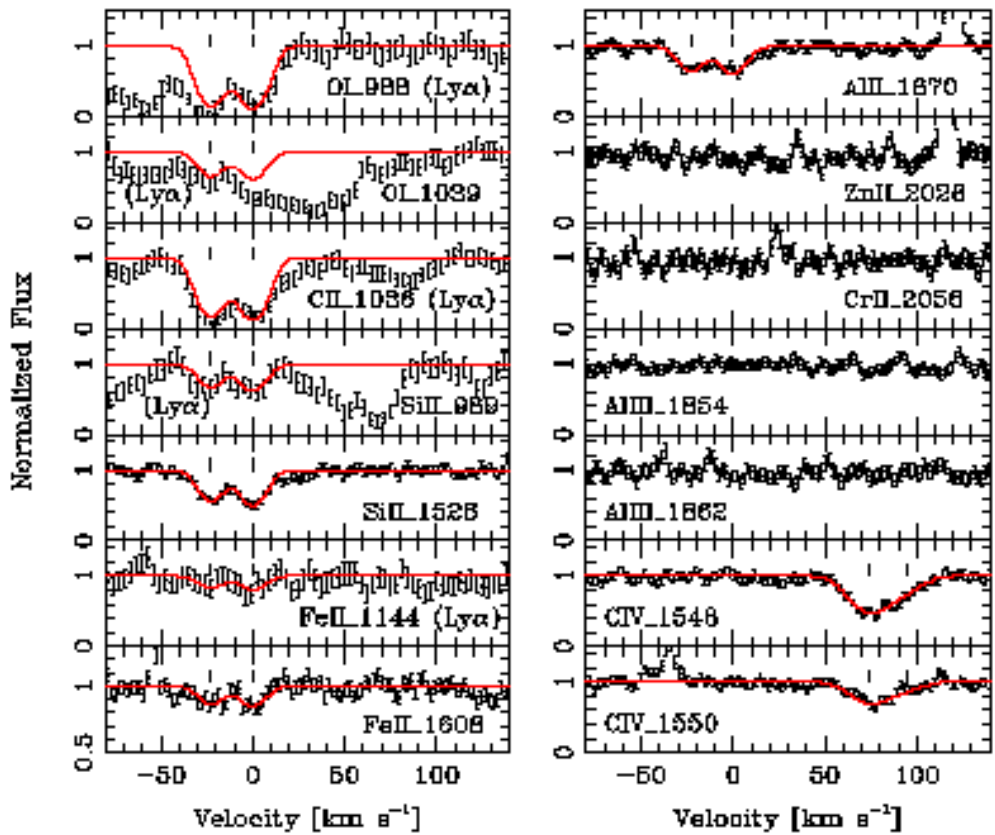}
\caption{Same as Fig.~\ref{Q1101-fits} for the sub-DLA towards Q1451+123. The zero velocity is
fixed at $z = 3.171109$. The vertical bars mark the positions of the fitted velocity components (see
Table~\ref{Q1451-3p171-values}).}
\label{Q1451-3p171-fits}
\end{figure*}
%

\subsubsection{$z_{\rm sub-DLA} = 3.171$}\label{Q1451-3p171}

The Ly$\alpha$ absorption line of this sub-DLA system is outside the available spectral coverage. We 
identified this sub-DLA system thanks to the \ion{H}{i} column density measurement reported by 
Petitjean, Srianand \& Ledoux (2000). They estimated the hydrogen column density of this system from 
the equivalent width measurement by Bechtold (1994) and give $\log N$(\ion{H}{i}) $\simeq 19.70$. We 
confirm this value with the measurements obtained from the fits of the Ly$\beta$, Ly$\gamma$ and Ly8 
lines (the redshift and the $b$-value were left as free parameters; see Fig.~\ref{Q1451-3p171-Ly}). 
The other members of the Lyman series were not taken into account, being heavily blended with 
Ly$\alpha$ forest absorptions.

This sub-DLA system shows a very simple velocity structure. The low-ion transition profiles are well
fitted with two components (see Table~\ref{Q1451-3p171-values} and Fig.~\ref{Q1451-3p171-fits}). 
We obtained the column density measurements of Si$^+$, C$^+$, Al$^+$ and Fe$^+$, and confirm the 
Petitjean, Srianand \& Ledoux (2000) [Fe/H] and [Si/H] abundance measurements. The \ion{C}{ii} 
column density was deduced from the \ion{C}{ii} $\lambda$1036 line located in the Ly$\alpha$ forest. 
We adopted a relatively large error on $N$(\ion{C}{ii}) to take into account possible \ion{H}{i} 
contaminations. We observe eight \ion{O}{i} lines in the available spectra, but the bulk of them are 
heavily blended with Ly$\alpha$ forest absorptions. Only the \ion{O}{i} $\lambda$988,1039 lines 
allowed us to obtain a significant upper limit on $N$(\ion{O}{i}). The \ion{Zn}{ii} and \ion{Cr}{ii} 
lines are not detected in the spectra, thus we provide only 4~$\sigma$ upper limits. No column 
density measurement of intermediate-ion transitions was obtained. The \ion{Fe}{iii} $\lambda$1122 
line is heavily blended in the Ly$\alpha$ forest, and the \ion{Al}{iii} lines are not detected, 
hence only a 4~$\sigma$ upper limit was derived.

Similarly to the sub-DLA system at $z_{\rm abs} = 2.087$ towards Q1444+014, the \ion{C}{iv} high-ion 
transition lines observed in this sub-DLA show a peculiar behaviour. They are shifted by $\sim 80$ 
km~s$^{-1}$ redwards of the low-ion transition lines. Although such a velocity difference between 
low- and high-ion lines is also observed in some DLA absorbers (Ledoux et~al. 1998), in this 
particular case, the low- and the high-ion profiles do {\it not} overlap at all in velocity space 
(see Table~\ref{Q1451-3p171-values} and Fig.~\ref{Q1451-3p171-fits}). This could suggest that the 
\ion{C}{iv} lines are not associated with the studied sub-DLA system. Under this assumption, it may 
mean that another metal-line system at $\sim +80$ km~s$^{-1}$ is blended with the absorber associated
with the low-ion transitions. However, this situation seems unlikely, since we should then observe 
the absorption lines of higher members of the Lyman series breaking into two components. A weak 
emission feature is observed at $\sim 40$ km~s$^{-1}$ in the Ly8 line (see 
Figure~\ref{Q1451-3p171-Ly}) and it may cast some doubt on the previous statement. However, given 
the low quality of the data, it is difficult to determine with confidence whether this feature is 
real or just noise. Indeed, the feature is detected at less than 2~$\sigma$ in the Ly8 line which 
is observed less than 15 \AA\ away from the blue edge of the quasar spectrum (the quasar flux cut-off 
is at $\sim 3835$ \AA), where the signal-to-noise ratio is very low (only $\sim 5$ FWHM). Therefore, 
with the available data we assume that the analysis of this absorption line system presented here is 
correct. However, we advise the reader that higher S/N spectra in the blue are required to determine 
whether the observed emission feature in the Ly8 line is real or not, since it will then imply a 
revision of the \ion{H}{i} column density measurement in this absorber. If in the end this system is 
not a sub-DLA, it has no impact on our unbiased sample of sub-DLAs and on the statistical work 
presented in Paper~II, since due to the non-observation of the Ly$\alpha$ line in this system, it 
was excluded from the sample of sub-DLAs used for the statistical work (see 
Section~\ref{sub-DLA-ident}).
%

\begin{figure}
   \includegraphics[width=85mm]{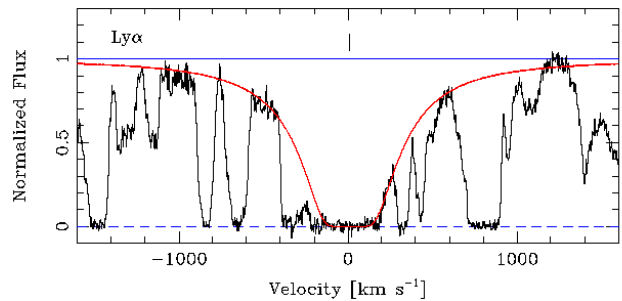}
\caption{Normalized UVES spectrum of Q1511+090 showing the sub-DLA Ly$\alpha$ profile with the Voigt
profile fit. The zero velocity is fixed at $z = 2.088223$. The vertical bar corresponds to the 
velocity centroid used for the best fit, $z = 2.088330$. The measured \ion{H}{i} column density is 
$\log N$(\ion{H}{i}) $= 19.47\pm 0.10$.}
\label{Q1511-Ly}
\end{figure}
%

\begin{figure*}
   \includegraphics[width=170mm]{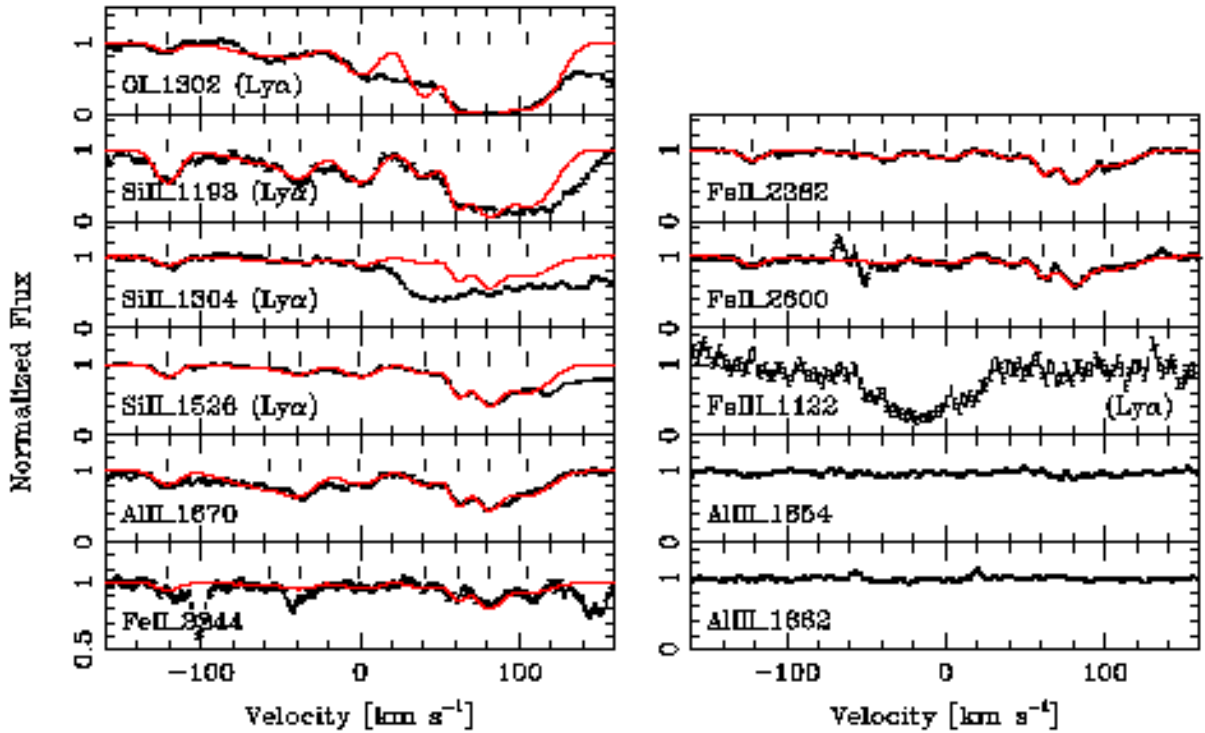}
\caption{Same as Fig.~\ref{Q1101-fits} for the sub-DLA towards Q1511+090. The zero velocity is
fixed at $z = 2.088223$. The vertical bars mark the positions of the fitted velocity components (see
Table~\ref{Q1511-values}).}
\label{Q1511-fits}
\end{figure*}
%

\subsection{Q1511+090, $z_{\rm sub-DLA} = 2.088$}

The absorption line systems towards this quasar were first analysed by Sargent, Steidel \& 
Boksenberg (1988). But, the sub-DLA system at $z_{\rm abs} = 2.088$ has never been previously 
studied, nor reported in the literature. It has been identified thanks to the Ly$\alpha$ absorption 
line, the only hydrogen line observed in the available spectra. The blue damping wing of the 
Ly$\alpha$ line is blended with other \ion{H}{i} clouds, and only the outer parts of the blue wing 
are visible. The red damping wing is on the contrary relatively well constrained and can be used to 
determine the hydrogen column density of the system (see Fig.~\ref{Q1511-Ly}). By fixing the 
$b$-value at 20 km~s$^{-1}$ and by leaving the redshift as a free parameter, we obtained a 
\ion{H}{i} column density of $\log N$(\ion{H}{i}) $= 19.47\pm 0.10$ as the best fit solution for the 
Ly$\alpha$ line. The derived redshift at $z = 2.08833$ is located between the components 4 and 5 of 
the low-ion transition profiles.

Table~\ref{Q1511-values} and Fig.~\ref{Q1511-fits} present the velocity structure of this sub-DLA 
system. The low-ion transition lines are well described by 8 components extended over a large 
velocity interval of $\sim 240$ km~s$^{-1}$. All the metal lines with rest-wavelengths up to 
$\lambda_{\rm rest} = 1527$ \AA\ are located in the Ly$\alpha$ forest, thus the column density 
determinations of a large number of elements have to be done with caution. We obtained the column 
density measurements of \ion{Si}{ii}, \ion{Al}{ii} and \ion{Fe}{ii}. The access to the strong 
\ion{Fe}{ii} $\lambda$2344,2382,2600 transitions allowed us to measure accurately $N$(\ion{Fe}{ii}) 
of this rather low metallicity system, [Fe/H] $= -1.71\pm 0.10$. The \ion{O}{i} $\lambda$1302 line 
is blended with Ly$\alpha$ forest lines, thus no reliable column density measurement can be 
determined, and we choose to consider the derived value as an upper limit. The case of the 
\ion{C}{ii} $\lambda$1334 line is even more difficult. The saturation of the line together with the 
strong blending, makes impossible the estimation of the column density of C$^+$. The Al$^{++}$ and 
Fe$^{++}$ intermediate-ion transition lines are not detected, hence we provide only 4~$\sigma$ upper 
limits on $N$(\ion{Al}{iii}) and $N$(\ion{Fe}{iii}) $-$ the \ion{Fe}{iii} $\lambda$1122 line is 
located in the Ly$\alpha$ forest, but the spectral region where the strongest components of the 
\ion{Fe}{iii} line should be detected is free from \ion{H}{i} blendings, and thus we can derive a 
reliable 4~$\sigma$ upper limit on $N$(\ion{Fe}{iii}). The observed high-ion transition lines, the 
\ion{C}{iv} and \ion{Si}{iv} doublets, are all heavily blended with Ly$\alpha$ forest lines 
preventing us from measuring their column densities.
%

\begin{table}
\caption{Component structure of the $z_{\rm abs} = 2.507$ sub-DLA system towards Q2059$-$360} 
\label{Q2059-values}
\begin{tabular}{l c c c l c}
\hline
No & $z_{\rm abs}$ & $v_{\rm rel}^*$ & $b (\sigma _b)$ & Ion & $\log N (\sigma_{\log N})$ \\
   &               & km~s$^{-1}$     & km~s$^{-1}$     &     & cm$^{-2}$                  
\\     
\hline
\multicolumn{4}{l}{Low-ion transitions} & & \\
\hline
1 & 2.507341 & \phantom{0}0 & 4.9{\scriptsize (0.2)} & \ion{Si}{ii} & 13.66{\scriptsize (0.08)} \\
  &	     &  	    &                        & \ion{O}{i}   & 15.52{\scriptsize (0.21)} \\
  &	     &  	    &                        & \ion{C}{ii}  & 14.82{\scriptsize (0.21)} \\
  &	     &  	    &                        & \ion{N}{i}   & 12.66{\scriptsize (0.18)} \\
  &	     &  	    &                        & \ion{Fe}{ii} & 13.41{\scriptsize (0.03)} \\
2 & 2.507535 &        $+$17 & 5.3{\scriptsize (0.8)} & \ion{Si}{ii} & 12.91{\scriptsize (0.25)} \\
  &	     &  	    &                        & \ion{O}{i}   & 14.11{\scriptsize (0.13)} \\
  &	     &  	    &                        & \ion{C}{ii}  & 13.87{\scriptsize (0.15)} \\
  &	     &  	    &                        & \ion{Fe}{ii} & 12.59{\scriptsize (0.04)} \\
\hline
\end{tabular}
\begin{minipage}{140mm}
$^*$ Velocity relative to $z = 2.507341$
\end{minipage}
\end{table}
%

\begin{figure}
   \includegraphics[width=85mm]{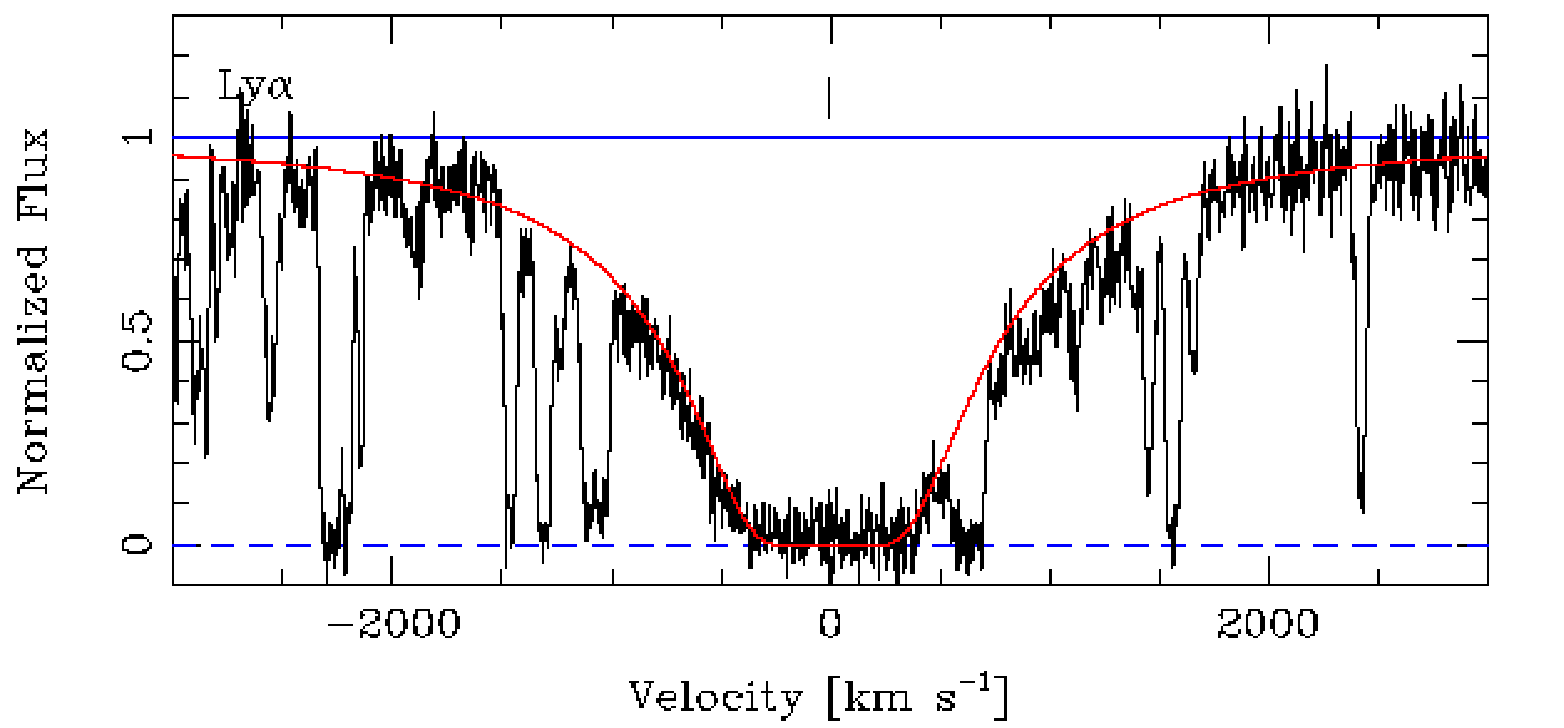}
\caption{Normalized UVES spectrum of Q2059$-$360 showing the sub-DLA Ly$\alpha$ profile with the 
Voigt profile fit. The zero velocity is fixed at $z = 2.507341$. The vertical bar corresponds to the 
velocity centroid of the strongest component of the low-ion transition lines used for the best fit, 
the component 1 at $z = 2.507341$. The measured \ion{H}{i} column density is $\log N$(\ion{H}{i}) 
$= 20.21\pm 0.10$.}
\label{Q2059-Ly}
\end{figure}
%

\begin{figure*}
   \includegraphics[width=140mm]{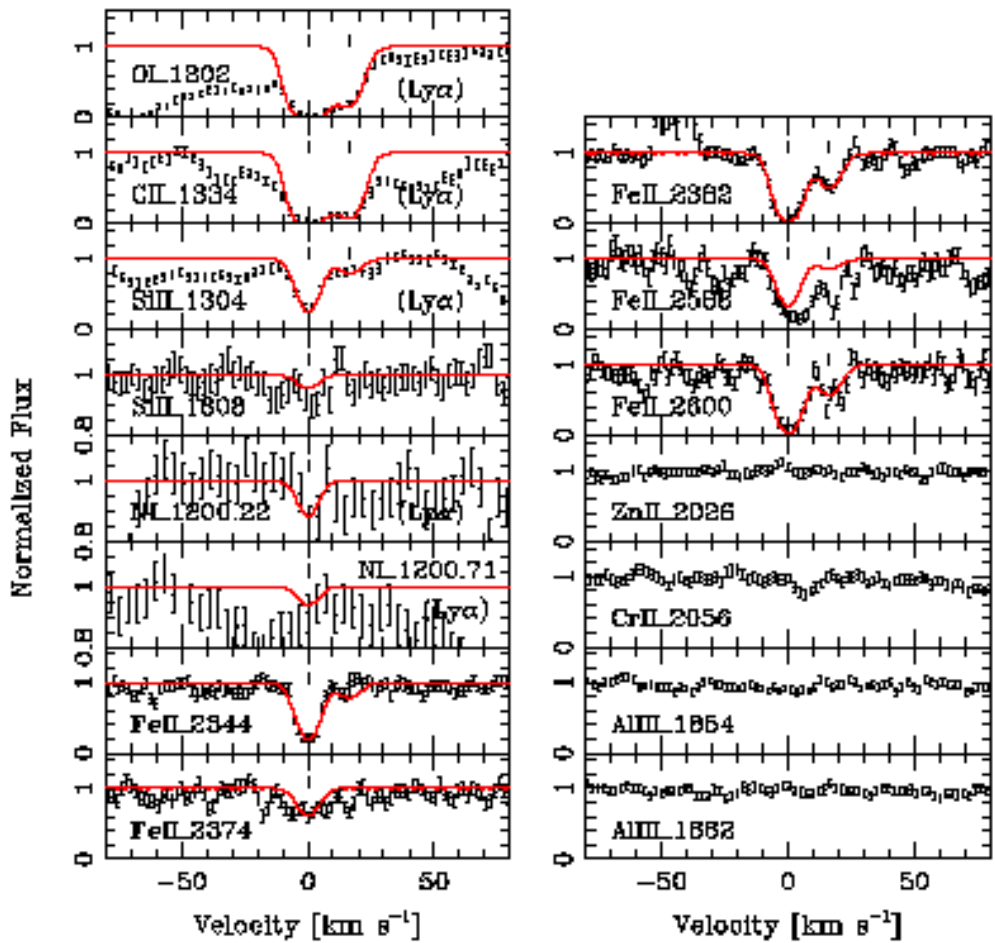}
\caption{Same as Fig.~\ref{Q1101-fits} for the sub-DLA towards Q2059$-$360. The zero velocity is
fixed at $z = 2.507341$. The vertical bars mark the positions of the fitted velocity components (see
Table~\ref{Q2059-values}).}
\label{Q2059-fits}
\end{figure*}
%

\subsection{Q2059$-$360, $z_{\rm sub-DLA} = 2.507$}

The quasar Q1451+123 was discovered by Warren, Hewett \& Osmer (1991), and it has been studied for 
its intervening DLA system at $z_{\rm abs} = 3.083$. The molecular content of this DLA has been 
analysed by Petitjean, Srianand \& Ledoux (2000). We have found in the spectra of Q2059$-$360 an 
additional high hydrogen column density absorption line system, a sub-DLA at $z_{\rm abs} = 2.507$. 
It has been identified thanks to the Ly$\alpha$ absorption line which shows well defined damping 
wings (see Fig.~\ref{Q2059-Ly}). No other hydrogen line is observed in the available spectra. The 
fit of the Ly$\alpha$ line was performed by fixing the $b$-value at 20 km~s$^{-1}$ and the redshift 
at the redshift of the strongest component of the low-ion transition lines, the component 1. The 
derived \ion{H}{i} column density is $\log N$(\ion{H}{i}) $= 20.21\pm 0.10$.

The low-ion transition profiles are characterized by a simple velocity structure composed of two 
components (see Table~\ref{Q2059-values} and Fig.~\ref{Q2059-fits}). In the weaker lines, only the 
strongest component at $v = 0$ is observed. We obtained an accurate Fe$^+$ column density measurement 
thanks to the numerous observed \ion{Fe}{ii} lines at $\lambda_{\rm rest} = 2344$, 2374, 2382, 2586 
and 2600, and despite the fact that the reddest parts of the available spectra at $\lambda > 8000$ 
\AA\ are noisy with a S/N per pixel lower than 10. The observed \ion{Si}{ii} $\lambda$1304, 
\ion{O}{i} $\lambda$1302 and \ion{C}{ii} $\lambda$1334 lines are all located in the Ly$\alpha$ 
forest. Large errors have been adopted on their column density measurements to take into account 
possible Ly$\alpha$ contaminations. In the case of Si$^{+}$, we could constrain the column density 
of the strongest component with the \ion{Si}{ii} $\lambda$1808 line located outside the Ly$\alpha$ 
forest. However, the derived O$^0$ and C$^+$ column density measurements might be considered as 
borderline cases between values and limits. We also obtained a column density measurement of N$^0$ 
from the \ion{N}{i} $\lambda$1200.223,1200.710 lines, the \ion{N}{i} $\lambda$1199.550 line being 
blended. This measurement is marginal, since the detected \ion{N}{i} lines are extremely weak, at 
the limit of the noise level. The \ion{Zn}{ii} and \ion{Cr}{ii} lines are not detected despite the 
relatively high \ion{H}{i} column density of the system, thus we provide 4~$\sigma$ upper limits. We 
also give the 4~$\sigma$ upper limit on the Al$^{++}$ intermediate-ion transition column density, 
the \ion{Al}{iii} lines being not detected. No high-ion transition column density has been measured, 
since the \ion{Si}{iv} lines are heavily blended with Ly$\alpha$ forest absorptions and the 
\ion{C}{iv} lines are not covered in the available spectra.
%

\begin{figure}
   \includegraphics[width=85mm]{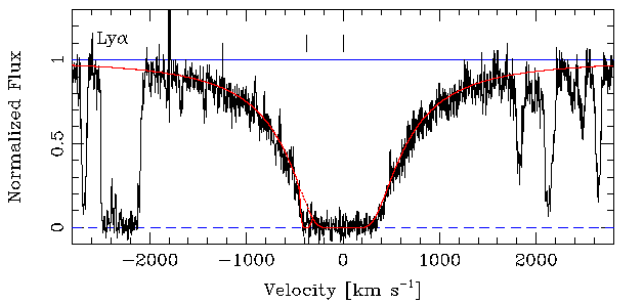}
\caption{Normalized UVES spectrum of Q2116$-$358 showing the sub-DLA Ly$\alpha$ profile with the Voigt
profile fit. The zero velocity is fixed at $z = 1.996119$. The vertical bars correspond from right to
left to the velocity centroids of the component 6 of the low-ion transition lines at $z = 1.996224$ 
and $z = 1.992317$ used for the best fit, belonging to the sub-DLA system and an additional absorber, 
respectively. The dotted line corresponds to the fit with the sub-DLA only. The measured \ion{H}{i} 
column density of the sub-DLA is $\log N$(\ion{H}{i}) $= 20.06\pm 0.10$.}
\label{Q2116-Ly}
\end{figure} 
%

\begin{figure*}
   \includegraphics[width=170mm]{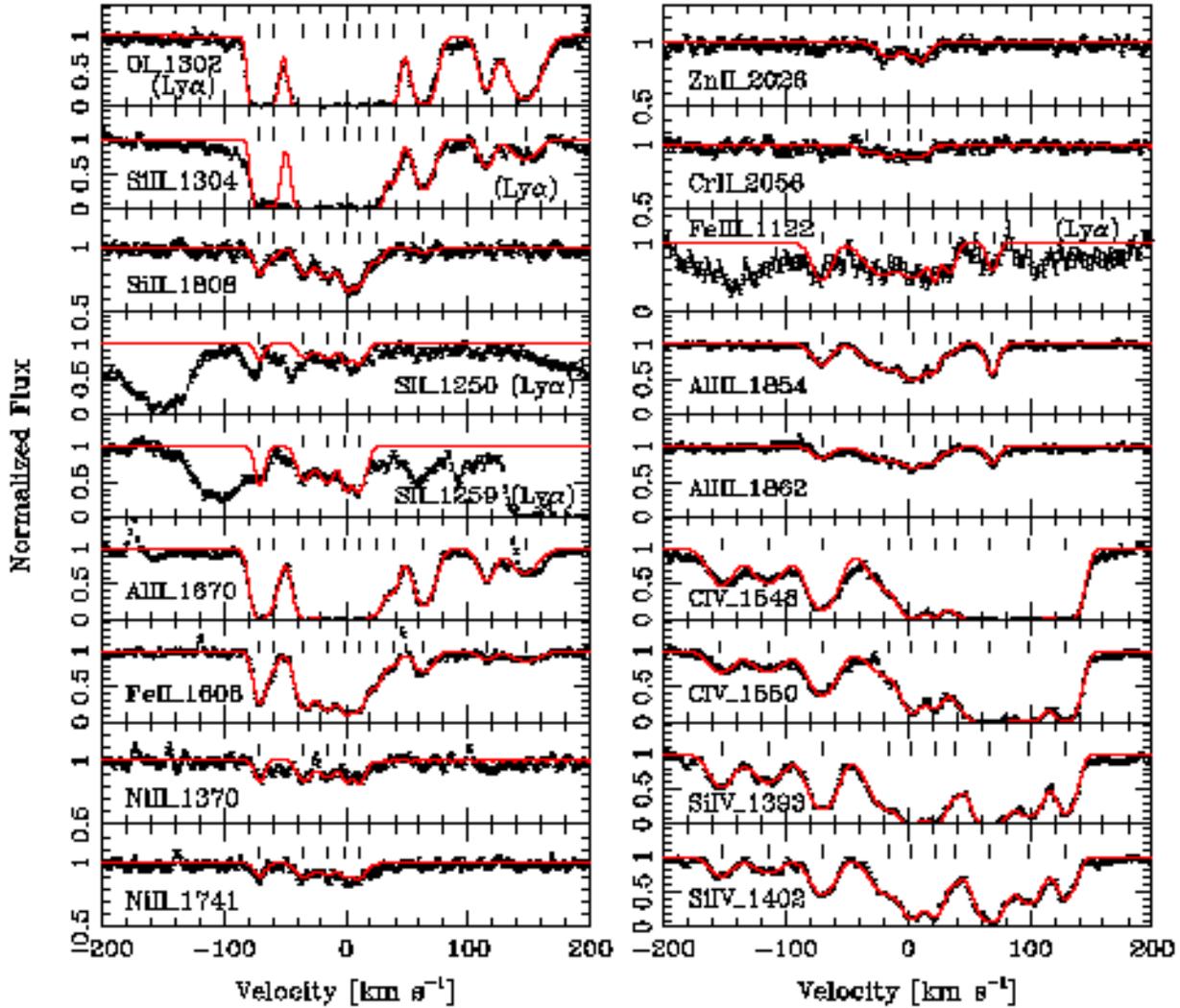}
\caption{Same as Fig.~\ref{Q1101-fits} for the sub-DLA towards Q2116$-$358. The zero velocity is
fixed at $z = 1.996119$. The vertical bars mark the positions of the fitted velocity components (see
Table~\ref{Q2116-values}).}
\label{Q2116-fits}
\end{figure*}
%

\subsection{Q2116$-$358, $z_{\rm sub-DLA} = 1.996$}

The quasar Q2116$-$358 was discovered by Adam (1985). Its intervening absorption metal-line system at 
$z_{\rm abs} = 1.996$ has previously been reported by Lanzetta, Wolfe \& Turnshek (1987), Wampler, 
Bergeron \& Petitjean (1993), M\o ller, Jakobsen \& Perryman (1994), and it has even been targeted in 
an imaging programme by Warren et~al. (2001). In this paper we provide a new complete abundance 
analysis of this system using the UVES high-resolution spectra. 
%

\begin{table*}
\caption{Component structure of the $z_{\rm abs} = 1.996$ sub-DLA system towards Q2116$-$358} 
\label{Q2116-values}
\begin{tabular}{l c c c l c | l c c c l c}
\hline
No & $z_{\rm abs}$ & $v_{\rm rel}^*$ & $b (\sigma _b)$ & Ion & $\log N (\sigma_{\log N})$ & No & $z_{\rm abs}$ & $v_{\rm rel}^*$ & $b (\sigma _b)$ & Ion & $\log N (\sigma_{\log N})$ \\
   &               & km~s$^{-1}$     & km~s$^{-1}$     &     & cm$^{-2}$                  &    &               & km~s$^{-1}$     & km~s$^{-1}$     &     & cm$^{-2}$                  
\\     
\hline
\multicolumn{4}{l}{Low-ion transitions} & & & & & & & & \\
\hline
1 & 1.995416 &         $-$70 & 5.0{\scriptsize (0.8)} & \ion{Si}{ii} & 14.39{\scriptsize (0.07)} &  6 & 1.996224 & \phantom{0}$+$10 & \phantom{0}7.5{\scriptsize (2.6)} & \ion{Si}{ii} & 14.71{\scriptsize (0.14)} \\
  &	     &  	     &                        & \ion{O}{i}   & $> 15.69$                 &    & 	 &		    &					& \ion{O}{i}   & $> 16.00$		   \\
  &          &               &                        & \ion{S}{ii}  & $< 14.12$                 &    & 	 &		    &					& \ion{S}{ii}  & $< 14.42$  \\
  &	     &  	     &                        & \ion{Al}{ii} & 13.29{\scriptsize (0.17)} &    & 	 &		    &					& \ion{Al}{ii} & $> 13.36$		   \\
  &	     &  	     &                        & \ion{Ni}{ii} & 12.79{\scriptsize (0.07)} &    & 	 &		    &					& \ion{Ni}{ii} & 12.99{\scriptsize (0.05)} \\
  &	     &  	     &                        & \ion{Fe}{ii} & 13.83{\scriptsize (0.03)} &    & 	 &		    &					& \ion{Fe}{ii} & 14.05{\scriptsize (0.13)} \\
2 & 1.995528 &         $-$59 & 4.8{\scriptsize (1.1)} & \ion{Si}{ii} & 13.93{\scriptsize (0.18)} &    & 	 &		    &					& \ion{Zn}{ii} & 11.94{\scriptsize (0.08)} \\ 
  &	     &  	     &                        & \ion{O}{i}   & 14.11{\scriptsize (0.16)} &    & 	 &		    &					& \ion{Cr}{ii} & 12.42{\scriptsize (0.11)} \\
  &	     &  	     &                        & \ion{Al}{ii} & 12.22{\scriptsize (0.15)} &  7 & 1.996373 & \phantom{0}$+$25 & \phantom{0}5.5{\scriptsize (1.8)} & \ion{Si}{ii} & 14.07{\scriptsize (0.13)} \\
  &	     &  	     &                        & \ion{Fe}{ii} & 13.06{\scriptsize (0.19)} &    & 	 &		    &					& \ion{O}{i}   & $> 16.00$		   \\
3 & 1.995777 &         $-$34 & 7.2{\scriptsize (1.3)} & \ion{Si}{ii} & 14.49{\scriptsize (0.09)} &    & 	 &		    &					& \ion{Al}{ii} & 12.51{\scriptsize (0.16)} \\
  &	     &  	     &                        & \ion{O}{i}   & $> 15.32$                 &    & 	 &		    &					& \ion{Fe}{ii} & 13.48{\scriptsize (0.07)} \\
  &          &               &                        & \ion{S}{ii}  & $< 14.20$                 &  8 & 1.996504 & \phantom{0}$+$38 & \phantom{0}5.6{\scriptsize (0.8)} & \ion{Si}{ii} & 13.50{\scriptsize (0.05)} \\
  &	     &  	     &                        & \ion{Al}{ii} & $> 12.94$                 &    & 	 &		    &					& \ion{O}{i}   & 14.20{\scriptsize (0.15)} \\
  &	     &  	     &                        & \ion{Ni}{ii} & 12.93{\scriptsize (0.06)} &    & 	 &		    &					& \ion{Al}{ii} & 12.17{\scriptsize (0.03)} \\
  &	     &  	     &                        & \ion{Fe}{ii} & 14.00{\scriptsize (0.08)} &    & 	 &		    &					& \ion{Fe}{ii} & 13.03{\scriptsize (0.07)} \\
  &	     &  	     &                        & \ion{Cr}{ii} & 12.15{\scriptsize (0.20)} &  9 & 1.996759 & \phantom{0}$+$64 & \phantom{0}9.2{\scriptsize (0.3)} & \ion{Si}{ii} & 13.82{\scriptsize (0.03)} \\
4 & 1.995963 &         $-$16 & 9.5{\scriptsize (2.4)} & \ion{Si}{ii} & 14.66{\scriptsize (0.12)} &    & 	 &		    &					& \ion{O}{i}   & 14.53{\scriptsize (0.10)} \\
  &	     &  	     &                        & \ion{O}{i}   & $> 16.00$                 &    & 	 &		    &					& \ion{Al}{ii} & 12.56{\scriptsize (0.01)} \\
  &          &               &                        & \ion{S}{ii}  & $< 14.35$                 &    & 	 &		    &					& \ion{Fe}{ii} & 13.41{\scriptsize (0.04)} \\
  &	     &  	     &                        & \ion{Al}{ii} & $> 13.14$                 & 10 & 1.997269 &	     $+$115 & \phantom{0}7.6{\scriptsize (0.6)} & \ion{Si}{ii} & 13.34{\scriptsize (0.06)} \\
  &	     &  	     &                        & \ion{Ni}{ii} & 12.95{\scriptsize (0.06)} &    & 	 &		    &					& \ion{O}{i}   & 14.05{\scriptsize (0.03)} \\
  &	     &  	     &                        & \ion{Fe}{ii} & 14.13{\scriptsize (0.11)} &    & 	 &		    &					& \ion{Al}{ii} & 12.06{\scriptsize (0.03)} \\
  &	     &  	     &                        & \ion{Zn}{ii} & 11.96{\scriptsize (0.08)} &    & 	 &		    &					& \ion{Fe}{ii} & 12.89{\scriptsize (0.11)} \\
  &	     &  	     &                        & \ion{Cr}{ii} & 12.49{\scriptsize (0.10)} & 11 & 1.997586 &	     $+$147 &		13.7{\scriptsize (0.5)} & \ion{Si}{ii} & 13.42{\scriptsize (0.10)} \\
5 & 1.996119 & \phantom{0-}0 & 4.0{\scriptsize (2.3)} & \ion{Si}{ii} & 14.46{\scriptsize (0.16)} &    & 	 &		    &					& \ion{O}{i}   & 14.50{\scriptsize (0.05)} \\
  &	     &  	     &                        & \ion{O}{i}   & $> 16.00$                 &    & 	 &		    &					& \ion{Al}{ii} & 12.12{\scriptsize (0.15)} \\
  &          &               &                        & \ion{S}{ii}  & $< 14.06$                 &    & 	 &		    &					& \ion{Fe}{ii} & 13.18{\scriptsize (0.12)} \\
  &	     &  	     &                        & \ion{Al}{ii} & $> 12.96$                 & & & & & & \\
  &	     &  	     &                        & \ion{Ni}{ii} & 12.61{\scriptsize (0.10)} & & & & & & \\
  &	     &  	     &                        & \ion{Fe}{ii} & 13.88{\scriptsize (0.10)} & & & & & & \\
  &	     &  	     &                        & \ion{Zn}{ii} & 11.23{\scriptsize (0.24)} & & & & & & \\
  &	     &  	     &                        & \ion{Cr}{ii} & 12.07{\scriptsize (0.20)} & & & & & & \\
\hline
\multicolumn{4}{l}{Intermediate-ion transitions} & & & & & & & & \\
\hline
1 & 1.995417 &           $-$70 & \phantom{0}9.8{\scriptsize (1.0)} & \ion{Al}{iii} & 12.38{\scriptsize (0.03)} & 4 & 1.996339 & $+$22 & 3.9{\scriptsize (1.9)} & \ion{Al}{iii} & 12.10{\scriptsize (0.20)} \\
  &	     &  	       &                                   & \ion{Fe}{iii} & $< 13.93$	               &   &	      &       & 		       & \ion{Fe}{iii} & $< 13.49$  \\
2 & 1.995903 &           $-$22 &           17.1{\scriptsize (2.9)} & \ion{Al}{iii} & 12.67{\scriptsize (0.10)} & 5 & 1.996458 & $+$34 & 6.4{\scriptsize (2.5)} & \ion{Al}{iii} & 11.87{\scriptsize (0.11)} \\
  &	     &  	       &                                   & \ion{Fe}{iii} & $< 14.06$	               &   &	      &       & 		       & \ion{Fe}{iii} & $< 13.62$  \\   
3 & 1.996172 & \phantom{0}$+$5 &           13.5{\scriptsize (3.0)} & \ion{Al}{iii} & 12.80{\scriptsize (0.11)} & 6 & 1.996814 & $+$69 & 5.5{\scriptsize (0.6)} & \ion{Al}{iii} & 12.40{\scriptsize (0.02)} \\
  &	     &  	       &                                   & \ion{Fe}{iii} & $< 13.99$	               &   &	      &       & 		       & \ion{Fe}{iii} & $< 13.51$  \\
\hline
\multicolumn{4}{l}{High-ion transitions} & & & & & & & & \\
\hline
1 & 1.994594 &           $-$153 &           12.5{\scriptsize (0.7)} & \ion{Si}{iv} & 12.88{\scriptsize (0.02)} &  6... & 1.996344 & \phantom{0}$+$22 & \phantom{0}9.7{\scriptsize (0.8)} & \ion{Si}{iv} & 13.53{\scriptsize (0.03)} \\
  &	     &  	        &                                   & \ion{C}{iv}  & 13.33{\scriptsize (0.06)} &       &	  &		     &  				 & \ion{C}{iv}  & 13.89{\scriptsize (0.14)} \\
2 & 1.994974 &           $-$115 &           16.3{\scriptsize (1.1)} & \ion{Si}{iv} & 12.90{\scriptsize (0.02)} &  7... & 1.996517 & \phantom{0}$+$40 & \phantom{0}6.4{\scriptsize (0.8)} & \ion{Si}{iv} & 12.65{\scriptsize (0.05)} \\
  &	     &  	        &                                   & \ion{C}{iv}  & 13.41{\scriptsize (0.04)} &       &	  &		     &  				 & \ion{C}{iv}  & 13.48{\scriptsize (0.16)} \\
3 & 1.995426 & \phantom{0}$-$69 &           13.6{\scriptsize (0.3)} & \ion{Si}{iv} & 13.29{\scriptsize (0.01)} &  8... & 1.996789 & \phantom{0}$+$67 &  	 13.7{\scriptsize (0.5)} & \ion{Si}{iv} & 13.81{\scriptsize (0.01)} \\
  &	     &  	        &                                   & \ion{C}{iv}  & 13.80{\scriptsize (0.02)} &       &	  &		     &  				 & \ion{C}{iv}  & $> 15.03$  \\
4 & 1.995957 & \phantom{0}$-$16 &           17.4{\scriptsize (1.3)} & \ion{Si}{iv} & 13.41{\scriptsize (0.05)} &  9... & 1.997112 & \phantom{0}$+$99 &  	 13.2{\scriptsize (0.5)} & \ion{Si}{iv} & 13.44{\scriptsize (0.01)} \\
  &	     &  	        &                                   & \ion{C}{iv}  & 13.55{\scriptsize (0.07)} &       &	  &		     &  				 & \ion{C}{iv}  & $> 14.43$  \\      
5 & 1.996150 & \phantom{00}$+$3 & \phantom{0}9.1{\scriptsize (0.6)} & \ion{Si}{iv} & 13.52{\scriptsize (0.04)} & 10... & 1.997414 &	      $+$130 & \phantom{0}9.7{\scriptsize (0.4)} & \ion{Si}{iv} & 13.22{\scriptsize (0.02)} \\
  &	     &  	        &                                   & \ion{C}{iv}  & 13.92{\scriptsize (0.13)} &       &	  &		     &  				 & \ion{C}{iv}  & 14.33{\scriptsize (0.12)} \\      
\hline
\end{tabular}
\begin{minipage}{140mm}
$^*$ Velocity relative to $z = 1.996119$
\end{minipage}
\end{table*}
%

Only the hydrogen Ly$\alpha$ line is observed in the available data. It shows well defined damping 
wings which allowed an accurate measurement of the \ion{H}{i} column density of the system. We 
obtained a value of $\log N$(\ion{H}{i}) $= 20.06\pm 0.10$, lower by $\sim 0.1$~dex than the value 
reported by M\o ller, Jakobsen \& Perryman (1994), and which indicates that the absorber is a 
sub-DLA system. Fig.~\ref{Q2116-Ly} shows the best fit solution of the Ly$\alpha$ line. The presence 
of an asymmetry in the profiles of the blue and red damping wings made it necessary to include, 
besides the contribution of the sub-DLA system, a second low column density absorption system 
bluewards of the sub-DLA in order to obtain a good fit of both wings of the Ly$\alpha$ line. During 
the fitting procedure, the redshift and the $b$-value of the sub-DLA system were fixed at the 
redshift of the component 6 of the low-ion transition lines and at 20 km~s$^{-1}$, respectively, and 
the redshift and the $b$-value of the second absorber were left as free parameters. We obtained the 
best fit solution with the second absorption line system at $z = 1.992317$ shifted by about $-391$ 
km~s$^{-1}$ from the sub-DLA system and with a \ion{H}{i} column density of $\log N$(\ion{H}{i}) 
$= 14.59\pm 0.10$.

Metal lines from a wide range of ionization species are present in this sub-DLA system. 
Fig.~\ref{Q2116-fits} shows the low-, intermediate- and high-ion transition line profiles, and their 
velocity structures are described in Table~\ref{Q2116-values}. The low-ion transition lines are 
characterized by a relatively complex velocity structure of 11 components spread over $\sim 220$ 
km~s$^{-1}$. Only the strongest components, the components 3 to 6, are observed in the weak 
metal lines, such as the \ion{Ni}{ii}, \ion{Cr}{ii} and \ion{Zn}{ii} lines. We obtained the column 
density measurements of Si$^+$, Ni$^+$, Cr$^+$, Zn$^+$ and Fe$^+$. Many components of the \ion{O}{i} 
$\lambda$1302 and \ion{Al}{ii} $\lambda$1670 lines are strongly saturated, while the weaker 
components are well matched by the model parameters defined from the \ion{Si}{ii} and \ion{Fe}{ii} 
lines. Since no reliable value of the total column density of O$^0$ and Al$^+$ can be deduced, we 
provide only lower limits. We detect also the \ion{S}{ii} $\lambda$1250,1259 lines. But, since they 
are blended with Ly$\alpha$ forest absorptions, we preferred to consider the measured S$^+$ column 
density as an upper limit.

Relatively strong \ion{Al}{iii} lines are observed in this sub-DLA system (see 
Fig.~\ref{Q2116-fits}). These intermediate-ion transition lines show profiles which are very similar 
to the low-ion profiles, but need to be fitted with slightly different model parameters given in 
Table~\ref{Q2116-values}. The \ion{Fe}{iii} $\lambda$1122 intermediate-ion line is also observed. 
Nevertheless, it is located in the Ly$\alpha$ forest and is blended, therefore we provide only an 
upper limit on its column density. The observed high-ion transition lines are very strong and show a 
complex velocity structure composed of 10 components. They extend over a wider velocity range of 
$\sim 290$ km~s$^{-1}$ than the low-ion profiles (see Fig.~\ref{Q2116-fits}). We obtained the column 
density measurement of \ion{Si}{iv} and a lower limit on $N$(\ion{C}{iv}) $-$ two components of the 
\ion{C}{iv} doublet are heavily saturated.
%

\subsection{PSS J2155+1358}\label{J2155}

High signal-to-noise ratio, 5 \AA\ resolution (FWHM) spectra of this $z_{\rm em} > 4$ quasar have 
been obtained by P\'eroux et~al. (2001). They identified a high \ion{H}{i} column density DLA system 
at $z_{\rm abs} = 3.316$. The metallicity of this DLA system was first reported by Dessauges-Zavadsky 
et~al. (2001b). Here we present the chemical analysis of three newly identified sub-DLA systems at 
$z_{\rm abs} = 3.142$, $3.565$ and $4.212$ towards this quasar.
%

\begin{figure}
   \includegraphics[width=85mm]{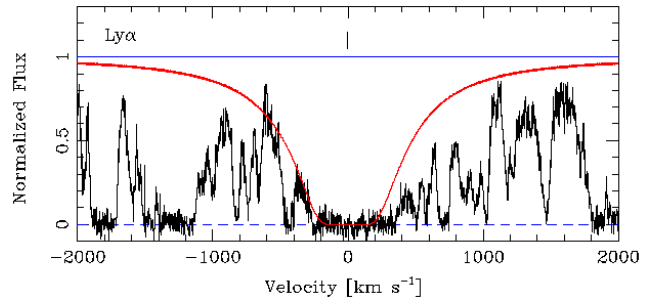}
\caption{Normalized UVES spectrum of PSS J2155+1358 showing the sub-DLA Ly$\alpha$ profile with the 
Voigt profile fit. The zero velocity is fixed at $z = 3.141998$. The vertical bar corresponds to the 
velocity centroid of the strongest component of the low-ion transition lines used for the best fit, 
the component 1 at $z = 3.141998$. The measured \ion{H}{i} column density is $\log N$(\ion{H}{i}) $= 
19.94\pm 0.10$.}
\label{J2155-3p142-Ly}
\end{figure}
%

\begin{table}
\caption{Component structure of the $z_{\rm abs} = 3.142$ sub-DLA system towards PSS J2155+1358} 
\label{J2155-3p142-values}
\begin{tabular}{l c c c l c}
\hline
No & $z_{\rm abs}$ & $v_{\rm rel}^*$ & $b (\sigma _b)$ & Ion & $\log N (\sigma_{\log N})$ \\
   &               & km~s$^{-1}$     & km~s$^{-1}$     &     & cm$^{-2}$                  
\\     
\hline
\multicolumn{5}{l}{Low- and intermediate-ion transitions} & \\
\hline
1 & 3.141998 & \phantom{0+}0 & \phantom{0}5.7{\scriptsize (1.0)} & \ion{Si}{ii}  & 13.36{\scriptsize (0.06)} \\
  &	     &  	     &                                   & \ion{O}{i}    & $< 14.22$ \\
  &	     &  	     &                                   & \ion{C}{ii}   & 14.05{\scriptsize (0.20)} \\
  &	     &  	     &                                   & \ion{Al}{ii}  & 12.07{\scriptsize (0.07)} \\
  &	     &  	     &                                   & \ion{Fe}{ii}  & 12.77{\scriptsize (0.20)} \\
  &	     &  	     &                                   & \ion{Al}{iii} & $< 11.57$ \\
2 & 3.142313 &         $+$23 &           14.5{\scriptsize (2.6)} & \ion{Si}{ii}  & 12.76{\scriptsize (0.10)} \\
  &	     &  	     &                                   & \ion{O}{i}	 & $< 13.91$ \\
  &	     &  	     &                                   & \ion{C}{ii}   & $< 13.66$ \\
  &	     &  	     &                                   & \ion{Al}{ii}  & 11.50{\scriptsize (0.18)} \\
3 & 3.142712 &         $+$52 &           10.0{\scriptsize (2.8)} & \ion{Si}{ii}  & 13.20{\scriptsize (0.07)} \\        
  &	     &  	     &                                   & \ion{O}{i}    & $< 14.42$ \\     
  &	     &  	     &                                   & \ion{C}{ii}   & $< 14.00$ \\     
  &	     &  	     &                                   & \ion{Al}{ii}  & 11.81{\scriptsize (0.13)} \\     
  &	     &  	     &                                   & \ion{Fe}{ii}  & 13.04{\scriptsize (0.18)} \\     
\hline
\end{tabular}
\begin{minipage}{140mm}
$^*$ Velocity relative to $z = 3.141998$
\end{minipage}
\end{table}
%

\begin{figure*}
   \includegraphics[width=140mm]{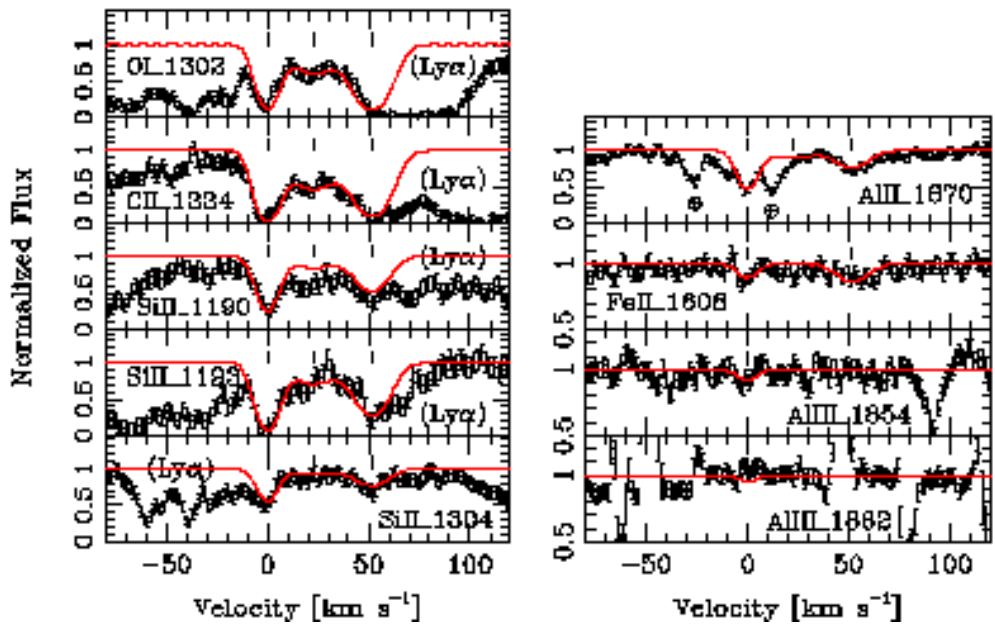}
\caption{Same as Fig.~\ref{Q1101-fits} for the sub-DLA towards PSS J2155+1358. The zero velocity is
fixed at $z = 3.141998$. The vertical bars mark the positions of the fitted velocity components (see
Table~\ref{J2155-3p142-values}).}
\label{J2155-3p142-fits}
\end{figure*}
%

\subsubsection{$z_{\rm sub-DLA} = 3.142$}

This sub-DLA system has been identified thanks to the Ly$\alpha$ absorption line. No other hydrogen
lines are observed in this sub-DLA, since they are beyond the quasar flux cut-off. Its \ion{H}{i} 
column density is thus poorly constrained, all the more since the Ly$\alpha$ line lies in the far 
blue end of the quasar spectrum. Due to the high number of \ion{H}{i} absorptions present in these 
spectral regions, the Ly$\alpha$ damping wings are strongly blended and the continuum placement is 
uncertain. Fig.~\ref{J2155-3p142-Ly} shows the best fit solution obtained by fixing the $b$-value at 
20 km~s$^{-1}$ and the redshift at the redshift of the strongest component, the component 1, of the 
low-ion transition lines. We derived a hydrogen column density of $\log N$(\ion{H}{i}) $= 19.94\pm 
0.10$.

Few metal lines were detected in this sub-DLA system, mainly because all of the lines with 
rest-wavelengths less than 1543 \AA\ are located in the Ly$\alpha$ forest. The velocity structure of 
the low-ion transition lines was deduced from the \ion{Si}{ii}, \ion{O}{i}, \ion{C}{ii}, 
\ion{Al}{ii} and \ion{Fe}{ii} lines, all observed in the Ly$\alpha$ forest apart from the 
\ion{Al}{ii} and \ion{Fe}{ii} lines. It is well described by three components (see 
Table~\ref{J2155-3p142-values} and Fig.~\ref{J2155-3p142-fits}). We were able to measure the column 
density of Si$^+$ thanks to the detection of three \ion{Si}{ii} lines at $\lambda_{\rm rest} = 
1190$, 1193 and 1304 \AA. We obtained the column density measurement of Al$^+$ from the \ion{Al}{ii} 
$\lambda$1670 line whose velocity components are just not blended with telluric lines, which we 
identified with the help of the spectrum of a hot, fast rotating star observed the same night as the 
quasar. We deduced the Fe$^+$ column density from the \ion{Fe}{ii} $\lambda$1608 line. The weakness 
of this \ion{Fe}{ii} line observed in this case just above the noise level and which is usually 
almost saturated in the DLA systems indicates that the metallicity in this sub-DLA is low, [Fe/H] 
$= -2.21\pm 0.21$. To be conservative, we preferred to consider the derived O$^0$ and C$^+$ column 
density measurements as upper limits due to the possible blends with Ly$\alpha$ forest lines. No 
intermediate-ion and high-ion transitions are observed in this sub-DLA system. We provide an upper 
limit on the Al$^{++}$ column density obtained from the fit of the very weak feature observed in the 
\ion{Al}{iii} $\lambda$1854 line at the redshift of the strongest component of the low-ion line 
profiles.
%

\subsubsection{$z_{\rm sub-DLA} = 3.565$}

\begin{figure}
   \includegraphics[width=85mm]{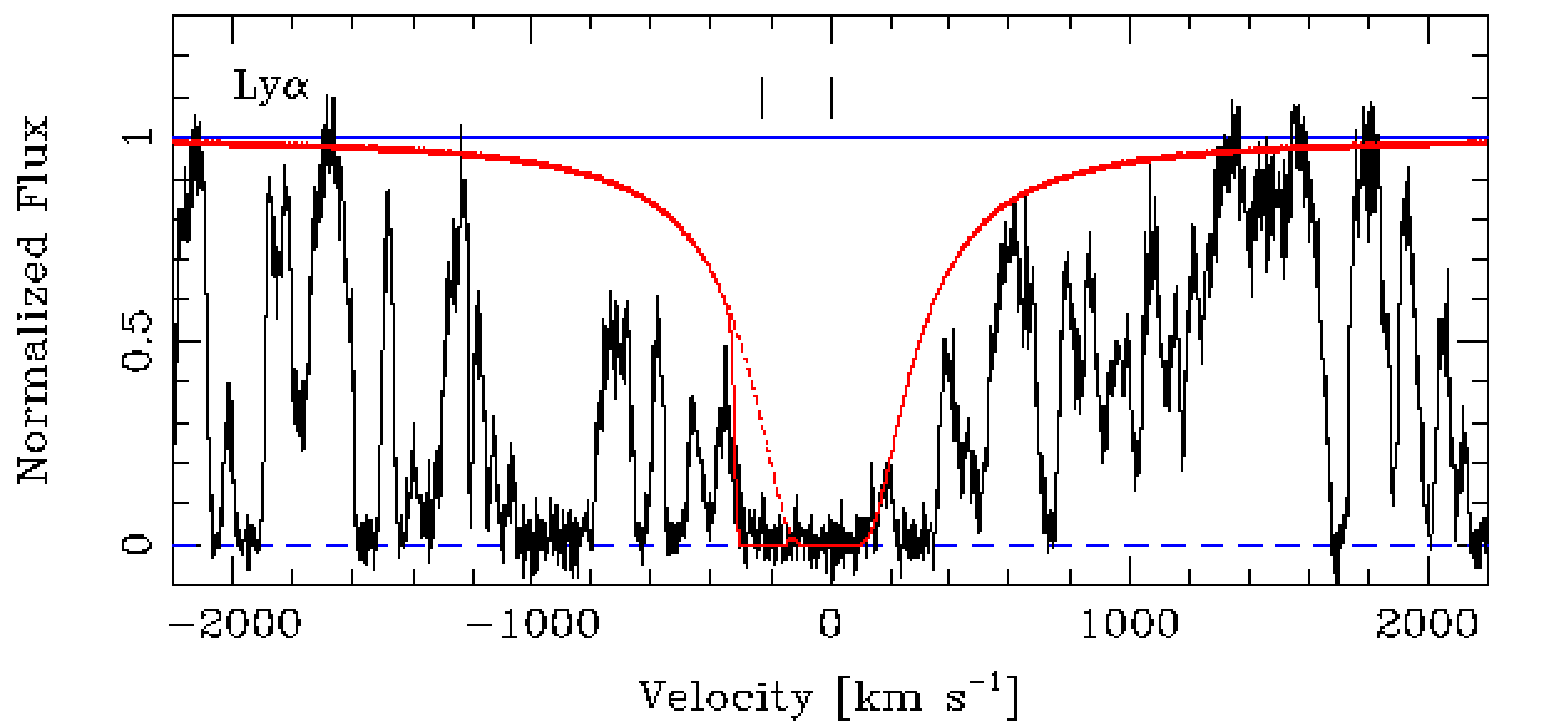}
\caption{Normalized UVES spectrum of PSS J2155+1358 showing the sub-DLA Ly$\alpha$ profile with the 
Voigt profile fit. The zero velocity is fixed at $z = 3.565272$. The vertical bars correspond from 
right to left to the velocity centroids of the single component of the low-ion transition lines at 
$z = 3.56527$ and $z = 3.561814$ used for the best fit, belonging to the sub-DLA system and an 
additional absorber, respectively. The dotted line corresponds to the fit with the sub-DLA only. The 
measured \ion{H}{i} column density of the sub-DLA is $\log N$(\ion{H}{i}) $= 19.37\pm 0.15$.}
\label{J2155-3p565-Ly}
\end{figure}

This sub-DLA system was also identified thanks to the Ly$\alpha$ absorption line. No other hydrogen 
line is observed in the available spectra. Given the high redshift of the quasar PSS J2155+1358, the 
Ly$\alpha$ forest shows many absorptions, and thus it is not surprising that the damping wings of 
the Ly$\alpha$ line are strongly blended. We made several fitting tests to find the best fit 
solution of the Ly$\alpha$ line. The first test was done by fixing in the fitting procedure the 
$b$-value at 20 km~s$^{-1}$ and by leaving the redshift as a free parameter. We obtained a good fit 
with $\log N$(\ion{H}{i}) $= 19.62\pm 0.15$, but it was not a satisfactory solution, since the 
deduced redshift was shifted by about 73 km~s$^{-1}$ bluewards of the redshift of the single 
component observed in the low-ion transition lines. The second and adopted test was made by fixing 
both the $b$-value at 20 km~s$^{-1}$ and the redshift at the redshift of the single component of the 
low-ion lines. We obtained the fit shown in Fig.~\ref{J2155-3p565-Ly}. But, in this case, to fill the 
whole observed Ly$\alpha$ absorption profile, we had to include, besides the sub-DLA contribution, a 
second low \ion{H}{i} column density absorption system bluewards of the sub-DLA system. In the 
fitting procedure, the $b$-value and the redshift of this second absorber were left as free 
parameters. We obtained the best fit solution with the second absorption line system located at $z = 
3.561814$, i.e. shifted by about $-228$ km~s$^{-1}$ from the sub-DLA system, and with a \ion{H}{i} 
column density of $\log N$(\ion{H}{i}) $= 16.46\pm 0.15$. The derived hydrogen column density of the 
sub-DLA is $\log N$(\ion{H}{i}) $= 19.37\pm 0.15$.
%

%

\begin{figure*}
   \includegraphics[width=140mm]{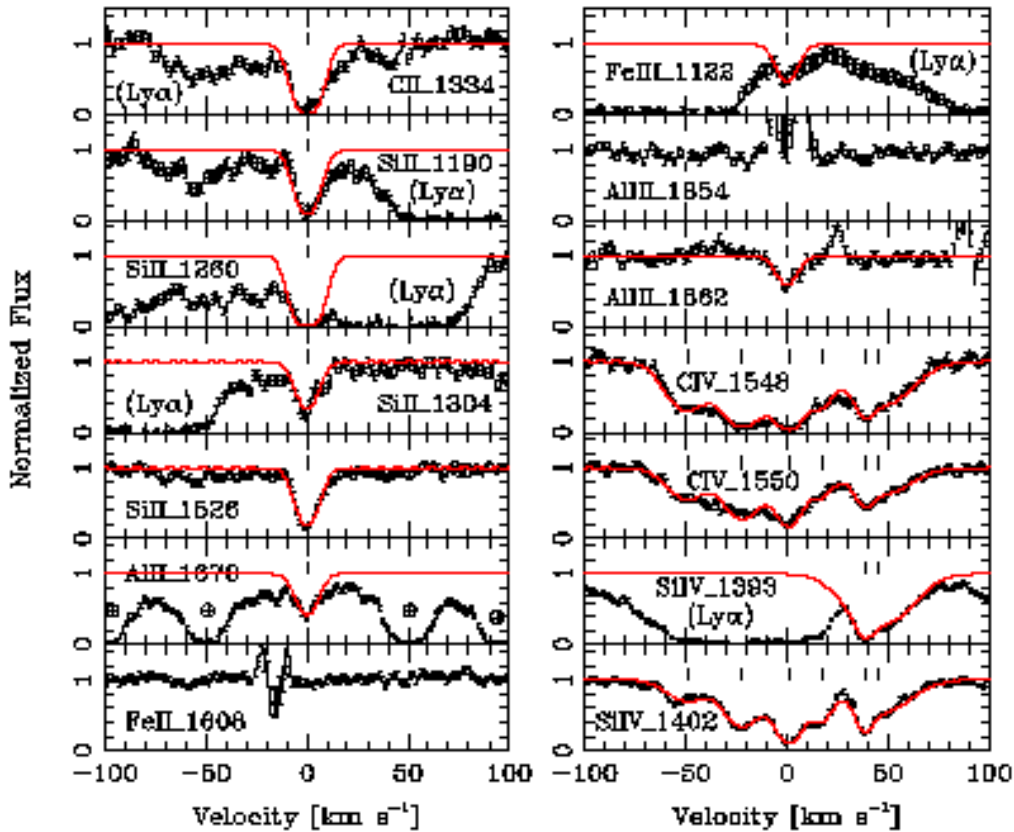}
\caption{Same as Fig.~\ref{Q1101-fits} for the sub-DLA towards PSS J2155+1358. The zero velocity is
fixed at $z = 3.565272$. The vertical bars mark the positions of the fitted velocity components (see
Table~\ref{J2155-3p565-values}).}
\label{J2155-3p565-fits}
\end{figure*}

As mentioned above, a very simple velocity structure with a single component is observed in the
low-ion line profiles (see Fig.~\ref{J2155-3p565-fits} and Table~\ref{J2155-3p565-values}). We 
obtained the column density measurements of Si$^+$, C$^+$ and Al$^+$. $N$(\ion{Si}{ii}) was deduced 
from four \ion{Si}{ii} lines at $\lambda_{\rm rest} = 1190$, 1260, 1304, 1526 \AA. $N$(\ion{C}{ii}) 
is a marginal measurement, since the \ion{C}{ii} $\lambda$1334 line is located in the Ly$\alpha$ 
forest, where blends are not excluded. Finally, $N$(\ion{Al}{ii}) was obtained from the \ion{Al}{ii} 
$\lambda$1670 line found in a spectral region heavily contaminated by telluric lines. The spectrum of 
a hot, fast rotating star observed the same night as the quasar allowed us to identify the telluric 
lines and to show that the Al$^+$ line is not blended, but the uncertainty relative to the continuum 
placement in this region remains. We therefore adopted a relatively large error on the measured
$N$(\ion{Al}{ii}). Only the \ion{Fe}{ii} $\lambda$1608 line is available to measure the Fe$^+$ 
column density, and no absorption feature at the redshift of the single component observed in the 
other low-ion transition lines is detected. We thus provide only a 4~$\sigma$ upper limit on the 
Fe$^+$ column density, which indicates that the metallicity is very low in this sub-DLA system, 
lower than [Fe/H] $< -2.31$.
%

\begin{table}
\caption{Component structure of the $z_{\rm abs} = 3.565$ sub-DLA system towards PSS J2155+1358} 
\label{J2155-3p565-values}
\begin{tabular}{l c c c l c}
\hline
No & $z_{\rm abs}$ & $v_{\rm rel}^*$ & $b (\sigma _b)$ & Ion & $\log N (\sigma_{\log N})$ \\
   &               & km~s$^{-1}$     & km~s$^{-1}$     &     & cm$^{-2}$                  
\\     
\hline
\multicolumn{5}{l}{Low- and intermediate-ion transitions} & \\
\hline
1 & 3.565272 & 0 & 5.9{\scriptsize (0.9)} & \ion{Si}{ii}  & 13.66{\scriptsize (0.04)} \\ 
  &	     &   &	                  & \ion{C}{ii}   & 14.34{\scriptsize (0.12)} \\ 
  &	     &   &	                  & \ion{Al}{ii}  & 12.16{\scriptsize (0.12)} \\ 
  &	     &   &	                  & \ion{Fe}{iii} & $< 13.81$ \\
  &	     &   &	                  & \ion{Al}{iii} & 12.71{\scriptsize (0.09)} \\ 
\hline 
\multicolumn{4}{l}{High-ion transitions} & & \\
\hline
1 & 3.564519 &           $-$49 &           12.5{\scriptsize (2.6)} & \ion{Si}{iv} & 12.95{\scriptsize (0.07)} \\
  &	     &  	       &                                   & \ion{C}{iv}  & 13.54{\scriptsize (0.07)} \\
2 & 3.564939 &           $-$22 &           10.8{\scriptsize (1.1)} & \ion{Si}{iv} & 13.38{\scriptsize (0.03)} \\
  &	     &  	       &                                   & \ion{C}{iv}  & 13.83{\scriptsize (0.17)} \\
3 & 3.565289 & \phantom{0}$+$1 & \phantom{0}7.4{\scriptsize (1.2)} & \ion{Si}{iv} & 13.57{\scriptsize (0.04)} \\
  &	     &  	       &                                   & \ion{C}{iv}  & 13.84{\scriptsize (0.06)} \\  
4 & 3.565531 &           $+$17 & \phantom{0}5.5{\scriptsize (0.9)} & \ion{Si}{iv} & 13.02{\scriptsize (0.04)} \\
  &	     &  	       &                                   & \ion{C}{iv}  & 13.15{\scriptsize (0.10)} \\
5 & 3.565859 &           $+$39 & \phantom{0}3.7{\scriptsize (1.1)} & \ion{Si}{iv} & 12.97{\scriptsize (0.08)} \\
  &	     &  	       &                                   & \ion{C}{iv}  & 13.01{\scriptsize (0.15)} \\
6 & 3.565966 &           $+$46 &           18.8{\scriptsize (1.0)} & \ion{Si}{iv} & 13.36{\scriptsize (0.03)} \\
  &	     &  	       &                                   & \ion{C}{iv}  & 13.64{\scriptsize (0.04)} \\
\hline
\end{tabular}
\begin{minipage}{140mm}
$^*$ Velocity relative to $z = 3.565272$
\end{minipage}
\end{table}
%

%

The column density measurements of the intermediate-ion transitions are crucial to constrain the
ionization parameters in this sub-DLA system which has among the lowest \ion{H}{i} column density in 
our sample. We detect only the \ion{Al}{iii} $\lambda$1862 line. The \ion{Al}{iii} $\lambda$1854 line 
is blended with emission lines from badly subtracted cosmic rays in the reduction procedure $-$ only 
two exposures were obtained for this quasar. The observed \ion{Al}{iii} $\lambda$1862 line shows a 
single component at exactly the redshift of the component of the low-ion transition lines (see 
Fig.~\ref{J2155-3p565-fits}). Assuming that this feature is real and is Al$^{++}$, we obtained a 
column density measurement using the same fitting parameters as those deduced for the low-ion 
lines. We also derived an upper limit on $N$(Fe$^{++}$) from the \ion{Fe}{iii} $\lambda$1122 line 
located in the Ly$\alpha$ forest. In addition, we detect the strong \ion{C}{iv} and \ion{Si}{iv} 
high-ion transition lines. They show a much more complex velocity structure, composed of 6 
components and spread over $\sim 120$ km~s$^{-1}$ than the low-ion lines (see 
Fig.~\ref{J2155-3p565-fits} and Table~\ref{J2155-3p565-values}). 
%

\begin{table}
\caption{Component structure of the $z_{\rm abs} = 4.212$ sub-DLA system towards PSS J2155+1358} 
\label{J2155-4p212-values}
\begin{tabular}{l c c c l c}
\hline
No & $z_{\rm abs}$ & $v_{\rm rel}^*$ & $b (\sigma _b)$ & Ion & $\log N (\sigma_{\log N})$ \\
   &               & km~s$^{-1}$     & km~s$^{-1}$     &     & cm$^{-2}$                  
\\     
\hline
\multicolumn{4}{l}{Low-ion transitions} & & \\
\hline
1 & 4.212229 & \phantom{0+}0 & 6.8{\scriptsize (0.5)} & \ion{Si}{ii} & 13.01{\scriptsize (0.04)} \\ 
  &	     &  	     &                        & \ion{O}{i}   & 14.26{\scriptsize (0.05)} \\
  &	     &  	     &                        & \ion{C}{ii}  & 13.70{\scriptsize (0.07)} \\  
  &	     &  	     &                        & \ion{Al}{ii} & 11.62{\scriptsize (0.16)} \\  
  &	     &  	     &                        & \ion{Fe}{ii} & 12.76{\scriptsize (0.20)} \\ 
2 & 4.212628 &         $+$23 & 6.3{\scriptsize (0.6)} & \ion{Si}{ii} & 12.88{\scriptsize (0.04)} \\
  &	     &  	     &                        & \ion{O}{i}   & 14.13{\scriptsize (0.05)} \\
  &	     &  	     &                        & \ion{C}{ii}  & 13.59{\scriptsize (0.05)} \\
  &	     &  	     &                        & \ion{Al}{ii} & 11.62{\scriptsize (0.17)} \\
  &	     &  	     &                        & \ion{Fe}{ii} & 12.43{\scriptsize (0.30)} \\
\hline
\multicolumn{4}{l}{High-ion transitions} & & \\
\hline
1 & 4.212394 & $+$10 &           22.5{\scriptsize (6.7)} & \ion{Si}{iv} & 12.87{\scriptsize (0.10)}  \\
2 & 4.212925 & $+$40 & \phantom{0}7.9{\scriptsize (4.9)} & \ion{Si}{iv} & 12.05{\scriptsize (0.25)} \\
\hline
\end{tabular}
\begin{minipage}{140mm}
$^*$ Velocity relative to $z = 4.212229$
\end{minipage}
\end{table}
%

\begin{figure}
   \includegraphics[width=85mm]{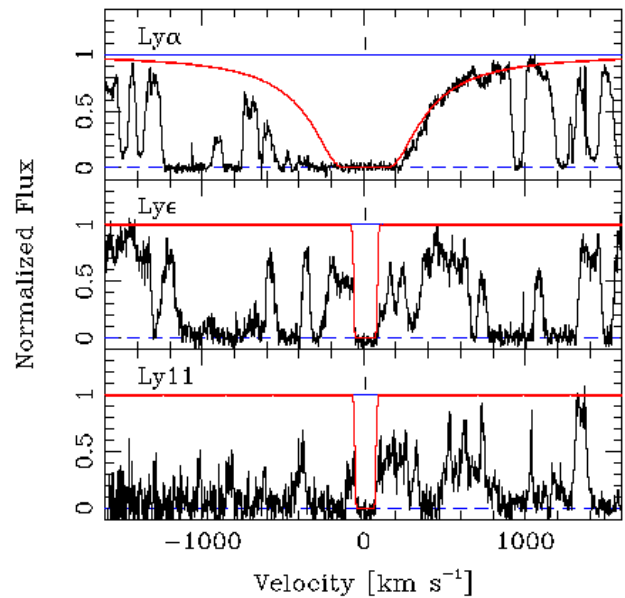}
\caption{Normalized UVES spectrum of PSS J2155+1358 showing the sub-DLA Ly$\alpha$, Ly$\epsilon$ and
Ly11 profiles with the Voigt profile fits. The zero velocity is fixed at $z = 4.212229$. The vertical 
bar corresponds to the velocity centroid used for the best fit, $z = 4.212442$. The measured 
\ion{H}{i} column density is $\log N$(\ion{H}{i}) $= 19.61\pm 0.10$.}
\label{J2155-4p212-Ly}
\end{figure}
%

\begin{figure*}
   \includegraphics[width=140mm]{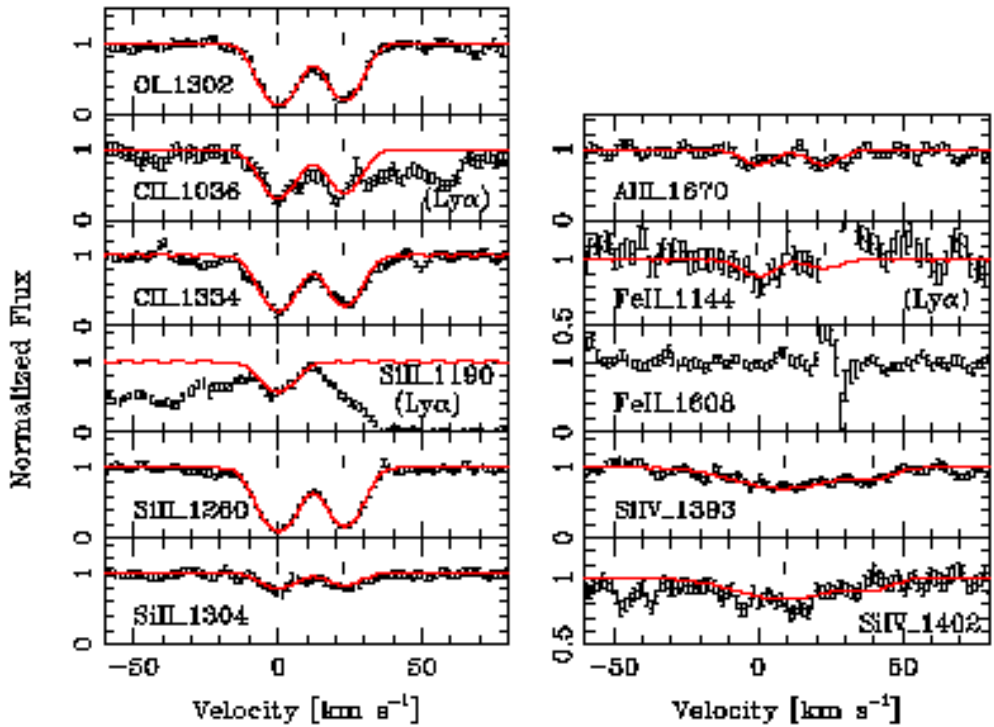}
\caption{Same as Fig.~\ref{Q1101-fits} for the sub-DLA towards PSS J2155+1358. The zero velocity is
fixed at $z = 4.212229$. The vertical bars mark the positions of the fitted velocity components (see
Table~\ref{J2155-4p212-values}).}
\label{J2155-4p212-fits}
\end{figure*}
%

\subsubsection{$z_{\rm sub-DLA} = 4.212$}

This high redshift sub-DLA system was easily identified with the Ly$\alpha$ absorption line which 
shows a very well defined red damping wing, the blue wing being partially blended with Ly$\alpha$ 
forest lines. The Lyman series is observed down to Ly13 in this sub-DLA and allowed us to obtain a 
very accurate \ion{H}{i} column density measurement by fitting simultaneously the Ly$\alpha$ line 
and the unblended Ly$\epsilon$ and Ly11 lines (see Fig.~\ref{J2155-4p212-Ly}). In the fitting 
procedure, we left the redshift as a free parameter and we fixed the $b$-value at 23 km~s$^{-1}$. 
The best fit solution was obtained with $z = 4.212442$ which fell just between the two components 
observed in the low-ion transition lines, and with a \ion{H}{i} column density of 
$\log N$(\ion{H}{i}) $= 19.61\pm 0.10$. However, it can be seen that the sub-DLA system is not 
sufficient to fill the whole observed Ly$\alpha$ absorption, so contributions from additional 
Ly$\alpha$ clouds have to be included.
%

\begin{table}
\caption{Component structure of the $z_{\rm abs} = 3.882$ sub-DLA system towards PSS J2344+0342} 
\label{J2344-values}
\begin{tabular}{l c c c l c}
\hline
No & $z_{\rm abs}$ & $v_{\rm rel}^*$ & $b (\sigma _b)$ & Ion & $\log N (\sigma_{\log N})$ \\
   &               & km~s$^{-1}$     & km~s$^{-1}$     &     & cm$^{-2}$                   
\\     
\hline
\multicolumn{4}{l}{Low-ion transitions} & & \\
\hline
 1 & 3.880511 &           $-$131 & \phantom{0}8.7{\scriptsize (2.0)} & \ion{Si}{ii} & 12.91{\scriptsize (0.11)} \\
   &	      &  	         &                                   & \ion{C}{ii}  & 13.55{\scriptsize (0.06)} \\
   &	      &  	         &                                   & \ion{Al}{ii} & 11.28{\scriptsize (0.15)} \\
 2 & 3.880811 &           $-$112 & \phantom{0}5.6{\scriptsize (1.3)} & \ion{Si}{ii} & 12.87{\scriptsize (0.10)} \\
   &	      &  	         &                                   & \ion{C}{ii}  & 13.48{\scriptsize (0.06)} \\
   &	      &  	         &                                   & \ion{Al}{ii} & 11.68{\scriptsize (0.25)} \\
 3 & 3.881785 & \phantom{0}$-$52 & \phantom{0}7.4{\scriptsize (1.1)} & \ion{Si}{ii} & 13.08{\scriptsize (0.05)} \\ 
   &	      &  	         &                                   & \ion{C}{ii}  & 13.79{\scriptsize (0.06)} \\
   &	      &  	         &                                   & \ion{Al}{ii} & 11.92{\scriptsize (0.09)} \\
 4 & 3.882032 & \phantom{0}$-$37 & \phantom{0}6.5{\scriptsize (1.9)} & \ion{Si}{ii} & 12.63{\scriptsize (0.07)} \\
   &	      &  	         &                                   & \ion{C}{ii}  & 13.35{\scriptsize (0.14)} \\
 5 & 3.882392 & \phantom{0}$-$15 & \phantom{0}4.2{\scriptsize (3.3)} & \ion{C}{ii}  & 12.88{\scriptsize (0.15)} \\
 6 & 3.882639 &    \phantom{0+}0 & \phantom{0}6.4{\scriptsize (1.1)} & \ion{Si}{ii} & 12.95{\scriptsize (0.10)} \\
   &	      &  	         &                                   & \ion{C}{ii}  & 13.59{\scriptsize (0.04)} \\
   &	      &  	         &                                   & \ion{Al}{ii} & 11.59{\scriptsize (0.15)} \\ 
 7 & 3.883733 & \phantom{0}$+$67 &           25.6{\scriptsize (2.8)} & \ion{C}{ii}  & 13.91{\scriptsize (0.04)} \\
 8 & 3.884153 & \phantom{0}$+$93 & \phantom{0}7.1{\scriptsize (3.4)} & \ion{C}{ii}  & 12.87{\scriptsize (0.32)} \\
 9 & 3.884580 &           $+$119 &           10.5{\scriptsize (0.9)} & \ion{Si}{ii} & 13.45{\scriptsize (0.04)} \\
   &	      &  	         &                                   & \ion{C}{ii}  & 14.12{\scriptsize (0.04)} \\
   &	      &  	         &                                   & \ion{Al}{ii} & 11.92{\scriptsize (0.09)} \\
10 & 3.885067 &           $+$149 &           12.1{\scriptsize (1.2)} & \ion{Si}{ii} & 13.35{\scriptsize (0.05)} \\
   &	      & 	         &                                   & \ion{C}{ii}  & 14.09{\scriptsize (0.03)} \\
   &	      & 	         &                                   & \ion{Al}{ii} & 12.16{\scriptsize (0.06)} \\
11 & 3.885412 &           $+$170 & \phantom{0}3.5{\scriptsize (2.0)} & \ion{C}{ii}  & 13.02{\scriptsize (0.09)} \\
\hline
\multicolumn{4}{l}{High-ion transitions} & & \\
\hline 
1 & 3.883235 & \phantom{0}$+$37 &           20.0{\scriptsize (5.7)} & \ion{C}{iv}  & 13.15{\scriptsize (0.10)} \\
2 & 3.883706 & \phantom{0}$+$66 &           11.8{\scriptsize (2.8)} & \ion{Si}{iv} & 12.91{\scriptsize (0.07)} \\
  &	     &  	        &                                   & \ion{C}{iv}  & 13.50{\scriptsize (0.04)} \\
3 & 3.884165 & \phantom{0}$+$94 &           10.7{\scriptsize (1.4)} & \ion{Si}{iv} & 13.00{\scriptsize (0.04)} \\
  &	     &  	        &                                   & \ion{C}{iv}  & 13.40{\scriptsize (0.05)} \\
4 & 3.884840 &           $+$135 &           18.8{\scriptsize (3.9)} & \ion{Si}{iv} & 13.28{\scriptsize (0.10)} \\
  &	     &  	        &                                   & \ion{C}{iv}  & 13.79{\scriptsize (0.04)} \\
5 & 3.885314 &           $+$164 & \phantom{0}7.7{\scriptsize (1.4)} & \ion{Si}{iv} & 12.81{\scriptsize (0.06)} \\
  &	     &  	        &                                   & \ion{C}{iv}  & 13.35{\scriptsize (0.06)} \\
\hline
\end{tabular}
\begin{minipage}{140mm}
$^*$ Velocity relative to $z = 3.882639$
\end{minipage}
\end{table}
%

Given the high signal-to-noise ratio of the available spectra in the regions where we detect the 
low-ion transition lines and the fact that they show a simple velocity structure composed of two 
well separated components (see Fig.~\ref{J2155-4p212-fits} and Table~\ref{J2155-4p212-values}), we 
obtained very accurate column density measurements of O$^0$, C$^+$, Si$^+$ and Al$^+$. Two
\ion{Fe}{ii} lines at $\lambda_{\rm rest} = 1144$ and 1608 \AA\ are observed in the available 
spectra. A very weak feature is detected in the \ion{Fe}{ii} $\lambda$1144 line at the redshift of 
the component 1. But, the second component is partially blended with an emission line from badly 
subtracted cosmic rays in the reduction procedure $-$ a single exposure was obtained for this quasar 
in the blue. No absorption feature is observed in the \ion{Fe}{ii} $\lambda$1608 lines, moreover the 
second component is blended with a badly subtracted strong sky line. We thus obtained only a 
marginal measurement of the Fe$^+$ column density, on which we adopted a large error. The weakness 
of these \ion{Fe}{ii} lines indicates that the metallicity in this sub-DLA system is very low, 
[Fe/H] $= -2.17\pm 0.25$. 

The \ion{Al}{iii} intermediate-ion transition lines are covered in the available spectra, but the
signal-to-noise ratio is so low in this particular region that no Al$^{++}$ line is detected, and 
no reliable 4~$\sigma$ upper limit can be derived. In addition, the \ion{Fe}{iii} $\lambda$1122, 
\ion{Si}{iii} $\lambda$1206 and \ion{C}{iii} $\lambda$977 lines are all blended with Ly$\alpha$ 
forest absorptions. Consequently, no intermediate-ion column density measurement was obtained in 
this sub-DLA system. On the other hand, the \ion{Si}{iv} high-ion transition lines are observed (the 
\ion{C}{iv} lines are outside the available spectral coverage). Again they show a different velocity 
structure from the low-ion line profiles, but are in this case extended over a similar velocity 
range (see Fig.~\ref{J2155-4p212-fits} and Table~\ref{J2155-4p212-values}).
%

\begin{figure}
   \includegraphics[width=85mm]{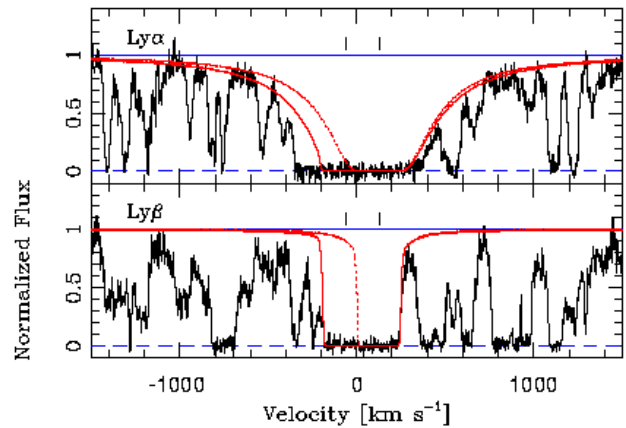}
\caption{Normalized UVES spectrum of PSS J2344+0342 showing the sub-DLA Ly$\alpha$ profile with the 
Voigt profile fit. The zero velocity is fixed at $z = 3.882639$. The vertical bars correspond from 
right to left to the velocity centroids $z = 3.884697$ and $z = 3.881604$ used for the best fit, 
belonging to the sub-DLA system and an additional absorber, respectively. The dotted line corresponds 
to the fit with the sub-DLA only. The measured \ion{H}{i} column density of the sub-DLA is $\log 
N$(\ion{H}{i}) $= 19.50\pm 0.10$.}
\label{J2344-Ly}
\end{figure}
%

\begin{figure*}
   \includegraphics[width=170mm]{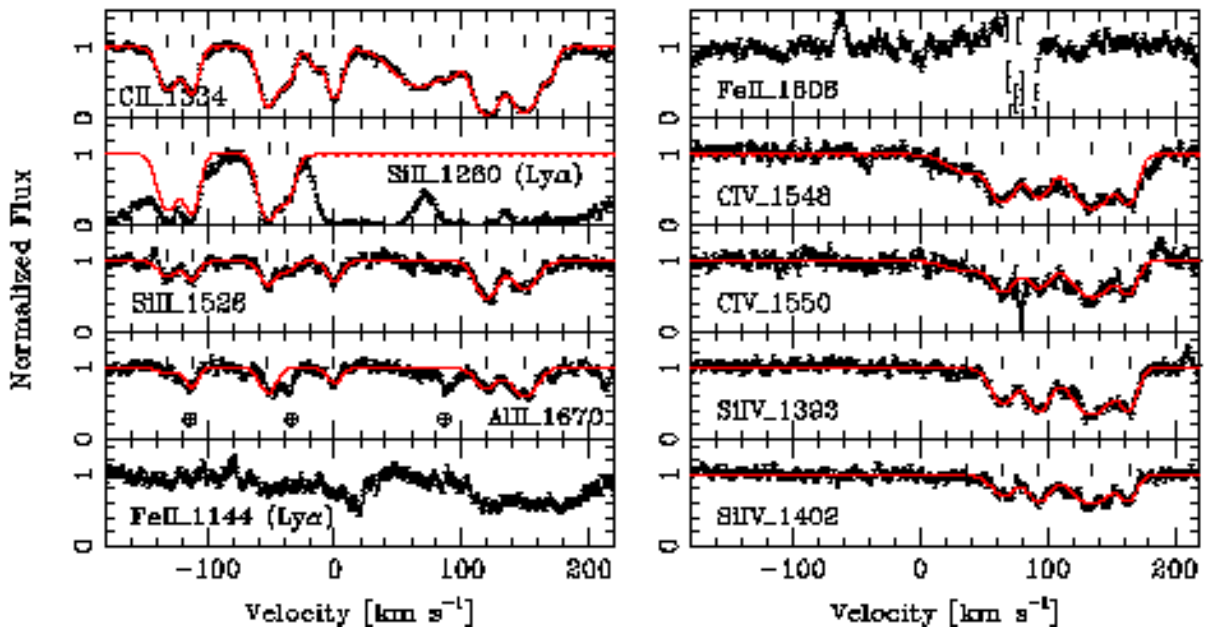}
\caption{Same as Fig.~\ref{Q1101-fits} for the sub-DLA towards PSS J2344+0342. The zero velocity is
fixed at $z = 3.882639$. The vertical bars mark the positions of the fitted velocity components (see
Table~\ref{J2344-values}).}
\label{J2344-fits}
\end{figure*}
%

\subsection{PSS J2344+0342, $z_{\rm sub-DLA} = 3.882$}\label{J2344}

P\'eroux et~al. (2001) obtained 5 \AA\ resolution (FWHM) spectra of this quasar and identified a DLA 
system at $z_{\rm abs} = 3.220$. The metallicity of this DLA system was first reported by 
Dessauges-Zavadsky et~al. (2001b). Here we present the chemical analysis of a newly discovered 
high-redshift sub-DLA system at $z_{\rm abs} = 3.882$. It has been identified thanks to the 
Ly$\alpha$ absorption line which shows partially blended damping wings. The Ly$\beta$ and Ly$\gamma$ 
lines are also observed in this sub-DLA system. The simultaneous fit of both Ly$\alpha$ and 
Ly$\beta$ $-$ the Ly$\gamma$ line is strongly blended~$-$ allowed us to obtain an accurate 
\ion{H}{i} column density measurement.

A first series of fitting tests of the Ly$\alpha$ and Ly$\beta$ lines was done by considering a 
single absorption line system. By leaving the redshift and the $b$-value as free parameters, the best 
fit solution was obtained with a sub-DLA system at $z = 3.884697$ $-$ very close to the redshift of 
the strongest component of the low-ion transition lines, the component 9~$-$ and a \ion{H}{i} column
density of $\log N$(\ion{H}{i}) $= 19.50\pm 0.10$. In this solution the red wings of the Ly$\alpha$
and Ly$\beta$ lines were well fitted, but the fit clearly shows that another strong absorption 
line system is present bluewards of the sub-DLA system (see Fig.~\ref{J2344-Ly}). We then made a 
second series of fitting tests by considering this time two absorption line systems. By leaving in 
the fitting procedure all the $b$-values and redshifts as free parameters, we obtained the best fit 
solution with a sub-DLA system characterized by the values described above plus a second strong 
absorber at $z = 3.881604$ shifted by about $-190$ km~s$^{-1}$ from the sub-DLA system with a 
\ion{H}{i} column density of $\log N$(\ion{H}{i}) $= 18.94\pm 0.15$. This absorption line system is 
a borderline case between the Lyman limit and sub-DLA systems, but we do not include it in the 
sub-DLA sample.

Table~\ref{J2344-values} and Fig.~\ref{J2344-fits} present the velocity structure of the sub-DLA
system. The low-ion transition lines show a complex velocity structure well described by 11 
components. We obtained accurate column density measurements of C$^+$ and Si$^+$. The \ion{Al}{ii} 
$\lambda$1670 line is located in a region strongly contaminated by telluric lines. We identified the 
telluric lines with the help of the spectrum of a hot, fast rotating star observed the same night as 
the quasar, and noticed that the components 2 and 7 are blended. We adopted a large error on the 
deduced $N$(Al$^+$) to take into account the contributions of these blends. Two Fe$^+$ lines, 
\ion{Fe}{ii} $\lambda$1144 and \ion{Fe}{ii} $\lambda$1608, are observed. However, the \ion{Fe}{ii} 
$\lambda$1144 line is partially blended with Ly$\alpha$ forest lines, and none of the 11 components 
is detected in the \ion{Fe}{ii} $\lambda$1608 line. We thus provide a 4~$\sigma$ upper limit on the 
Fe$^+$ column density which indicates that the metallicity in this sub-DLA system is low, lower than 
[Fe/H] $< -1.98$. No intermediate-ion transition line is clearly observed. The \ion{Fe}{iii} 
$\lambda$1122, \ion{C}{iii} $\lambda$977, \ion{N}{ii} $\lambda$1083 and \ion{Si}{iii} $\lambda$1206 
lines are strongly blended with Ly$\alpha$ forest lines, and the \ion{Al}{iii} lines are not 
detected. We provide a 4~$\sigma$ upper limit on the Al$^{++}$ column density.
%

\begin{table*}
\caption{Spearman's tests and linear regressions} 
\label{correlations}
\begin{tabular}{l c c c c c}
\hline
Correlations for the whole sample of & \# of systems & $r$ & $\alpha$ & slope & intercept \\
both the DLA and sub-DLA systems     &               &     &          &       &
\\     
\hline
$\log N$(Si$^+$) vs $\log N$(Fe$^+$) & 39 & \phantom{$-$}0.95 & $3.2\times 10^{-20}$ & \phantom{$-$}$1.04\pm 0.04$ & \phantom{1}$-0.10\pm 0.58$ \\
$\log N$(Al$^+$) vs $\log N$(Si$^+$) & 19 & \phantom{$-$}0.92 & $3.4\times 10^{-08}$ & \phantom{$-$}$0.97\pm 0.08$ & \phantom{1}$-0.97\pm 1.19$ \\
$\log N$(Al$^+$) vs $\log N$(Fe$^+$) & 17 & \phantom{$-$}0.82 & $4.9\times 10^{-05}$ & \phantom{$-$}$0.89\pm 0.13$ & \phantom{1$-$}$0.58\pm 1.83$ \\
$\log N$(Al$^{++}$)/$N$(Al$^+$) vs $\log N$(H$^0$) $^*$ & 30 & $-$0.67 & $5.7\times 10^{-05}$ & $-0.59\pm 0.13$ & \phantom{$-$}$11.49\pm 2.84$ \\ 
\hline
\end{tabular}
\begin{minipage}{140mm}
$^*$ Computed for the sample of DLAs only. \\
{\bf References:} This paper; Dessauges-Zavadsky et~al. (2001a); Lopez et~al. (1999,2002); Lu et~al. 
(1995,1996b); Prochaska \& Wolfe (1996,1997,1999); Prochaska et~al. (2001); P\'eroux et~al. (2002b)
\end{minipage}
\end{table*}
%

It should be emphasized that the observed low-ion transition line profiles extend over a very large 
velocity interval of $\sim 320$ km~s$^{-1}$, from $-140$ to $+180$ km~s$^{-1}$ in velocity space. It 
implies that the low-ion transition lines overlap over the two identified clustered absorption line 
systems $-$ the sub-DLA and the absorber at $z = 3.881604$ shifted by only $\sim -190$ km~s$^{-1}$ 
from the sub-DLA (see above). It may thus be possible that some of the components of the detected 
low-ion metal lines are associated with this second absorber and not with the sub-DLA system. Two 
distinct clumps in the low-ion profiles are not yet observed, but the two high \ion{H}{i} column 
density absorption line systems are so close that it would not be surprising to find their metal 
lines blended. Moreover, the observed high-ion transition lines, \ion{C}{iv} and \ion{Si}{iv}, are 
surprisingly confined to a much narrower velocity interval from $+25$ to $+170$ km~s$^{-1}$ (see 
Fig.~\ref{J2344-fits} and Table~\ref{J2344-values}). This may suggest that the detected high-ion 
lines are associated only with the sub-DLA system. If this is the case, the components of the low-ion 
transition lines belonging to the sub-DLA could be only the ones which are located in the same 
velocity interval as the high-ion lines, i.e. the components 7 to 11. The components 1 to 6 could 
then be associated with the second absorption line system. It is difficult to be categorical on the 
real situation, therefore it would perhaps be more sensible to speak of an interval of values for the 
column density measurements in this sub-DLA system.
%

\section{Ionization correction study}\label{ionization}

In the 12 sub-DLA systems and 1 borderline case between the DLA and sub-DLA systems studied, we 
obtained ionic column density measurements of low-ion transitions, O$^0$, C$^+$, Si$^+$, N$^0$, 
S$^+$, Mg$^+$, Al$^+$, Fe$^+$, Ni$^+$, Zn$^+$ and Cr$^+$, intermediate-ion transitions, Al$^{++}$ and 
Fe$^{++}$, and high-ion transitions, C$^{3+}$ and Si$^{3+}$. To derive the {\it intrinsic} abundance 
measurements of the 11 detected elements, it is necessary to take into account both the ionization 
and dust depletion effects, since we are measuring gas-phase abundances. Depletion is discussed in 
the wider context of Paper~II. Here we analyse the ionization status of these systems.

Several authors have investigated the ionization effects in the DLA systems (Viegas 1995; Howk \& 
Sembach 1999; Izotov, Schaerer \& Charbonnel 2001; Vladilo et~al. 2001; Prochaska et~al. 2002). Many 
of them were motivated by the observations of the intermediate-ion Al$^{++}$, a tracer of moderately 
ionized gas, with the same velocity profiles as the low-ion transitions (Lu et~al. 1996b; Prochaska 
\& Wolfe 1999; Prochaska et~al. 2001), which suggest that ionization effects may be present in these
systems. Indeed, to explain the similarity of \ion{Al}{iii} and other low-ionization species line 
profiles, Howk \& Sembach (1999) and Izotov \& Thuan (1999) proposed that these lines originate in 
the same ionized region or in a mix of neutral and ionized clouds, and stressed the importance of 
abundance corrections for ionization effects. Although the different approaches used to deal with the problem of photoionization in DLAs 
have led to slightly different conclusions, it is generally accepted that ionization corrections in 
DLAs are negligible, being below the measurement errors.

In the present study, we are concerned about the ionization corrections for sub-DLA, i.e. systems
with \ion{H}{i} column densities lower than those of DLA systems, between $10^{19} < N$(\ion{H}{i}) 
$< 2\times 10^{20}$ cm$^{-2}$. The check of ionization effects in these systems is very important, 
since their relatively low \ion{H}{i} column density implies that some of the gas might be ionized.
%

\subsection{Al$^{++}$/Al$^+$ as Indicator of the Ionization State}\label{obs-hints}

In six out of the 12 sub-DLA systems studied we detect the \ion{Al}{iii} $\lambda$1854,1862 lines, 
and in four we also detect the \ion{Fe}{iii} $\lambda$1122 line. While neutral and singly ionized 
species with ionization potential $< 13.6$ eV can be present in the inner parts of H$^0$ regions, 
the production of Fe$^{++}$ and Al$^{++}$ requires photons with $h \nu > 16.2$ and 18.8~eV, 
respectively, which cannot so easily penetrate large \ion{H}{i} column densities. The analysis of 
these intermediate-ion transition profiles, however, shows that in all of the sub-DLAs but one at 
$z_{\rm abs} = 1.838$ towards Q1101$-$264 (see Section~\ref{Q1101}), their velocity structure is 
almost indistinguishable from that of low-ion transition profiles, while the high-ion \ion{C}{iv} 
and \ion{Si}{iv} transitions have very different velocity profiles. This suggests that the Fe$^{++}$ 
and Al$^{++}$ and regions of neutral H in these systems are physically connected.
%

\begin{figure}
   \includegraphics[width=80mm]{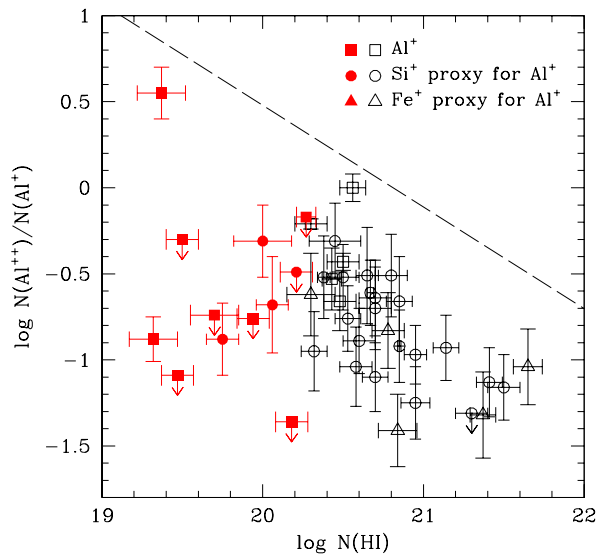}
\caption{Al$^{++}$/Al$^+$ column density ratios versus \ion{H}{i} column densities in sub-DLAs 
(filled symbols) and DLAs (open symbols) for all the systems found in the literature and studied in
this paper. The squares correspond to the data points obtained with the Al$^+$ column density 
measurements, while the circles and the triangles correspond to the data points computed from the 
Si$^+$ and Fe$^+$ column density measurements, respectively, according to the deduced linear 
regression parameters given in Table~\ref{correlations}. The dashed line represents a possible upper 
band limit of an increasing dispersion in the $\log N$(Al$^{++}$)/$N$(Al$^+$) ratios with lower 
\ion{H}{i} column densities. The slope of this upper band limit is equal to the slope of the observed 
$\log N$(Al$^{++}$)/$N$(Al$^+$) versus $\log N$(H$^0$) anti-correlation in DLAs.}
\label{AlIII-AlII}
\end{figure}
%

The $N$(Al$^{++}$)/$N$(Al$^+$) and $N$(Fe$^{++}$)/$N$(Fe$^+$) ratios can then be used as indicators 
of the ionization state. Vladilo et~al. (2001) studied the DLA ionization properties from the 
analysis of the Al$^{++}$/Al$^+$ ratios in a data sample of 20 DLAs, and found a $\log 
N$(Al$^{++}$)/$N$(Al$^+$) versus $\log N$(H$^0$) anti-correlation. The existence of such an empirical 
anti-correlation implies that the neutral hydrogen column density is an indirect tracer of the 
ionization state of the gas. Consequently, if this anti-correlation is further observed in the 
sub-DLA \ion{H}{i} column density range, it would indicate that the sub-DLA systems require higher 
ionization corrections than absorbers in the DLA \ion{H}{i} column density range.

Therefore, we decided to investigate the extension of the different correlations found by Vladilo 
et~al. (2001) in the DLA systems to the \ion{H}{i} column density range of the sub-DLA systems. For 
this purpose we have collected all the DLA and sub-DLA systems found in the literature for which 
measurements of the Al$^+$ and/or Al$^{++}$ column densities exist and was performed on high 
resolution spectra obtained at the 8-10\,m class telescopes. Most of them were derived from the 
HIRES/Keck spectra. We have then considered only the systems which show similarities in their 
Al$^{++}$ and low-ionization species line profiles, since it is only in this case that the neutral 
and ionized clouds might be physically related. Finally, we have adopted the values given by Vladilo 
et~al. (2001) for the few cases of Al$^+$ column density measurements that they have revisited due 
to previously unquoted saturations or contaminations by telluric lines. The results and the 
references for the adopted sample of both DLAs and sub-DLAs found in the literature and analysed in 
this paper are given in Table~\ref{correlations}. This table contains the Spearman's test 
correlation coefficients, $r$, the probabilities of null hypothesis (of no correlation), $\alpha$, 
the slopes and the intercepts obtained for the linear regressions of $\log N$(Si$^+$) versus 
$\log N$(Fe$^+$), $\log N$(Al$^+$) versus $\log N$(Si$^+$), $\log N$(Al$^+$) versus $\log N$(Fe$^+$), 
and $\log N$(Al$^{++}$)/$N$(Al$^+$) versus $\log N$(H$^0$). 

We confirm the existence of the $\log N$(Si$^+$) versus $\log N$(Fe$^+$), $\log N$(Al$^+$) versus 
$\log N$(Si$^+$), and $\log N$(Al$^+$) versus $\log N$(Fe$^+$) correlations found by Vladilo et~al. 
(2001) in the DLA systems (see Table~\ref{correlations}). These correlations are also satisfied by 
the sub-DLA systems. We then used the Si$^+$ (in priority, because the correlation with Al$^+$ has 
lower slope and intercept errors) or Fe$^+$ column densities as a proxy for $N$(Al$^+$), when no 
$N$(Al$^+$) measurement was available in the DLA or sub-DLA systems. In four sub-DLAs and several 
DLAs we took advantage of the correlations Si$^+$ versus Al$^+$ and Fe$^+$ versus Al$^+$ to 
indirectly estimate $N$(Al$^+$) in order to increase the number of the 
$\log N$(Al$^{++}$)/$N$(Al$^+$) ratio measurements.

Figure~\ref{AlIII-AlII} shows the Al$^{++}$/Al$^+$ column density ratios versus H$^0$ column 
densities both for the DLA (open symbols) and sub-DLA (filled symbols) systems. We confirm the 
$\log N$(Al$^{++}$)/$N$(Al$^+$) versus $\log N$(H$^0$) anti-correlation for the enlarged sample of 
DLA systems (30 objects instead of 17 used by Vladilo et~al. 2001) at the probability of null 
hypothesis of $5.7\times 10^{-5}$ (correlation coefficient of $r = -0.67$; see 
Table~\ref{correlations}). However, we show that at a lower \ion{H}{i} column density regime, this 
correlation is no longer valid. The sub-DLA systems have Al$^{++}$/Al$^+$ ratios which are similar 
to those of the DLAs and not higher, as shown in Fig.~\ref{AlIII-AlII} with five measured values and 
seven upper limits (filled symbols). Only one system at $z_{\rm abs} = 3.565$ towards PSS J2155+1358 
with $\log N$(\ion{H}{i}) $= 19.37\pm 0.15$ is really in agreement with the extrapolation of the DLA 
correlation to lower \ion{H}{i} column densities (filled square in the upper left corner in 
Fig.~\ref{AlIII-AlII}).

This implies that the neutral hydrogen column density is {\it not} a reliable indicator of the 
ionization state of the gas in absorption line systems as suggested by Vladilo et~al. (2001), since 
it does not scale with the Al$^{++}$/Al$^+$ column density ratio which is at a first approximation an 
indicator of the ionization level in the gas. On the contrary, Figure~\ref{AlIII-AlII} shows that
since the sub-DLA systems, despite their lower \ion{H}{i} column densities, span approximately the 
same range of the Al$^{++}$/Al$^+$ column density ratio values as the DLA systems, they are likely 
to be affected by ionization corrections on average at the same level as the DLAs and not higher. 
However, we cannot claim that all of the absorption line systems with \ion{H}{i} column densities 
between $10^{19}$ and $2\times 10^{20}$ cm$^{-2}$ (the sub-DLA \ion{H}{i} column density definition) 
satisfy the same trend, as already shown by the example of the sub-DLA system at $z_{\rm abs} = 
3.565$ towards PSS J2155+1358. We can rather interpret the present data as showing an increasing 
dispersion in the $\log N$(Al$^{++}$)/$N$(Al$^+$) ratios with lower H$^0$ column densities. The 
plotted dashed line in Figure~\ref{AlIII-AlII} with the slope of the $\log N$(Al$^{++}$)/$N$(Al$^+$) 
versus $\log N$(H$^0$) anti-correlation observed in DLAs can be considered as {\it an upper band 
limit} of this increasing dispersion trend. If this trend is further confirmed by additional 
measurements, this would suggest that at high $N$(H$^0$) $> 2\times 10^{20}$ cm$^{-2}$ only low 
ionization effects are expected, while at lower $N$(H$^0$) between $10^{19}$ and $2\times 10^{20}$ 
cm$^{-2}$ one can find systems affected by low as well as high ionization corrections.
%

\subsection{Photoionization Models}\label{methodology}

To investigate the ionization corrections in our sub-DLAs, we have used the CLOUDY software package 
(version 94.00, Ferland 1997), and we have computed photoionization models assuming ionization 
equilibrium. In the models we have adopted a plane-parallel geometry for the gas cloud, the solar 
abundance pattern, and two different ionizing spectra, since the major uncertainty in determining 
the ionization effects in absorption line systems is the unknown shape of the ionizing spectrum. The 
two most likely origins for ionizing UV photons are external background sources and internal stellar 
sources $-$ it is reasonable to expect some star formation in these systems. For the external 
ionizing source, we adopted the Haardt-Madau (hereafter HM) ionizing UV spectrum from background 
quasars with contributions from galaxies (Haardt \& Madau 1996; Madau, Haardt \& Rees 1999). For the 
stellar ionizing source, we adopted the spectrum of a local ionization which we consider to be a 
composite stellar spectrum plus a power law (hereafter SSP; see D'Odorico \& Petitjean 2001). In the 
former case the ionizing photons have a harder spectrum.

With each of these two ionizing spectra, we computed a series of CLOUDY models for each sub-DLA 
system in our sample by varying the \ion{H}{i} column density of the plane-parallel `cloud', the 
metallicity and the redshift according to the system considered, and by varying the volume density 
of hydrogen, $n_{\rm H}$, through the variation of the ionization parameter, $U$, defined by
$$U\equiv \frac{\Phi_{912}}{c n_{\rm H}} = \frac{4 \pi J_{912}}{h c n_{\rm H}} = 2\times 10^{-5}
\frac{J_{912}/10^{-21.5}}{n_{\rm H}/cm^{-3}}\,,$$
where $\Phi_{912}$ is the surface flux of ionizing photons with $h\nu > 1$~Ryd, and $J_{912}$ is the 
intensity of the incident radiation at 1~Ryd. In practice, we varied $\log U$ from $-6$ to 0, and
obtained the theoretical column density predictions for all of the observed ions as a function of the 
ionization parameter.

We then determined the ionization state of a given sub-DLA system by measuring adjacent ions of 
elements (e.g. Al$^{++}$/Al$^+$, Fe$^{++}$/Fe$^+$). Indeed, the value of the ionization parameter,
$U$, at which the observed column density ratio of adjacent ions matches the predicted ratio, 
corresponds to the ionization parameter of the system studied. Therefore, the more numerous the 
measurements of column density ratios of adjacent ions of different elements, the more accurate is 
the determination of the ionization state of a system, all the more since the reliability of the 
derived photoionization models depends not only on the assumed hypothesis, but also on the accuracy 
of the input atomic data for the CLOUDY code (e.g. dielectric recombination coefficients).
%

\begin{figure*}
   \mbox{\includegraphics[width=85mm]{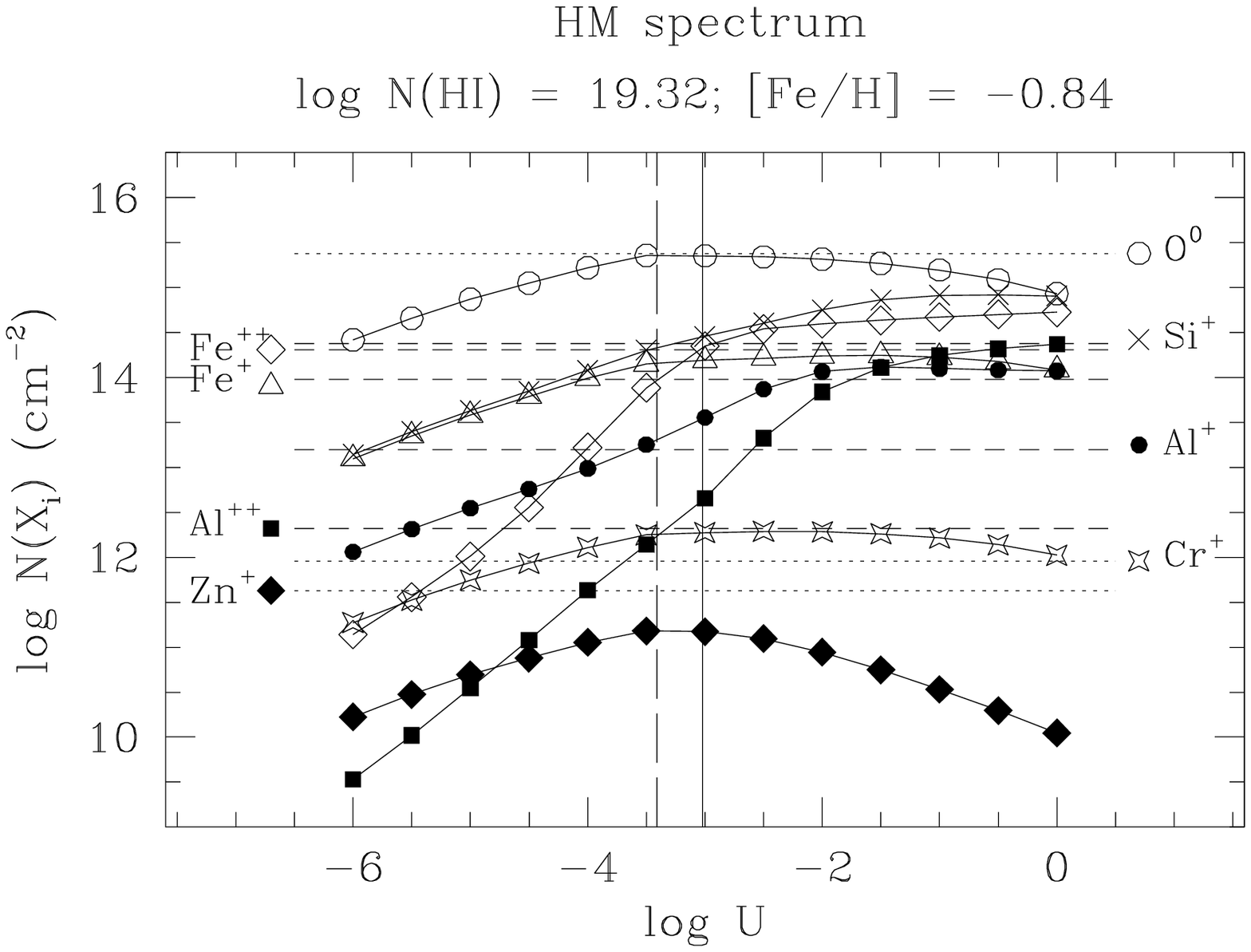}\quad
         \includegraphics[width=85mm]{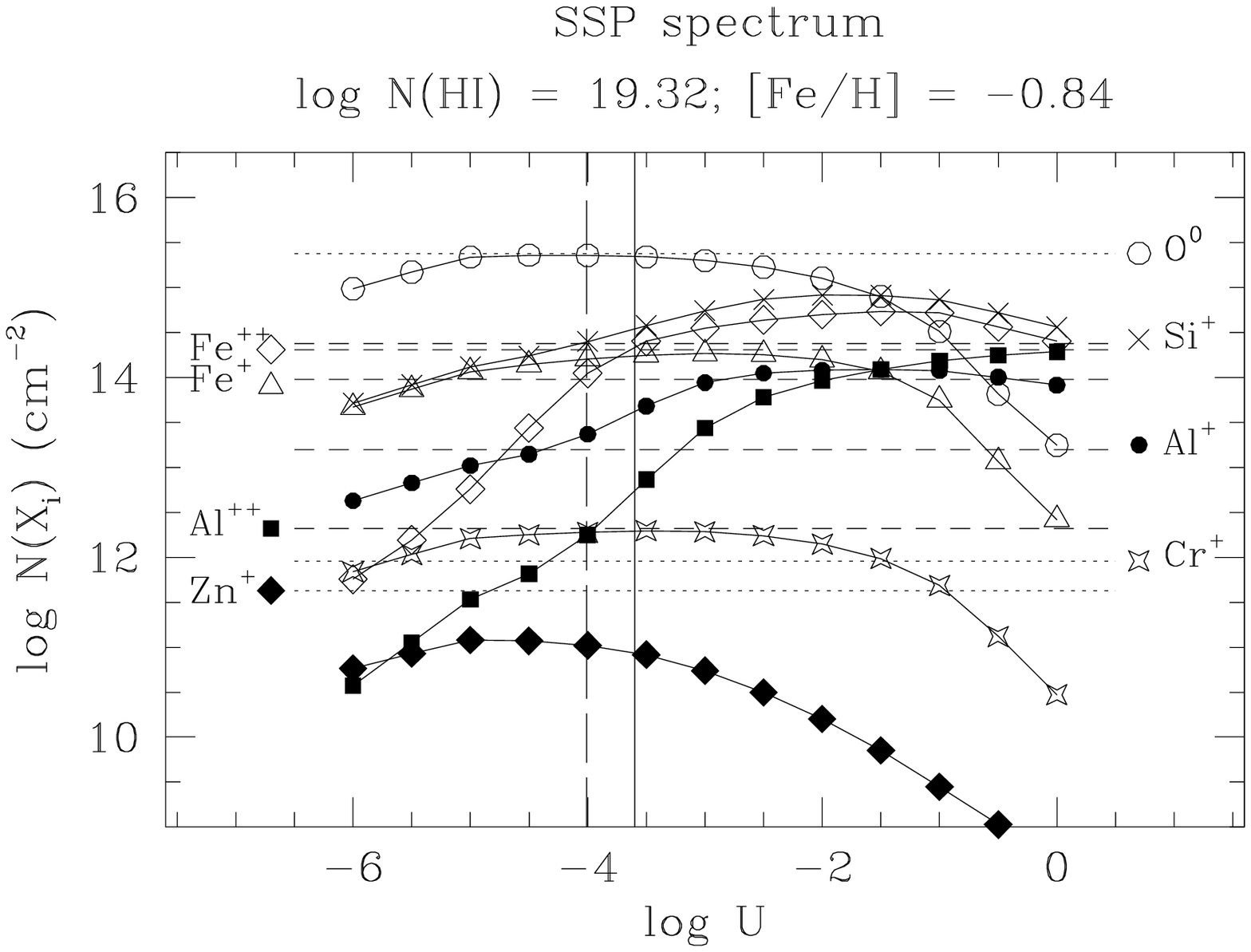}}
   \mbox{\includegraphics[width=85mm]{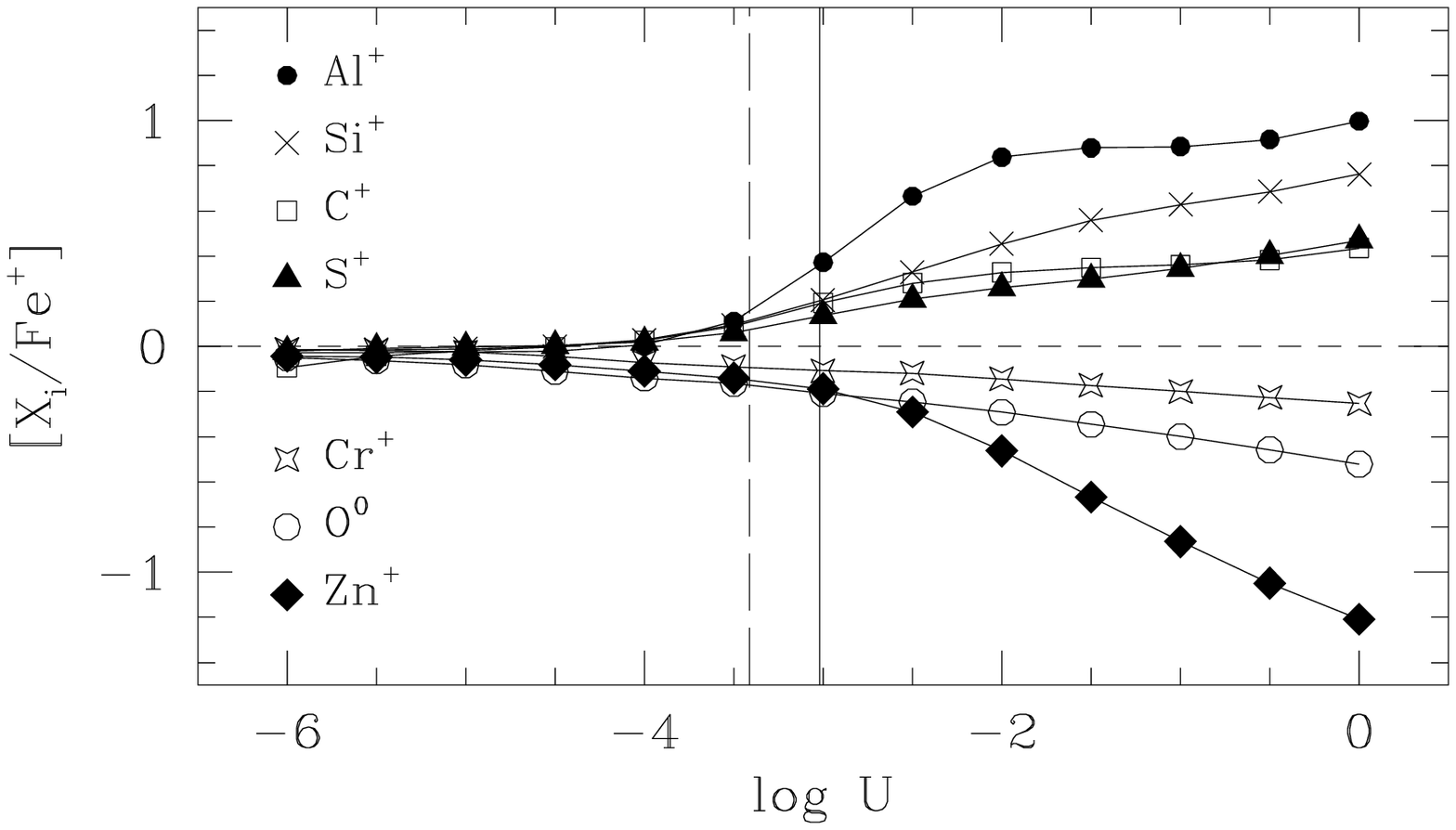}\quad
         \includegraphics[width=85mm]{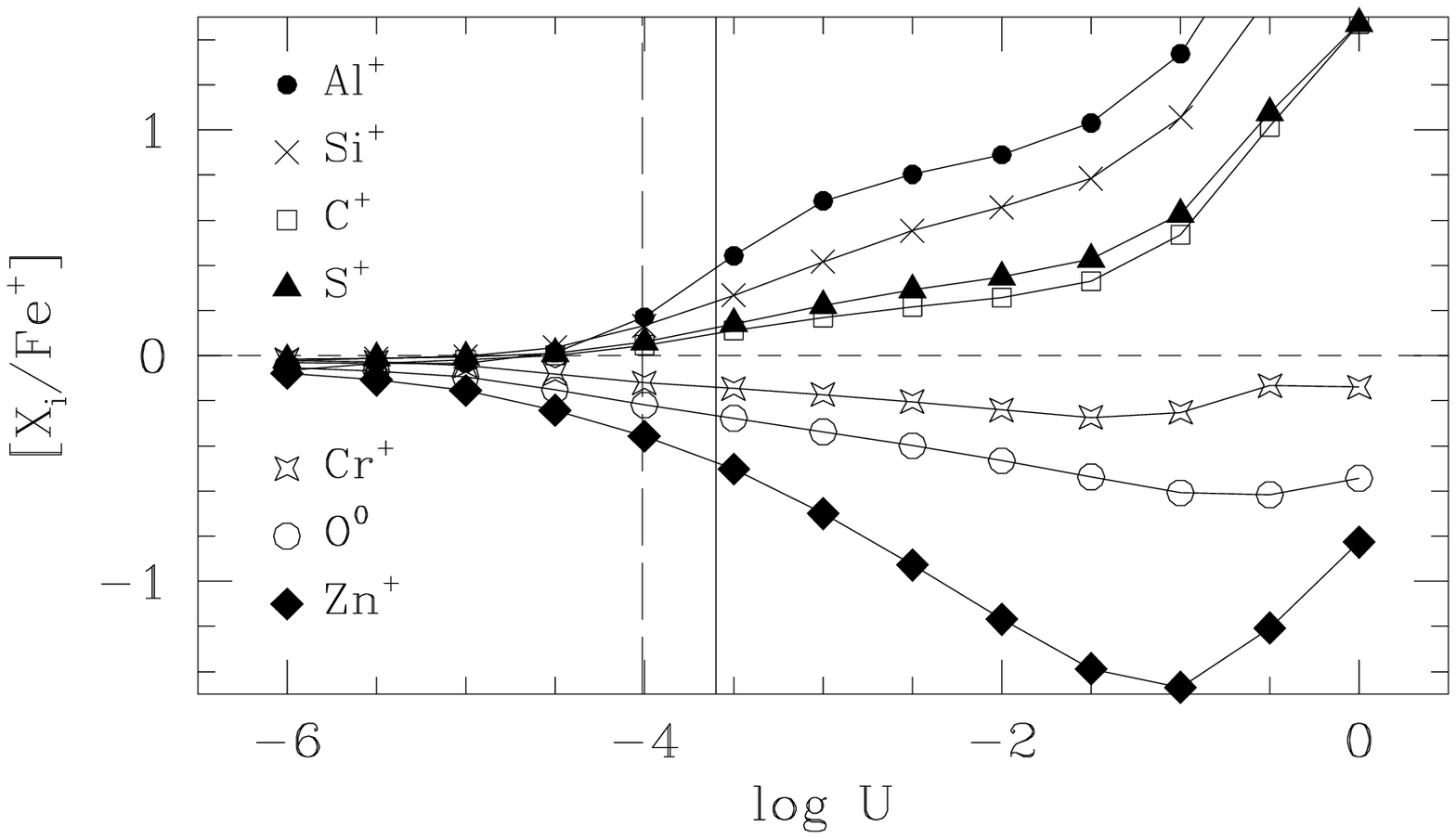}}
   \mbox{\includegraphics[width=85mm]{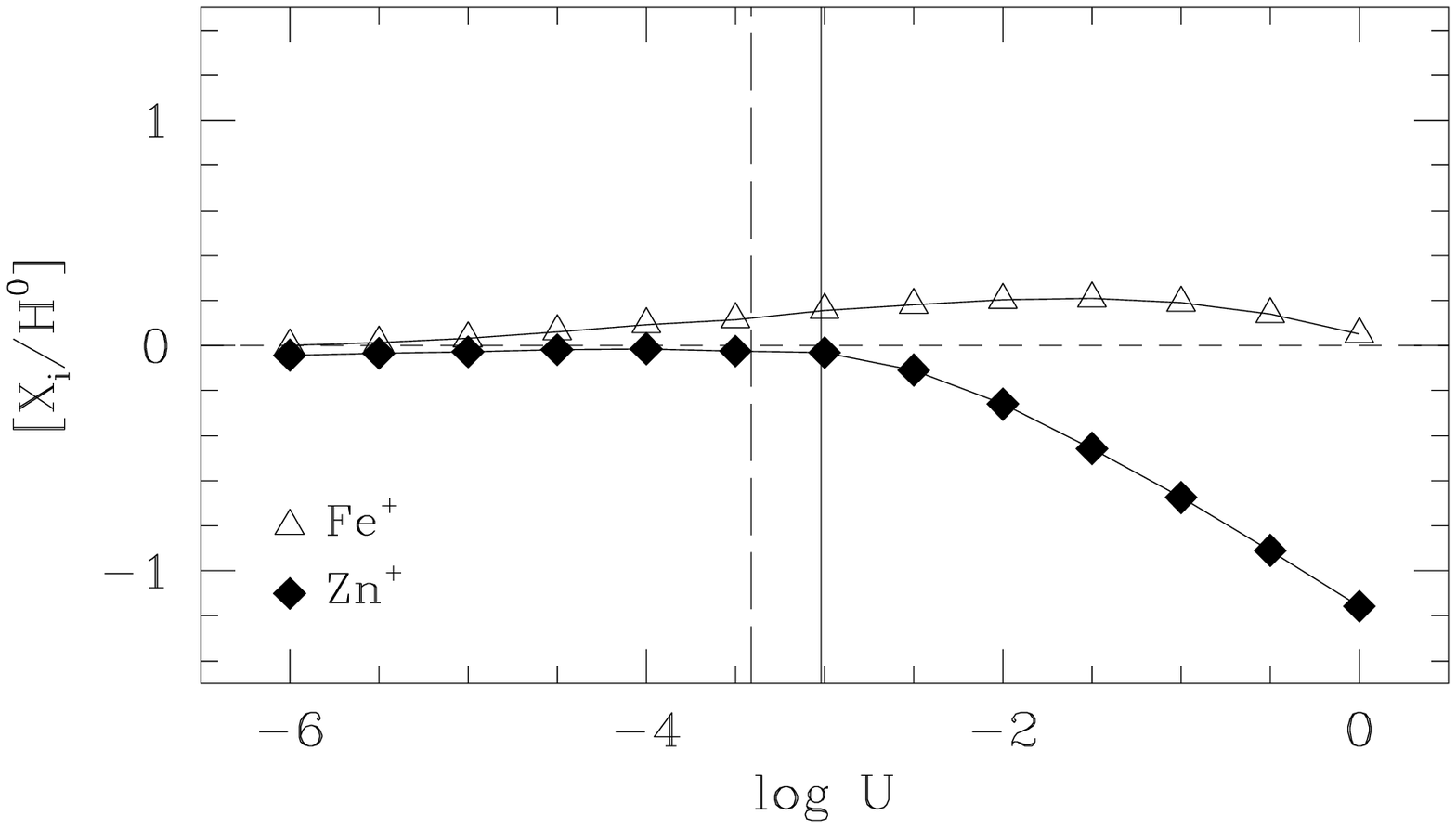}\quad
         \includegraphics[width=85mm]{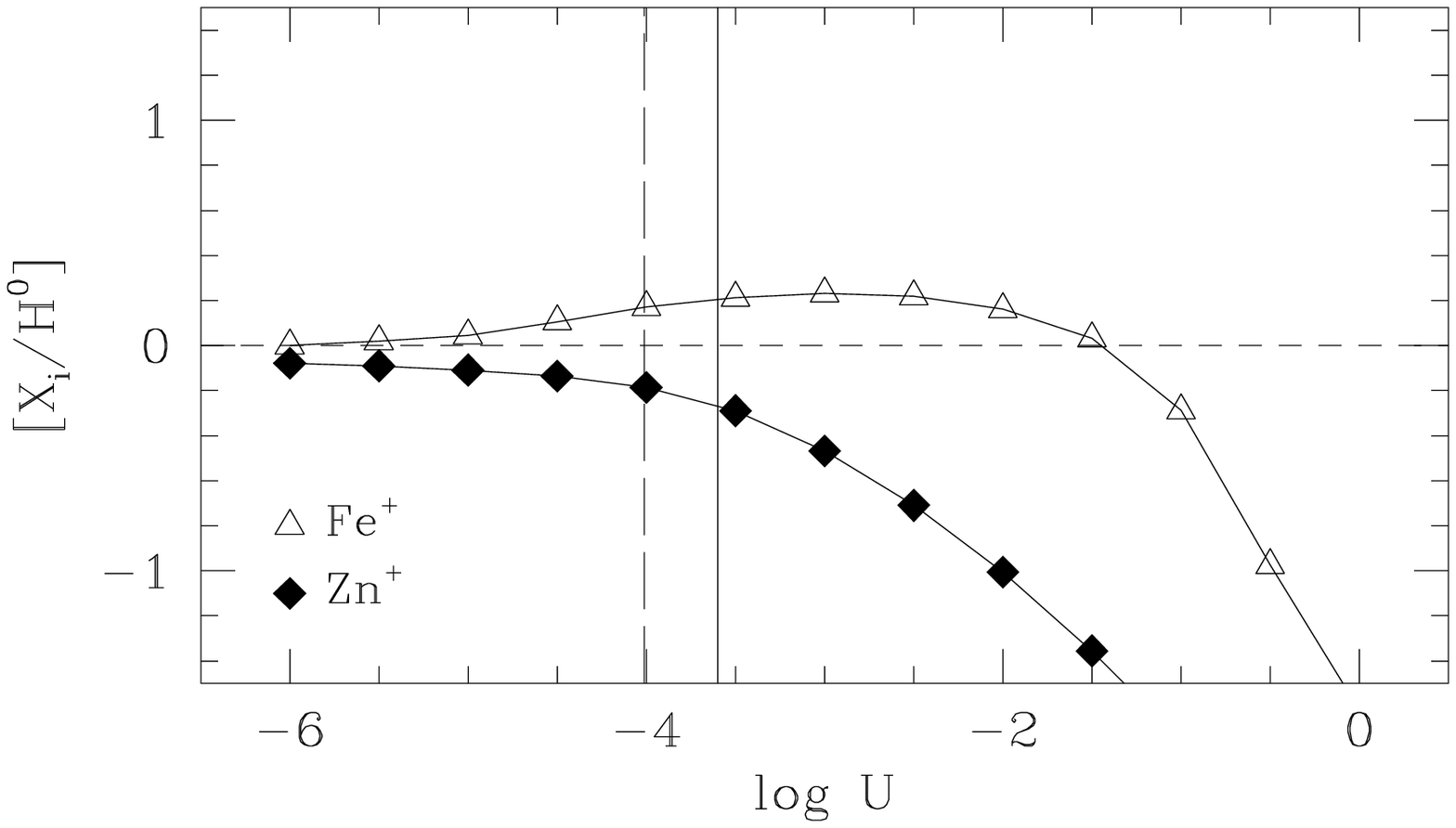}}
\caption{Photoionization models computed with the CLOUDY software package for the sub-DLA at $z_{\rm
abs} = 2.557$ towards Q1223+1753. This system is characterized by a relatively high metallicity [Fe/H]
$= -0.84\pm 0.15$ and has one of the lowest \ion{H}{i} column density ($\log N$(\ion{H}{i}) $=
19.32\pm 0.15$) among the sub-DLA systems in our sample. On the {\it left side}, the models are the 
result obtained with the Haardt-Madau (HM) UV ionizing source, and on the {\it right side}, a stellar 
ionizing source (SSP) is assumed. In {\it all panels} the vertical lines indicate the observational 
constraint on $U$ obtained from the Al$^{++}$/Al$^+$ (solid lines) and Fe$^{++}$/Fe$^+$ (long-dashed 
lines) ratios. The {\it top panels} show the predicted ionic column densities, $N$(X$_{\rm i}$), as 
a function of the ionization parameter, $U$. The horizontal dashed lines correspond to the observed 
column density values and the horizontal dotted lines correspond to the observed column density upper 
limits. The {\it middle} and {\it bottom panels} show the predicted abundance of low-ion X$_{\rm i}$ 
over Fe$^+$ and H$^0$ relative to the intrinsic abundance of these two elements, as defined in 
Section~\ref{methodology}, versus $U$, respectively. Departures of [X$_{\rm i}$/Fe] and 
[X$_{\rm i}$/H] indicate that ionization corrections have to be added to/subtracted from the observed 
relative and absolute abundances, respectively.}
\label{Q1223-ionization}
\end{figure*}
%

\begin{figure*}
   \mbox{\includegraphics[width=85mm]{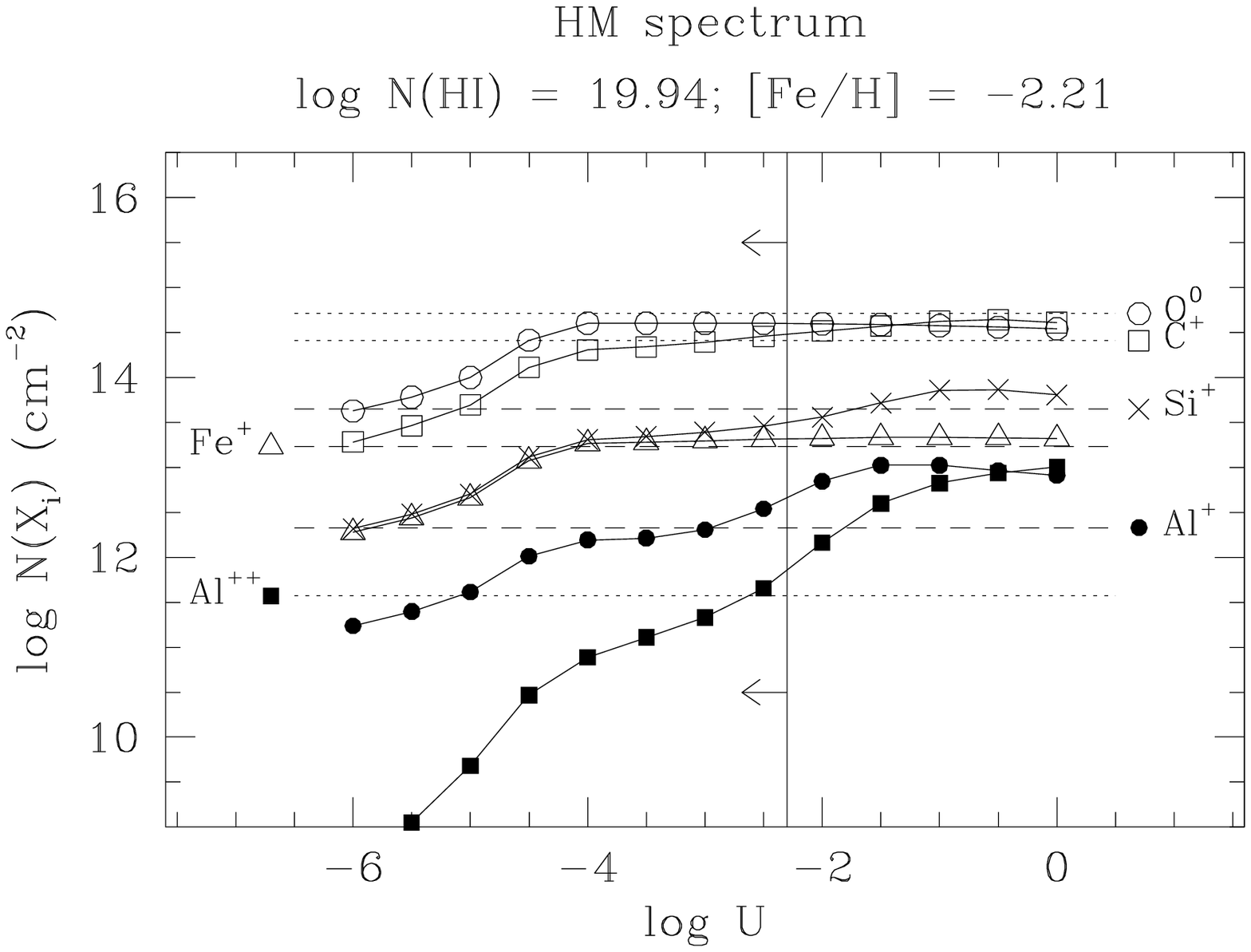}\quad
         \includegraphics[width=85mm]{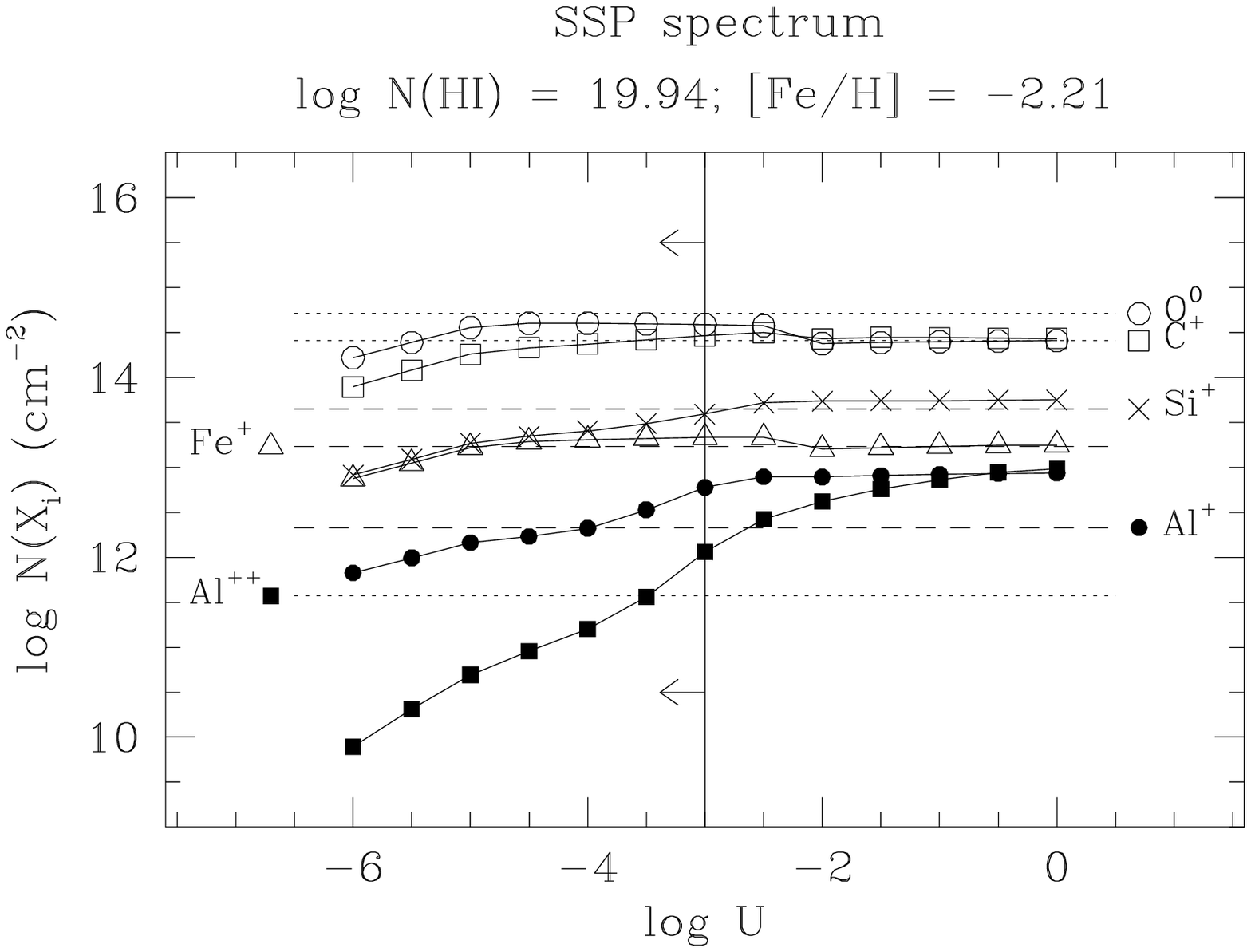}}
   \mbox{\includegraphics[width=85mm]{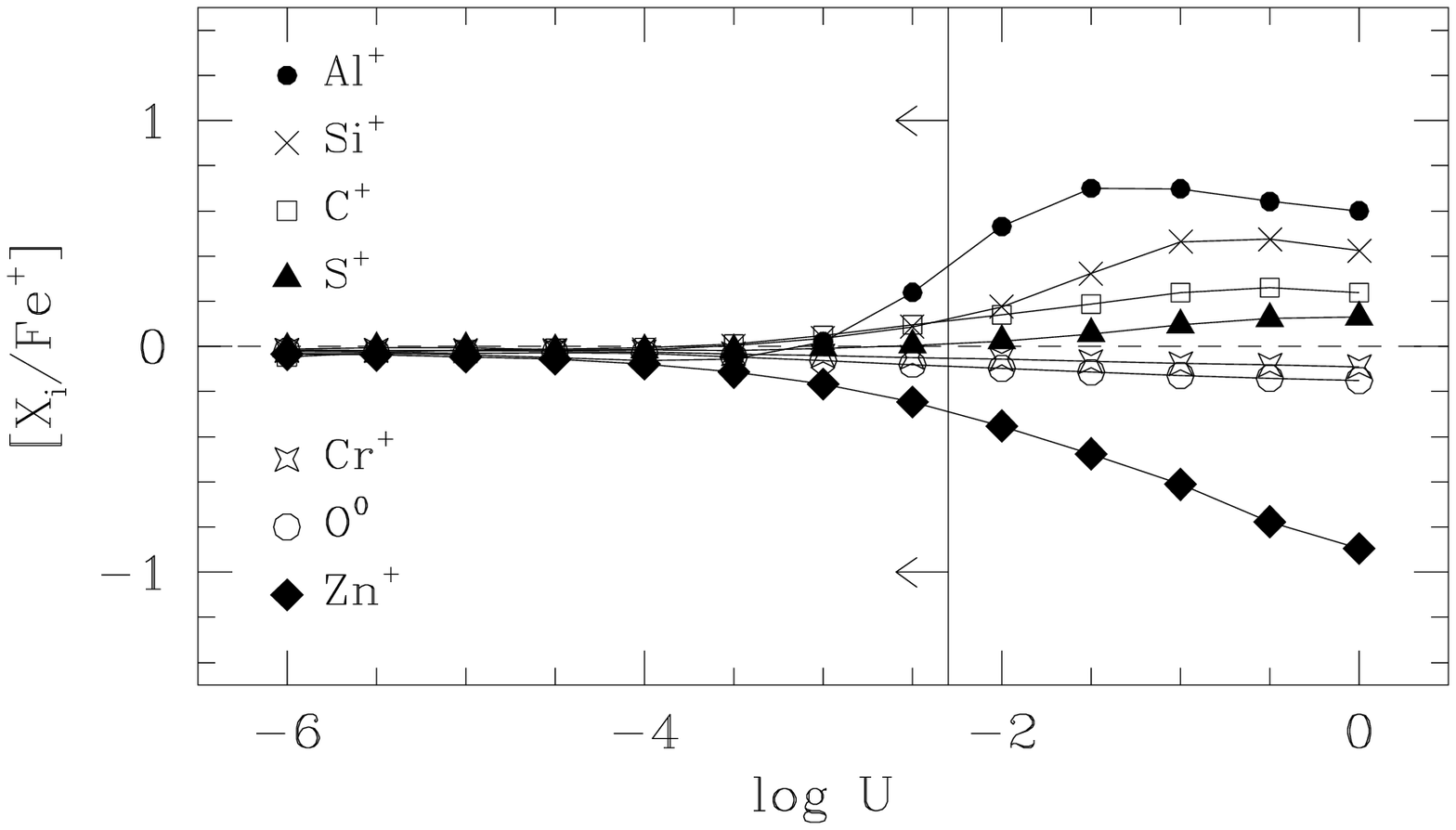}\quad
         \includegraphics[width=85mm]{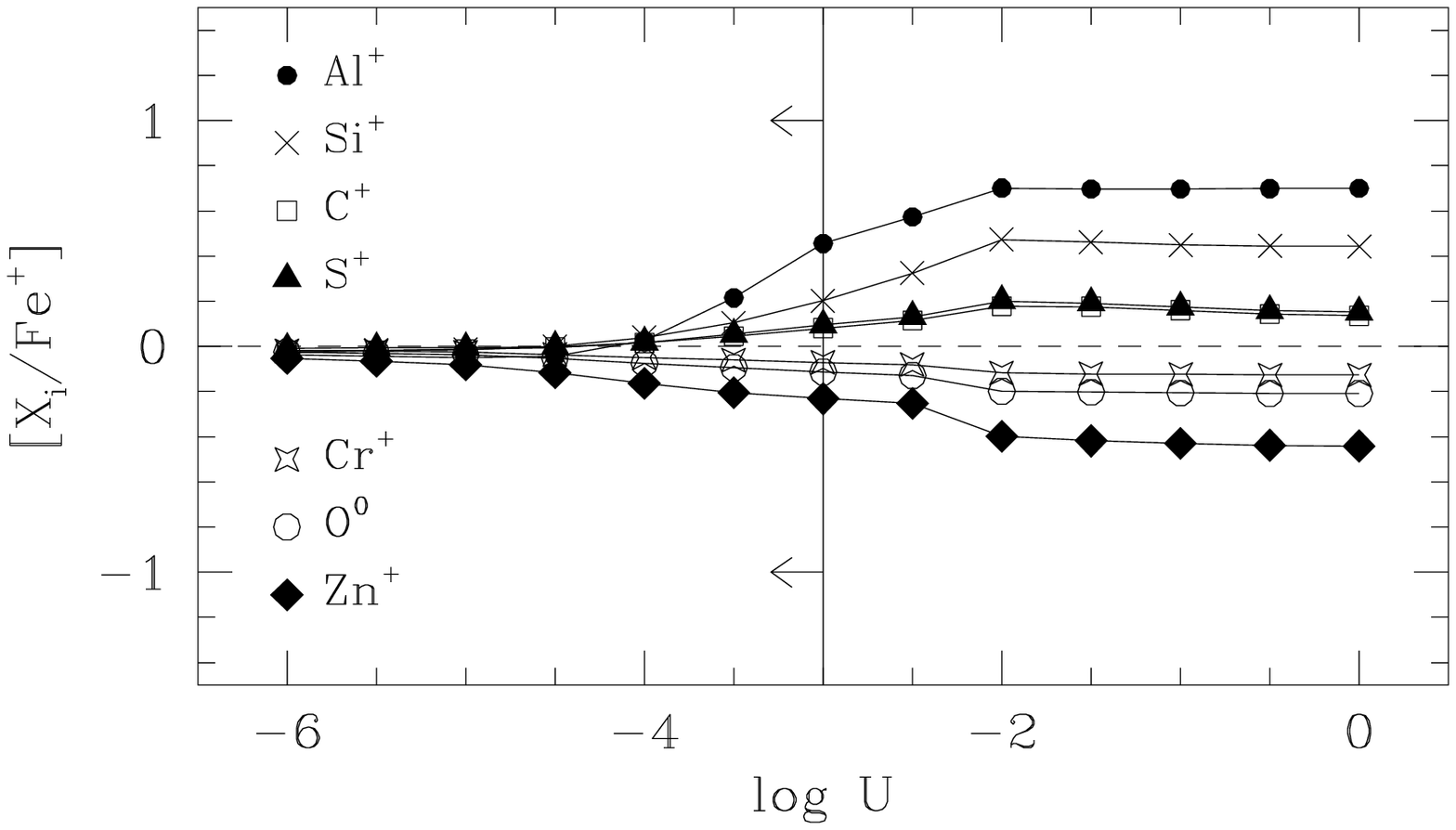}}
   \mbox{\includegraphics[width=85mm]{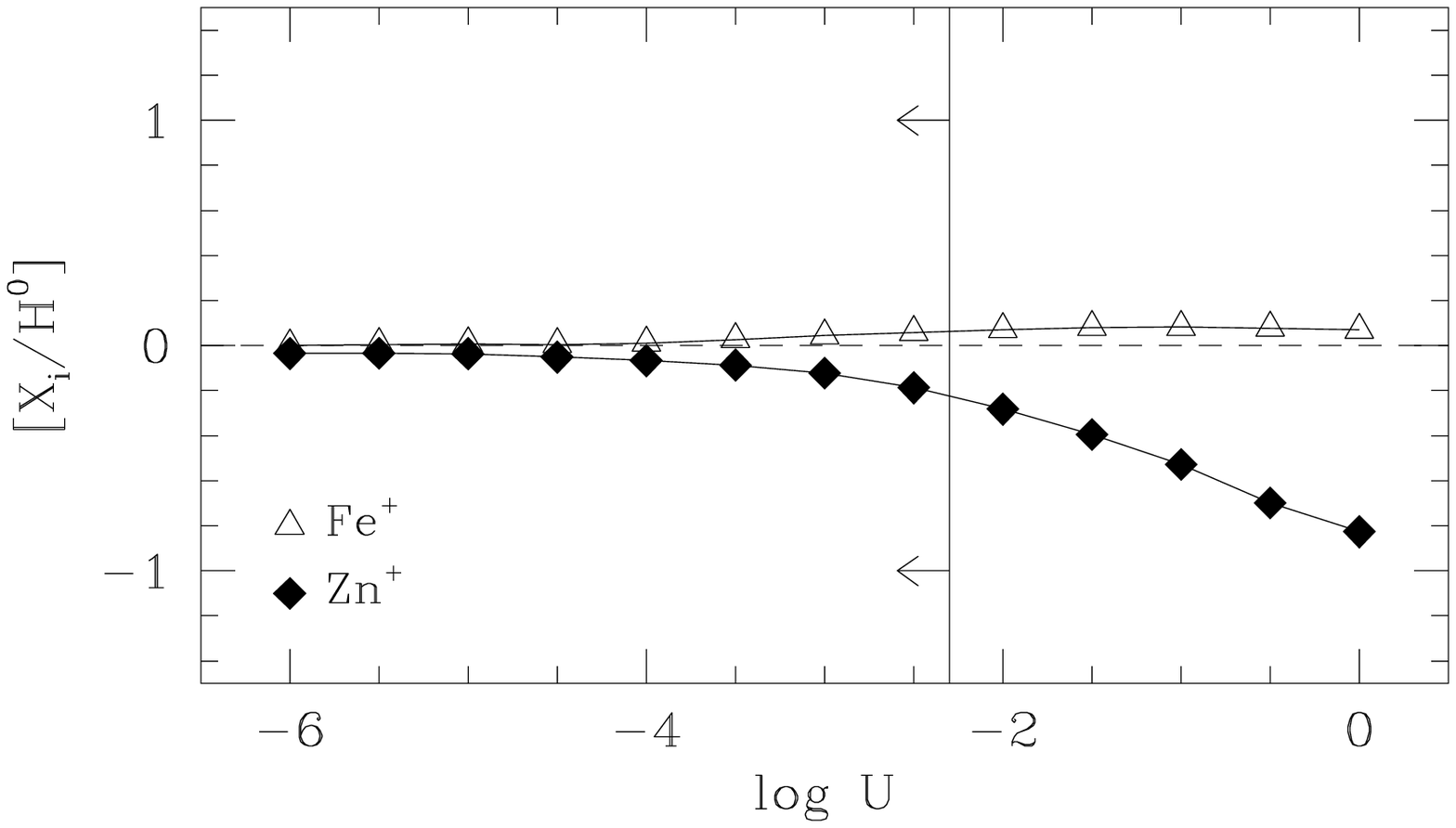}\quad
         \includegraphics[width=85mm]{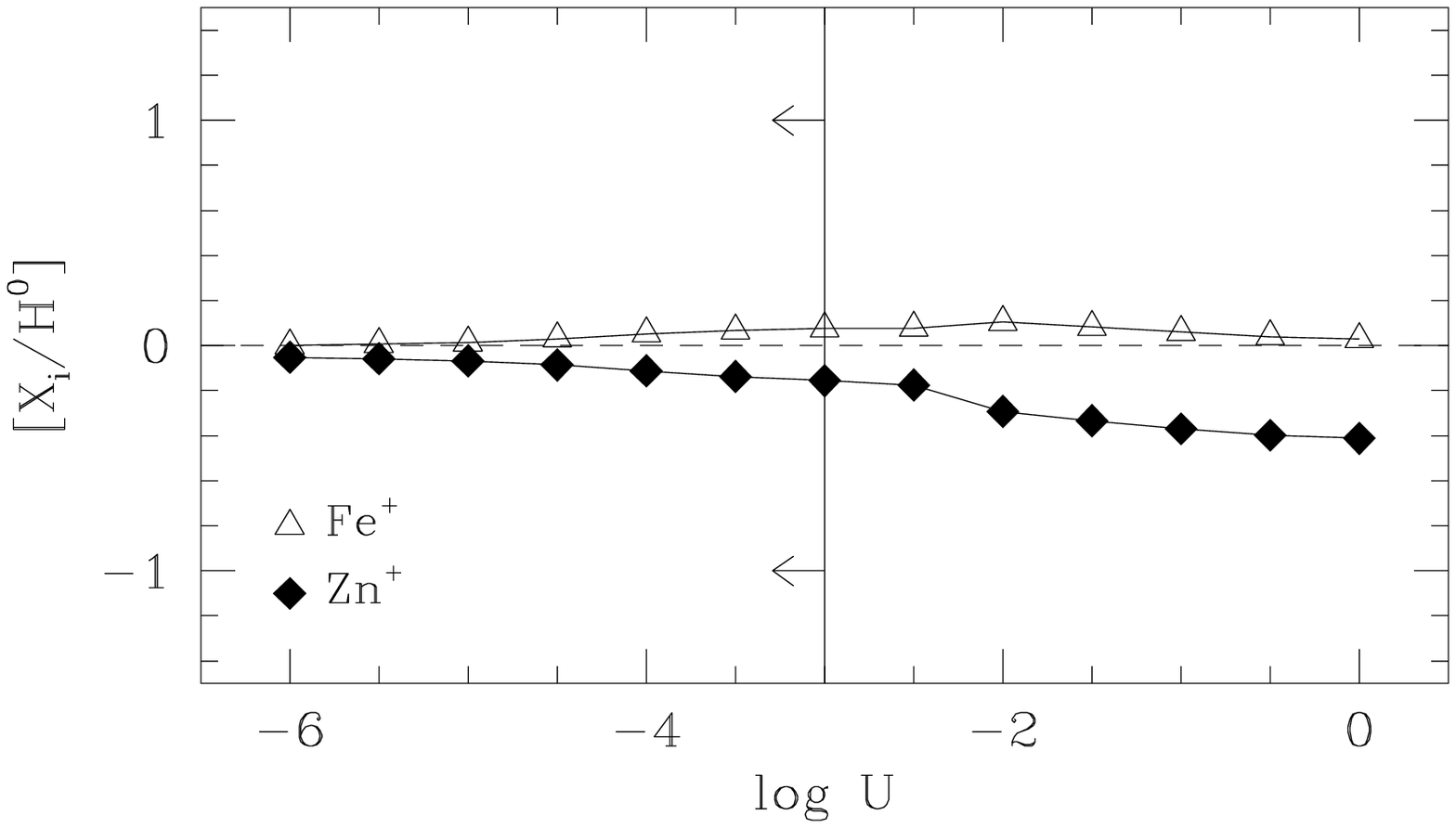}}
\caption{Same as Fig.~\ref{Q1223-ionization} for the sub-DLA at $z_{\rm abs} = 3.142$ towards
PSS J2155+1358. This system is characterized by a low metallicity [Fe/H] $= -2.21\pm 0.21$ and has 
one of the highest \ion{H}{i} column densities ($\log N$(\ion{H}{i}) $= 19.94\pm 0.10$) among the 
sub-DLA systems in our sample. In all panels the vertical line indicates the observational constraint 
on $U$ obtained from the Al$^{++}$/Al$^+$ upper limit.}
\label{Q2116-ionization}
\end{figure*}
%

Following the Prochaska et~al. (2002) ionization analysis, we have further computed models giving the 
predicted abundance of low-ion X$_{\rm i}$ over H$^0$ and Fe$^+$ relative to the intrinsic abundance 
of these two elements,
\begin{itemize}
\item[] ${\rm [X_i/H^0]} = \log (N({\rm X_i})/N({\rm H}^0)) - \log ({\rm X/H})_{\odot}$,

\item[] ${\rm [X_i/Fe^+]} = \log (N({\rm X_i})/N({\rm Fe}^+)) - \log ({\rm X/Fe})_{\odot}$,
\end{itemize}
as a function of the ionization parameter, $U$. Departures of [X$_{\rm i}$/H$^0$] and 
[X$_{\rm i}$/Fe$^+$] from zero indicate that ionization corrections have to be added to the observed 
absolute and relative elemental abundances, respectively. When the corrections to be applied are 
negative (positive), this implies that the  measured [X/H] or [X/Fe] are underestimated 
(overestimated). The main advantage of these models compared to the models giving the predicted 
X$_{\rm i}$ column densities as a function of $U$ is that they allow a direct quantification of the 
effects of photoionization on the absolute and relative abundances. 
%

\subsection{Results: Ionization Corrections}

In the sample of the 12 sub-DLA systems and one borderline case between the DLA and sub-DLA systems 
studied here, the \ion{H}{i} column densities range from $\log N$(\ion{H}{i}) $= 19.32$ to 20.30 
cm$^{-2}$, and the metallicities spread over $1.5$~dex (from [Fe/H] $< -2.40$ to [Fe/H] $= -0.79$). 
This large range of input parameters for the CLOUDY models allows us to highlight some general 
trends. Indeed, the computed CLOUDY models specific to each sub-DLA system in our sample show that 
for a given ionization parameter, $U$, the derived [X$_{\rm i}$/H$^0$] and [X$_{\rm i}$/Fe$^+$] 
ionization corrections (as defined in Section~\ref{methodology}) increase with lower \ion{H}{i} 
column densities of systems for all low-ions X$_{\rm i}$ and for both the HM or SSP ionizing spectra, 
while no particular trend is observed with the metallicity of the systems. In addition, we notice 
that the ionization corrections are on the average more important in the models with a SSP radiation 
field than with a HM spectrum for all of the systems studied. We show in Figs.~\ref{Q1223-ionization} 
and \ref{Q2116-ionization} the CLOUDY models for two sub-DLA systems, one at $z_{\rm abs} = 2.557$ 
towards Q1223+1753 with a very low \ion{H}{i} column density $\log N$(\ion{H}{i}) $= 19.32\pm 0.15$ 
and a high [Fe/H] $= -0.84\pm 0.15$, and the second at $z_{\rm abs} = 3.142$ towards PSS J2155+1358 
with a high \ion{H}{i} column density $\log N$(\ion{H}{i}) $= 19.94\pm 0.10$ and a low [Fe/H] $= 
-2.21\pm 0.21$, respectively. They illustrate these general trends.

Once we have computed the CLOUDY models, the determination of the ionization state is crucially 
dependent on the measurements of column density ratios of adjacent ions of elements, and thus on the 
detections of intermediate-ion transitions. In the sample of 13 systems studied, we were able to 
constrain the ionization parameter, $U$, for 10 systems. However, we obtained for only four systems  
the $N$(Al$^{++}$)/$N$(Al$^+$) and/or $N$(Fe$^{++}$)/$N$(Fe$^+$) measurements, while for the six 
remaining systems we derived only $N$(Al$^{++}$)/$N$(Al$^+$) and/or $N$(Fe$^{++}$)/$N$(Fe$^+$) upper 
limits. The analysis of the ionization state of our sub-DLA systems is thus severely limited by the 
fact that {\it i)}~we do not have a measurement of a column density ratio of adjacent ions of an 
element for all of the sub-DLAs, and {\it ii)}~for the majority of sub-DLAs we have only a 
measurement of the Al$^{++}$/Al$^+$ ratio or even only an upper limit, and as has already been 
pointed out by Petitjean, Rauch \& Carswell (1994), D'Odorico \& Petitjean (2001) and Vladilo et~al. 
(2001), the Al$^{++}$/Al$^+$ ratio is not a very reliable indicator of the ionization parameter. 
Indeed, the recombination coefficient of Al$^+$ is likely overestimated (Nussbaumer \& Storey 1986), 
and in this case the predicted Al$^{++}$/Al$^+$ ratio in the CLOUDY models is underestimated. This 
prevents us from obtaining a very accurate determination of the ionization parameter in our sub-DLA 
systems. In addition, with our set of data, we are not able to constrain the nature of the radiation 
field, since we cannot clearly differentiate which from the HM or SSP ionizing spectrum best 
represents the observations.

In this context we choose to extract from our analysis general conclusions on the ionization
corrections in the sub-DLA systems studied, rather than a discussion on individual systems which is 
limited by the partial information we have. Several important general remarks can be inferred from 
our study. {\it First}, as can be observed in Figs.~\ref{Q1223-ionization} and 
\ref{Q2116-ionization}, the ionization corrections to apply to absolute and relative abundances are 
different ion to ion. Indeed, in the case of C$^+$, O$^0$, S$^+$ or Cr$^+$, for example, the 
ionization corrections [X$_{\rm i}$/Fe$^+$] increase smoothly with the ionization parameter, $U$, 
and remain relatively small for all of the $U$ values, while for the Si$^+$, Al$^+$ and Zn$^+$ ions 
they already become important at relatively low $U$ values. Similarly, for the metallicities, the 
[X$_{\rm i}$/H$^0$] corrections are not significant for Fe$^+$ up to high $U$ values, while they are 
much higher for Zn$^+$ at high $U$ values. {\it Secondly, for 9 systems out of 10 for which we were 
able to constrain the ionization parameter the ionization corrections are lower than 0.2~dex for all 
of the elements but \ion{Al}{ii} and \ion{Zn}{ii}} (in some cases; mainly if we assume that the SSP 
radiation field is the dominant ionizing source in our systems) down to \ion{H}{i} column densities 
of $\log N$(\ion{H}{i}) $= 19.3$ cm$^{-2}$ (see Figs.~\ref{Q1223-ionization} and 
\ref{Q2116-ionization}). In the case of Al$^+$ and Zn$^+$ the ionization corrections may be 
overestimated, since the Al$^+$ and Zn$^+$ recombination coefficients and ionization cross sections 
are rather uncertain (see Nussbaumer \& Storey 1986 and Howk, Savage \& Fabian 1999, respectively). 
Only in the sub-DLA system at $z_{\rm abs} = 3.565$ towards PSS~J2155+1358 are the ionization 
corrections important (filled square in the upper left corner in Fig.~\ref{AlIII-AlII}). For three 
sub-DLAs in our sample we could not constrain the ionization parameter, $U$.
%

\section{Overview of the Chemical Abundances in Sub-DLAs}

The constructed sample of 12 sub-DLA systems and one borderline case between the DLA and sub-DLA
systems, and the homogeneous chemical abundance analysis of each system presented in this paper 
result in the first sub-DLA chemical abundance database ideal for the study of a number of important 
issues. Indeed, 16 different ions of low-, intermediate- and high-ionization and 11 different 
elements $-$ O, C, Si, N, S, Mg, Al, Fe, Ni, Zn and Cr~$-$ have been detected in the 13 systems 
studied, showing a variety of elements as diverse as that observed in the DLAs. 
Tables~\ref{N-summary}~a) and b) present a summary of the column density measurements of the low-ion 
transitions and the intermediate-/high-ion transitions, respectively, obtained in the 13 systems 
analysed. These column densities were derived by summing the column densities of all of the detected 
velocity components in each ion as given in Tables~\ref{Q1101-values}$-$\ref{J2344-values}. The 
cases in which we provide only lower limits correspond to the cases where the observed metal lines 
are saturated, and the cases in which we provide only upper limits correspond to cases where the 
observed metal lines are blended with Ly$\alpha$ forest absorption lines. 4~$\sigma$ upper limits
were derived in the cases where the metal lines are not detected and are identified by an asterisk.

In Table~\ref{absolute-summary} we provide a summary of the absolute abundance measurements of the 
12 sub-DLA systems and one borderline case between the DLA and sub-DLA systems, defined with respect 
to the solar meteoritic abundances from Grevesse \& Sauval (1998). No ionization correction has 
been adopted since, as was shown in Section~\ref{ionization} for all of the systems studied for 
which we were able to constrain the ionization parameter (10 systems out of 13) except one at 
$z_{\rm abs} = 3.565$ towards PSS J2155+1358, the ionization corrections are almost negligible, 
being at the level of the errors. Indeed, they are lower than 0.2~dex for all of the elements except 
\ion{Al}{ii} and \ion{Zn}{ii} (in some cases, but likely overestimated because the Al$^+$ and Zn$^+$ 
recombination coefficients are very uncertain) down to \ion{H}{i} column densities of 
$\log N$(\ion{H}{i}) $= 19.3$ cm$^{-2}$.

\begin{figure}
   \includegraphics[width=85mm]{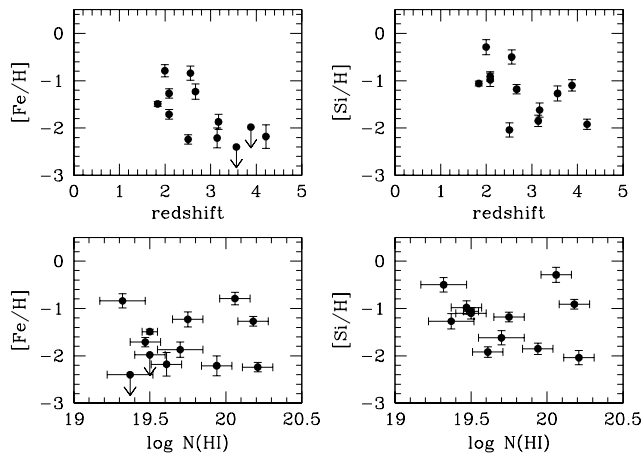}
\caption{The four plots show the distribution of the abundances of Fe and Si as a function of 
redshift and \ion{H}{i} column density, to illustrate the values sampled by the 12 sub-DLA systems 
in our sample. The overall abundances are presented in Table~\ref{absolute-summary}.}
\label{abundance-plots}
\end{figure}

Fig.~\ref{abundance-plots} shows the distribution of iron and silicon absolute abundances, [Fe/H] 
and [Si/H], as a function of redshift and \ion{H}{i} column density for the 12 sub-DLAs in our 
sample. The studied sub-DLAs span a range of redshift from 1.8 to 4.3 and of [Fe/H] metallicities 
from $-0.79$ to $<-2.40$, which is very similar to that observed in DLAs. We observe a trend of 
increasing metallicity with redshift, but no correlation between metallicities and \ion{H}{i} column 
densities is visible. An overall discussion on the chemical abundances of sub-DLAs, and a detailed 
comparison of the chemical, kinematic and clustering properties between the sub-DLAs and DLAs is 
given in Paper~II.
%

%

The lower \ion{H}{i} column densities of the sub-DLA systems relative to the DLA systems give us the 
opportunity to have access to some elements which absorption lines are usually saturated in DLAs, but 
prevent us from measuring abundances of some other elements whose lines are already weak in DLAs. As 
examples of the former case, one can quote the carbon and oxygen. Indeed, C is almost not accessible 
in DLAs, since the \ion{C}{ii} $\lambda$1036,1334 lines are heavily saturated. Only very few 
tentative measurements have been obtained (Lopez et~al. 1999,2002; Levshakov et~al. 2002; Songaila 
\& Cowie 2002). The O abundance is also very difficult to measure in DLAs, since the most commonly 
detected \ion{O}{i} line at $\lambda_{\rm rest} = 1302$ \AA\ is saturated and other \ion{O}{i} lines 
are located in the Ly$\alpha$ forest. On the contrary, in the sub-DLA systems one can more easily 
measure the abundances of these two important elements $-$ we obtained four O and six C abundance 
measurements~$-$ which may provide insight into the understanding of the nature and the chemical 
evolution history of these systems (see Paper~II). On the other hand, the useful Zn abundance used 
as an indicator of the dust depletion in DLAs is very hard to measure in sub-DLAs given the weakness 
of the \ion{Zn}{ii} lines. While numerous Zn abundance measurements exist in DLAs, we obtained only 
two Zn measurements in our sample of sub-DLAs. 
%

\section{Conclusion}

The findings by P\'eroux et~al. (2003a) that the `sub-DLA' systems with \ion{H}{i} column 
densities between $10^{19} < N$(\ion{H}{i}) $< 2\times 10^{20}$ cm$^{-2}$ contain a large fraction 
of the neutral hydrogen gas mass especially at $z > 3.5$, have motivated us to investigate 
systematically quasar spectra in the ESO UVES-VLT archives searching for these absorbers. We have 
obtained the following results:

\begin{itemize}
\item Out of 22 high-resolution quasar spectra available to us in July 2001 in the ESO UVES-VLT 
archives and suitable for the construction of a sample of sub-DLA systems, reduced and analysed with 
an identical approach, we have built the first sample of 12 sub-DLA systems in the redshift interval 
$z = 1.8-4.3$. These systems were identified in the quasar spectra both via an automated selection 
algorithm and visual criteria, like the presence of damping wings in the Ly$\alpha$ absorption line. 
Ten of them constitute a sample that we believe to be homogeneous and unbiased.

\item The detailed metal line analysis of each of the 12 sub-DLA systems and one borderline case
between the DLA and sub-DLA systems resulted in the column density measurements of 16 ions of low-, 
intermediate- and high-ionization $-$ \ion{O}{i}, \ion{C}{ii}, \ion{C}{iv}, \ion{Si}{ii}, 
\ion{Si}{iv}, \ion{N}{i}, \ion{S}{ii}, \ion{Mg}{i}, \ion{Mg}{ii}, \ion{Al}{ii}, \ion{Al}{iii}, 
\ion{Fe}{ii}, \ion{Fe}{iii}, \ion{Ni}{ii}, \ion{Zn}{ii} and \ion{Cr}{ii}. The detection of adjacent 
ions of the same element, Fe$^{++}$/Fe$^+$ and Al$^{++}$/Al$^+$, allowed us to estimate the 
ionization parameters in 10 systems. Using the CLOUDY software package, we computed photoionization 
models and found that for all of the 10 systems except one at $z_{\rm abs} = 3.565$ towards PSS 
J2155+1358, the ionization corrections are almost negligible, being at the level of the errors. 
Indeed, they are lower than 0.2~dex for all of the elements except \ion{Al}{ii} and \ion{Zn}{ii} (in 
some cases, but likely overestimated because the Al$^+$ and Zn$^+$ recombination coefficients are 
very uncertain) down to \ion{H}{i} column densities of $\log N$(\ion{H}{i}) $= 19.3$ cm$^{-2}$.

\item We have obtained the first sub-DLA chemical abundance database which contains the abundance 
measurements of 11 different elements $-$ O, C, Si, N, S, Mg, Al, Fe, Ni, Zn and Cr. The lower 
\ion{H}{i} column densities of sub-DLAs allowed us, in particular, to easily measure the O and C 
abundances, the \ion{O}{i} and \ion{C}{ii} lines being usually saturated in DLAs. We obtained O 
abundance measurements in four systems and C in six systems.
\end{itemize}

The constructed homogeneous and unbiased sample of the sub-DLA systems and the obtained sub-DLA 
chemical abundance database are ideal for the analysis of a number of important issues related both 
to the neutral gas mass and the metal content evolutions in the Universe. These issues are discussed 
in Paper~II of this series, where we also present a detailed study of the sub-DLA kinematic, 
chemical and clustering properties, and their comparison with the properties of the well studied DLA 
systems. We plan to extend the current sample of sub-DLA systems in order to increase the statistical 
weight of the obtained results both from the newly accessible data in the ESO UVES archives and from 
observations of pre-selected high-redshift quasars, since it is at high redshift $z > 3.5$ that the 
sub-DLAs are expected to contain a large fraction of the neutral hydrogen gas mass (P\'eroux et~al. 
2002a). 
%

\section{Acknowledgments}

We are very grateful to Paolo Molaro and Miriam Centuri\'on for providing us with the reduced and
normalized spectrum of the quasar Q0000$-$2620. We particularly thank Miriam Centuri\'on for her
help in the computation of photoionization models. We express our gratitude to the ESO staff in
charge of the VLT archive for their efficient work to the benefit of the community. We thank an
anonymous referee for helpful comments. M.D.-Z. is supported by the Swiss National Funds and C.P. is 
funded by a European Marie Curie Fellowship.
%

\begin{landscape}
\begin{table}
\caption{a) Low-ion transition column density summary} 
\label{N-summary}
\vspace{-0.2cm}
{\scriptsize
\begin{tabular}{l c c c c c c c c c c c c c}
\hline
Name                   & $z_{\rm abs}$          & $\log N$(\ion{H}{i})   & $\log N$(\ion{O}{i})   & $\log N$(\ion{C}{ii})  & 
$\log N$(\ion{Si}{ii}) & $\log N$(\ion{N}{i})   & $\log N$(\ion{S}{ii})  & $\log N$(\ion{Mg}{ii}) & 
$\log N$(\ion{Al}{ii}) & $\log N$(\ion{Fe}{ii}) & $\log N$(\ion{Ni}{ii}) & $\log N$(\ion{Zn}{ii}) & $\log N$(\ion{Cr}{ii}) 
\\
\hline
Q1101$-$264$^a$........... & 1.838 & 19.50{\tiny (0.05)} & 14.55{\tiny (0.09)} & 15.06{\tiny (0.07)} &
14.00{\tiny (0.01)} & ... & 13.66{\tiny (0.11)} & 14.07{\tiny (0.03)} &
12.85{\tiny (0.06)} & 13.51{\tiny (0.02)} & ... & $< 11.39$$^*$ & $< 12.04$$^*$ \\
Q1223$+$1753............ & 2.557 & 19.32{\tiny (0.15)} & $< 15.38$ & ... &
14.38{\tiny (0.02)} & ... & ... & ... &
13.20{\tiny (0.03)} & 13.98{\tiny (0.03)} & ... & $< 11.63$$^*$ & $< 11.96$$^*$ \\
Q1409$+$095............. & 2.668 & 19.75{\tiny (0.10)} & 15.31{\tiny (0.11)} & $> 16.09$ & 
14.13{\tiny (0.03)} & $< 13.49$ & 13.54{\tiny (0.06)} & ... &
... & 14.02{\tiny (0.13)} & ... & $< 11.34$$^*$ & 12.49{\tiny (0.15)} \\
Q1444$+$014............. & 2.087 & 20.18{\tiny (0.10)} & $> 16.08$ & $> 16.30$ &
14.83{\tiny (0.02)} & ... & ... & ... &
13.42{\tiny (0.03)} & 14.41{\tiny (0.03)} & ... & ... & ... \\
Q1451$+$123............. & 2.255 & 20.30{\tiny (0.15)} & ... & ... &
$> 15.46$ & ... & ... & ... &
... & 14.33{\tiny (0.07)} & $< 13.53$ & 11.85{\tiny (0.11)} & 12.88{\tiny (0.14)} \\
Q1451$+$123 ............ & 3.171 & 19.70{\tiny (0.15)} & $< 14.80$ & 14.30{\tiny (0.20)} & 
13.64{\tiny (0.03)} & ... & ... & ... &
12.31{\tiny (0.03)} & 13.33{\tiny (0.06)} & ... & $< 12.01$$^*$ & $< 12.65$$^*$ \\
Q1511$+$090 ............ & 2.088 & 19.47{\tiny (0.10)} & $< 15.30$ & ... &
14.05{\tiny (0.10)} & ... & ... & ... &
12.93{\tiny (0.14)} & 13.26{\tiny (0.03)} & ... & ... & ... \\
Q2059$-$360............. & 2.507 & 20.21{\tiny (0.10)} & 15.54{\tiny (0.21)} & 14.87{\tiny (0.19)} &
13.73{\tiny (0.11)} & 12.66{\tiny (0.18)} & ... & ... &
... & 13.47{\tiny (0.03)} & ... & $< 11.54$$^*$ & $< 12.19$$^*$ \\
Q2116$-$358............. & 1.996 & 20.06{\tiny (0.10)} & $> 16.68$ & ... &
15.33{\tiny (0.12)} & ... & $< 14.95$ & ... &
$> 13.94$ & 14.77{\tiny (0.09)} & 13.57{\tiny (0.06)} & 12.29{\tiny (0.09)} & 12.92{\tiny (0.13)} \\
PSS J2155$+$1358..... & 3.142 & 19.94{\tiny (0.10)} & $< 14.71$ & $< 14.41$ &
13.65{\tiny (0.07)} & ... & ... & ... &
12.33{\tiny (0.10)} & 13.23{\tiny (0.19)} & ... & ... & ... \\
PSS J2155$+$1358..... & 3.565 & 19.37{\tiny (0.15)} & ... & 14.34{\tiny (0.12)} &
13.66{\tiny (0.04)} & ... & ... & ... &
12.16{\tiny (0.12)} & $< 12.47$$^*$ & ... & ... & ... \\
PSS J2155$+$1358..... & 4.212 & 19.61{\tiny (0.10)} & 14.50{\tiny (0.05)} & 13.95{\tiny (0.06)} &
13.25{\tiny (0.04)} & ... & ... & ... &
11.92{\tiny (0.17)} & 12.93{\tiny (0.23)} & ... & ... & ... \\
PSS J2344$+$0342$^b$... & 3.882 & 19.50{\tiny (0.10)} & ... & 14.74{\tiny (0.05)} &
13.96{\tiny (0.06)} & ... & ... & ... &
12.62{\tiny (0.11)} & $< 13.02$$^*$ & ... & ... & ... \\
\hline
\multicolumn{14}{l}{{\footnotesize $^a$ The \ion{Mg}{i} column density is also measured in this sub-DLA: 
$\log N$(\ion{Mg}{i}) $= 11.89\pm 0.02$.}} \\
\multicolumn{14}{l}{{\footnotesize $^b$ If the components 1$-$6 do not belong to the sub-DLA (see Section~\ref{J2344}), then:
$\log N$(\ion{C}{ii}) $= 14.55\pm 0.04$, $\log N$(\ion{Si}{ii}) $= 13.70\pm 0.04$ and $\log N$(\ion{Al}{ii}) $= 12.36\pm 0.07$.}} 
\end{tabular}
}
{\scriptsize
\begin{tabular}{l c c c c c c}
\multicolumn{7}{l}{} \\
\multicolumn{7}{l}{{\footnotesize \phantom{{\bf Table16.}}b) Intermediate- and high-ion transition column density summary}} \\
\hline
Name & $z_{\rm abs}$ & $\log N$(\ion{H}{i}) & $\log N$(\ion{Al}{iii}) & $\log N$(\ion{Fe}{iii}) & $\log N$(\ion{C}{iv}) & 
$\log N$(\ion{Si}{iv}) 
\\
\hline
Q1101$-$264............. & 1.838 & 19.50{\tiny (0.05)} & 12.34{\tiny (0.04)} & ... & 14.22{\tiny (0.04)} &
13.84{\tiny (0.05)} \\
Q1223$+$1753............ & 2.557 & 19.32{\tiny (0.15)} & 12.32{\tiny (0.13)} & 14.31{\tiny (0.13)} & 14.44{\tiny (0.04)} &
13.99{\tiny (0.02)} \\
Q1409$+$095............. & 2.668 & 19.75{\tiny (0.10)} & 11.89{\tiny (0.05)} & 13.17{\tiny (0.12)} & ... &
... \\
Q1444$+$014............. & 2.087 & 20.18{\tiny (0.10)} & $< 12.06$$^*$ & $< 13.90$$^*$ & 13.22{\tiny (0.05)} &
12.94{\tiny (0.04)} \\
Q1451$+$123............. & 2.255 & 20.30{\tiny (0.15)} & 12.70{\tiny (0.05)} & ... & ... &
... \\
Q1451$+$123 ............ & 3.171 & 19.70{\tiny (0.15)} & $< 11.57$$^*$ & ... & 13.48{\tiny (0.13)} &
... \\
Q1511$+$090 ............ & 2.088 & 19.47{\tiny (0.10)} & $< 11.84$$^*$ & $< 13.75$$^*$ & ... & 
... \\
Q2059$-$360............. & 2.507 & 20.21{\tiny (0.10)} & $< 11.89$$^*$ & ... & ... & 
... \\
Q2116$-$358............. & 1.996 & 20.06{\tiny (0.10)} & 13.25{\tiny (0.09)} & $< 14.60$ & $> 15.28$ &
14.38{\tiny (0.02)} \\
PSS J2155$+$1358..... & 3.142 & 19.94{\tiny (0.10)} & $< 11.57$ & ... & ... 
& ... \\
PSS J2155$+$1358..... & 3.565 & 19.37{\tiny (0.15)} & 12.71{\tiny (0.09)} & $< 13.81$ & 14.38{\tiny (0.10)} &
14.05{\tiny (0.04)} \\
PSS J2155$+$1358..... & 4.212 & 19.61{\tiny (0.10)} & ... & ... & ... &
12.93{\tiny (0.12)} \\
PSS J2344$+$0342..... & 3.882 & 19.50{\tiny (0.10)} & $< 12.32$$^*$ & ... & 14.19{\tiny (0.05)} & 
13.64{\tiny (0.07)} \\
\hline
\multicolumn{7}{l}{{\footnotesize $^*$ 4~$\sigma$ upper limit (corresponding to non-detections).}} 
\end{tabular}
}
\end{table}

\begin{table}
\vspace{-0.1cm}
\caption{Absolute abundance summary} 
\label{absolute-summary}
\vspace{-0.2cm}
{\scriptsize
\begin{tabular}{l c c c c c c c c c c c c c}
\hline
Name                      & $z_{\rm abs}$  & $\log N$(\ion{H}{i})  & [O/H]  & [C/H]  & 
[Si/H]                    & [N/H]          & [S/H]                 & [Mg/H] & 
[Al/H]                    & [Fe/H]         & [Ni/H]                & [Zn/H] & [Cr/H]
\\
\hline
Q1101$-$264.............. & 1.838 & 19.50{\tiny (0.05)} & $-1.78${\tiny (0.12)} & $-0.96${\tiny (0.10)} &
$-1.06${\tiny (0.05)} & ... & $-1.04${\tiny (0.13)} & $-1.01${\tiny (0.06)} &
$-1.14${\tiny (0.08)} & $-1.49${\tiny (0.05)} & ... & $< -0.78$$^*$ & $< -1.15$$^*$ \\
Q1223$+$1753............ & 2.557 & 19.32{\tiny (0.15)} & $< -0.77$ & ... &
$-0.50${\tiny (0.15)} & ... & ... & ... &
$-0.61${\tiny (0.15)} & $-0.84${\tiny (0.15)} & ... & $< -0.36$$^*$ & $< -1.05$$^*$ \\
Q1409$+$095............. & 2.668 & 19.75{\tiny (0.10)} & $-1.27${\tiny (0.16)} & $> -0.18$ & 
$-1.18${\tiny (0.10)} & $< -2.18$ & $-1.41${\tiny (0.13)} & ... &
... & $-1.23${\tiny (0.16)} & ... & $< -1.08$$^*$ & $-0.95${\tiny (0.18)} \\
Q1444$+$014............. & 2.087 & 20.18{\tiny (0.10)} & $> -0.93$ & $> -0.40$ &
$-0.91${\tiny (0.10)} & ... & ... & ... &
$-1.25${\tiny (0.10)} & $-1.27${\tiny (0.10)} & ... & ... & ... \\
Q1451$+$123............. & 2.255 & 20.30{\tiny (0.15)} & ... & ... &
$> -0.40$ & ... & ... & ... &
... & $-1.47${\tiny (0.17)} & $< -1.02$ & $-1.12${\tiny (0.19)} & $-1.11${\tiny (0.20)} \\
Q1451$+$123 ............ & 3.171 & 19.70{\tiny (0.15)} & $< -1.73$ & $-1.92${\tiny (0.26)} & 
$-1.62${\tiny (0.15)} & ... & ... & ... &
$-1.88${\tiny (0.15)} & $-1.87${\tiny (0.16)} & ... & $< -0.36$$^*$ & $< -0.74$$^*$ \\
Q1511$+$090 ............ & 2.088 & 19.47{\tiny (0.10)} & $< -1.00$ & ... &
$-0.98${\tiny (0.14)} & ... & ... & ... &
$-1.03${\tiny (0.17)} & $-1.71${\tiny (0.10)} & ... & ... & ... \\
Q2059$-$360............. & 2.507 & 20.21{\tiny (0.10)} & $-1.50${\tiny (0.24)} & $-1.86${\tiny (0.22)} &
$-2.04${\tiny (0.15)} & $-3.47${\tiny (0.21)} & ... & ... &
... & $-2.24${\tiny (0.10)} & ... & $< -1.34$$^*$ & $< -1.71$$^*$ \\
Q2116$-$358............. & 1.996 & 20.06{\tiny (0.10)} & $> -0.21$ & ... &
$-0.29${\tiny (0.16)} & ... & $< -0.31$ & ... &
$> -0.61$ & $-0.79${\tiny (0.13)} & $-0.74${\tiny (0.12)} & $-0.44${\tiny (0.14)} & $-0.83${\tiny (0.16)} \\
PSS J2155$+$1358..... & 3.142 & 19.94{\tiny (0.10)} & $< -2.06$ & $< -2.05$ &
$-1.85${\tiny (0.12)} & ... & ... & ... &
$-2.10${\tiny (0.14)} & $-2.21${\tiny (0.21)} & ... & ... & ... \\
PSS J2155$+$1358$^c$... & 3.565 & 19.37{\tiny (0.15)} & ... & $-1.55${\tiny (0.20)} &
$-1.27${\tiny (0.16)} & ... & ... & ... &
$-1.70${\tiny (0.19)} & $< -2.40$$^*$ & ... & ... & ... \\
PSS J2155$+$1358..... & 4.212 & 19.61{\tiny (0.10)} & $-1.94${\tiny (0.13)} & $-2.18${\tiny (0.13)} &
$-1.92${\tiny (0.11)} & ... & ... & ... &
$-2.18${\tiny (0.20)} & $-2.18${\tiny (0.25)} & ... & ... & ... \\
PSS J2344$+$0342..... & 3.882 & 19.50{\tiny (0.10)} & ... & $-1.28${\tiny (0.13)} &
$-1.10${\tiny (0.12)} & ... & ... & ... &
$-1.37${\tiny (0.15)} & $< -1.98$$^*$ & ... & ... & ... \\
\hline
\multicolumn{14}{l}{{\footnotesize $^*$ 4~$\sigma$ upper limit (corresponding to non-detections).}} \\
\multicolumn{14}{l}{{\footnotesize $^c$ This system is very likely affected by high ionization corrections 
(see Section~\ref{ionization}).}} \\
\multicolumn{14}{l}{{\footnotesize Abundances relative to the solar values of Grevesse \& Sauval (1998).}}
\end{tabular}
}
\end{table}
\end{landscape}
%

%

\bsp

\label{lastpage} 

\end{document}